\documentclass{egpubl}
\usepackage{eg2026}
 
\STAR                   %

\CGFccby

\usepackage[T1]{fontenc}
\usepackage{dfadobe}  
\usepackage{dsfont}

\usepackage{cite}  %
\BibtexOrBiblatex
\electronicVersion
\PrintedOrElectronic
\ifpdf \usepackage[pdftex]{graphicx} \pdfcompresslevel=9
\else \usepackage[dvips]{graphicx} \fi

\usepackage{egweblnk} 

\usepackage{amsmath}                                        %
\usepackage{amssymb}                                        %
\usepackage{mathtools}
\usepackage{wrapfig}
\usepackage{xcolor}
\usepackage{svg}
\usepackage{enumitem}

\usepackage{float}                                          %
\usepackage{graphicx} 
\usepackage[labelfont=bf,textfont=it,labelsep=colon]{caption}             %
\usepackage{subcaption}                                     %
\usepackage{booktabs}                                       %
\usepackage{multirow}                                       %
\usepackage{pifont} 
\usepackage{tikzsymbols}
\usepackage[normalem]{ulem}
\usepackage{tabularx}
\definecolor{forestgreen}{rgb}{0.13, 0.55, 0.13}
\newcommand{\cmark}{{\color{forestgreen} \ding{51}}}%
\newcommand{\xmark}{{\color{red} \ding{55}}}%
\definecolor{teal}{rgb}{0.00, 0.50, 0.50}
\usepackage[capitalize]{cleveref}
\usepackage{bm}
\usepackage{adjustbox}

\crefname{section}{Sec.}{Secs.}
\Crefname{section}{Sec.}{Secs.}

\setkeys{Gin}{keepaspectratio}

\newcommand{\SUP}[0]{\textcolor{orange!90!black}{\textbf{S}}}
\newcommand{\UNSUP}[0]{\textcolor{green!50!black}{\textbf{U}}}
\newcommand{\NONE}[0]{\textcolor{violet}{\textbf{N}}}

\newcommand{\Energy}{\mathcal{E}}
\newcommand{\Loss}{\mathcal{L}}

\title[Non-Rigid 3D Shape Correspondences]{Non-Rigid 3D Shape Correspondences:~\\~From Foundations to Open Challenges and Opportunities}

\author[A.~Zhuravlev and L.~Bastian et al.]
{\parbox{\textwidth}{\centering\small
    A.~Zhuravlev$^{1 \star}$\orcid{0000-0003-4282-3702} \quad
    L.~Bastian$^{6,7,8\star}$\orcid{0000-0001-8088-3920} \quad
    D.~Cao$^{2,3}$\orcid{0000-0002-6505-6465} \quad
    N.~El Amrani$^{2,3}$\orcid{0009-0004-9961-2855} \quad
    P.~Roetzer$^{2,3}$\orcid{0009-0005-6698-6663} \quad
    V.~Ehm$^{7,8}$\orcid{0009-0009-0142-5442} \quad
    R.~Marin$^{7,8}$ \orcid{0000-0003-2392-4612} \quad
    H.~Nishizawa$^{4}$\orcid{0009-0009-0802-9047} \quad \\ 
    S.~Morishima$^{5}$\orcid{0000-0001-8859-6539} \quad
    C.~Theobalt$^{1}$\orcid{0000-0001-6104-6625} \quad
    N.~Navab$^{7,8}$\orcid{0000-0002-6032-5611} \quad
    D.~Cremers$^{7,8}$\orcid{0000-0002-3079-7984} \quad
    F.~Bernard$^{2,3}$\orcid{0009-0008-1137-0003} \quad
    Z.~Lähner$^{2,3}$\orcid{0000-0003-0599-094X} \quad
    V.~Golyanik$^{1}$\orcid{0000-0003-1630-2006} 
}
\\
\parbox{\textwidth}{\centering\footnotesize
    $^1$MPI for Informatics \quad
    $^2$University of Bonn \quad
    $^3$Lamarr Institute \quad
    $^4$Waseda University \quad
    $^5$Waseda Research Institute for Science and Engineering \quad \\
    $^6$Imperial College London \quad    
    $^7$Technical University of Munich \quad 
    $^8$Munich Center for Machine Learning \quad
    $^\star$Equal Contribution 
}
\\\\
\parbox{\textwidth}{\centering
    \url{https://nonrigid-shape-correspondences.github.io}
}
}

\begin{document}

\teaser{
  \centering
  \begin{subfigure}{0.32\linewidth}\centering
    \includegraphics[width=0.85\linewidth]{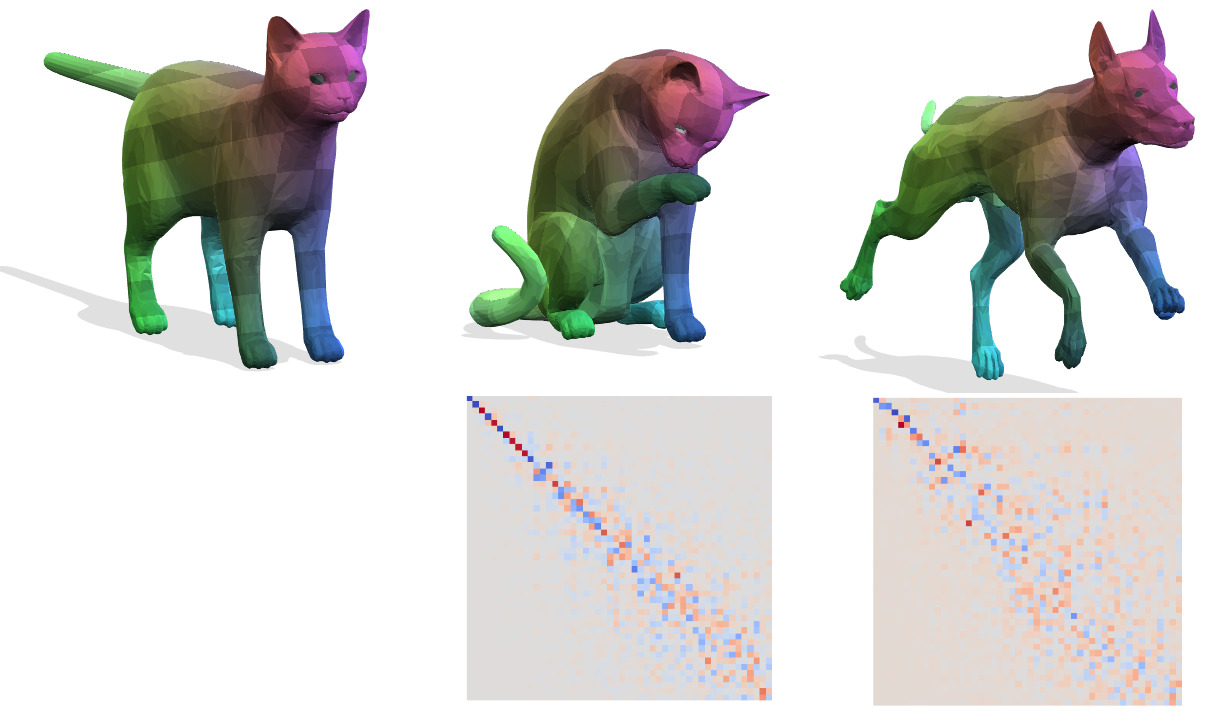}
  \end{subfigure}\hfill
  \begin{subfigure}{0.32\linewidth}\centering
    \includegraphics[trim={0 2cm 0 10cm}, width=0.6\linewidth]{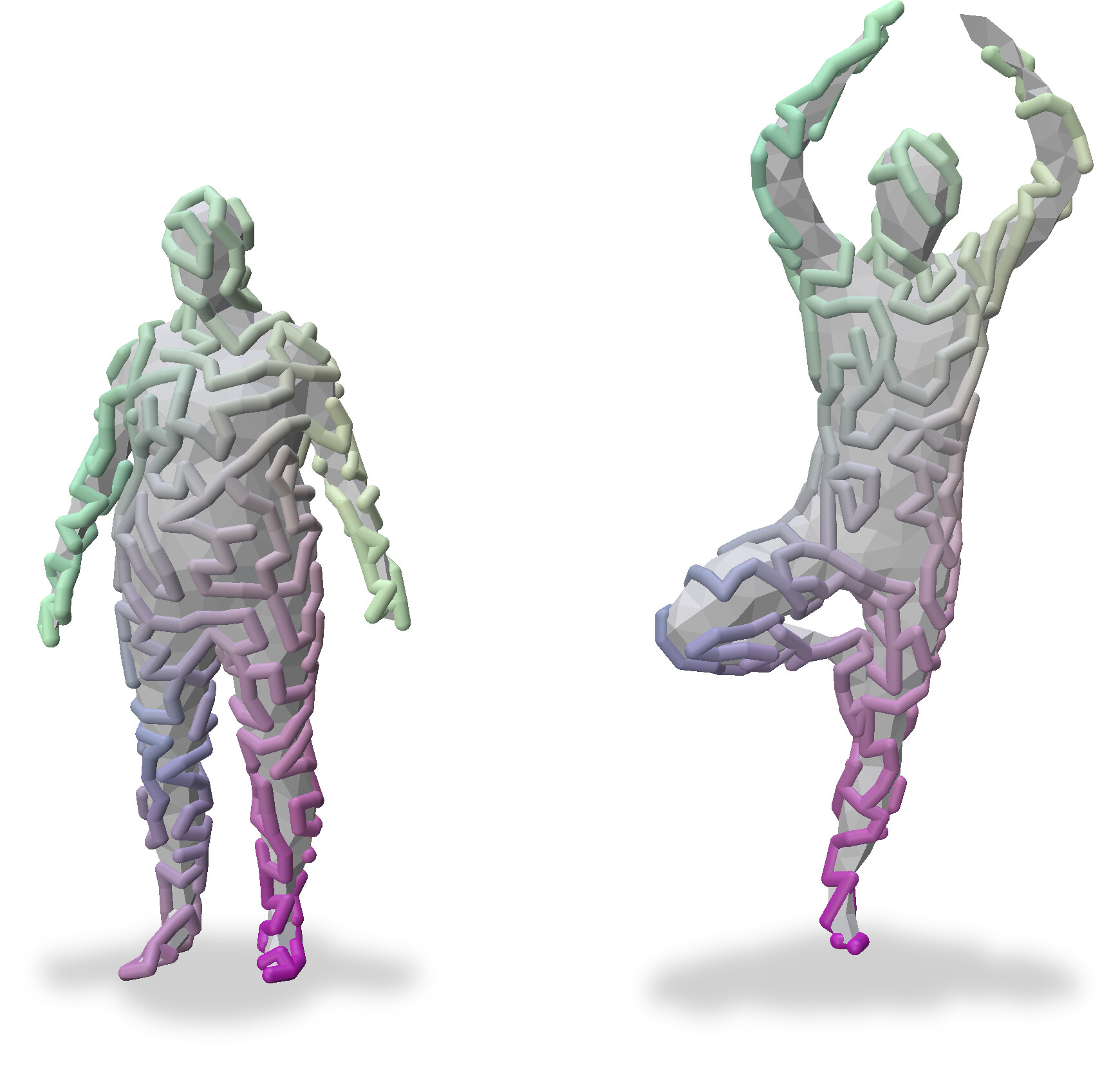}
  \end{subfigure}\hfill
  \begin{subfigure}{0.32\linewidth}\centering
    \includegraphics[trim=2cm 0cm 2cm 0.5cm, clip, width=0.85\linewidth]{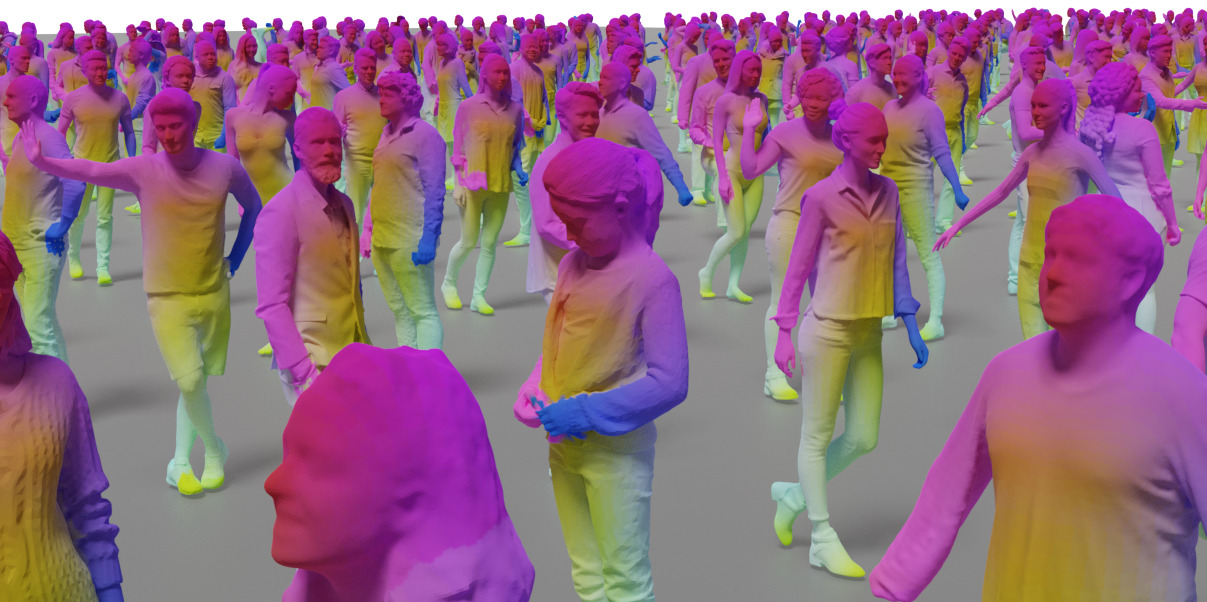}
  \end{subfigure}
\begin{tabularx}{\textwidth}{>{\centering\arraybackslash}X
                            >{\centering\arraybackslash}X
                            >{\centering\arraybackslash}X}
\textbf{Spectral methods} & \textbf{Combinatorial methods} & \textbf{Deformation-based methods} \\
\end{tabularx}
  \vspace{1.5mm}

  \begin{subfigure}{0.32\linewidth}\centering
    \includegraphics[width=0.85\linewidth]{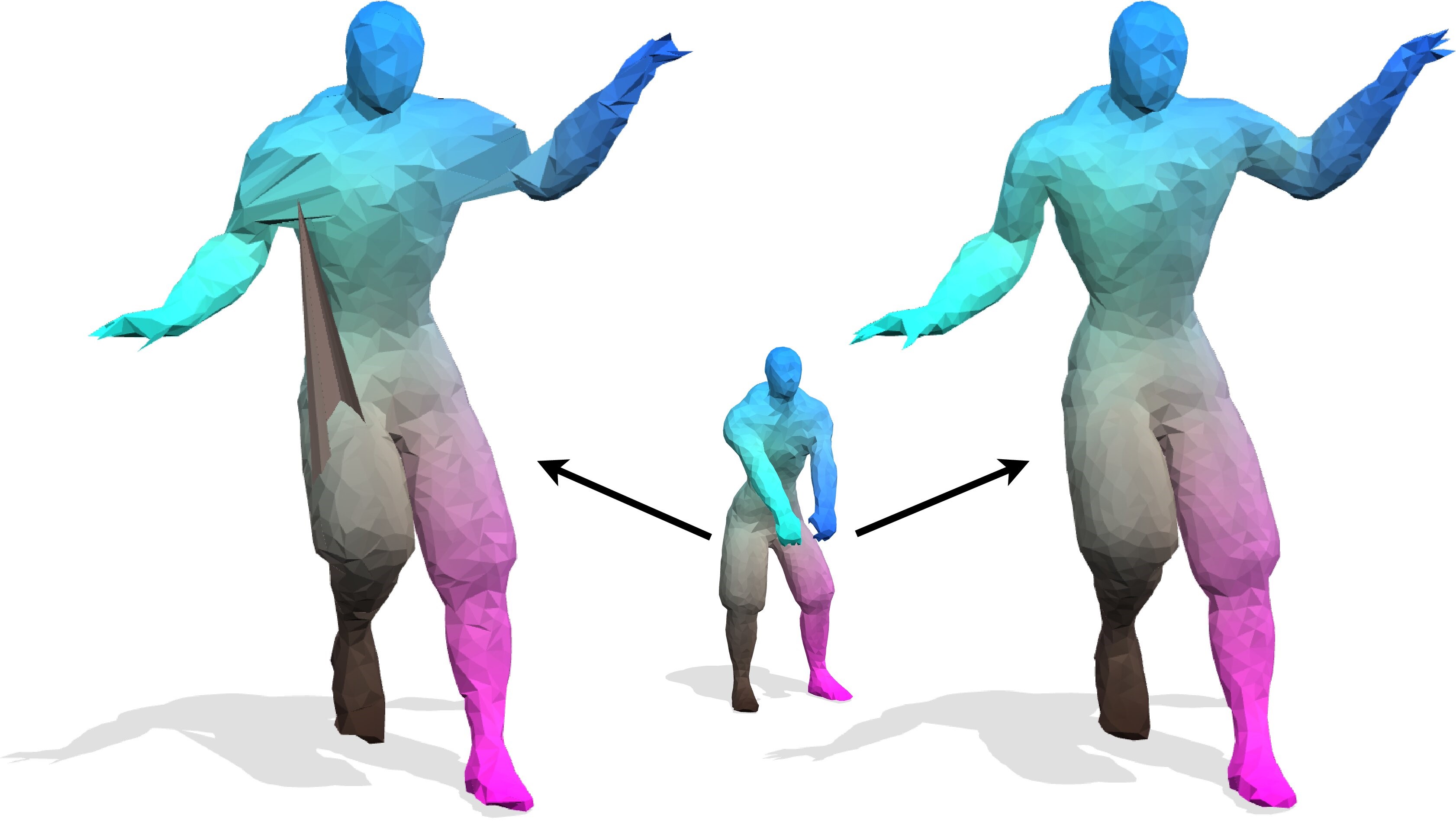}
  \end{subfigure}\hfill
  \begin{subfigure}{0.32\linewidth}\centering
    \adjincludegraphics[trim={{0.28\width} 0 {0.28\width} 0}, clip, width=0.85\linewidth]{figures/nafie/teaser_datasets_compressed}
  \end{subfigure}\hfill
  \begin{subfigure}{0.32\linewidth}\centering
    \includegraphics[width=0.85\linewidth
    ,clip
    ]{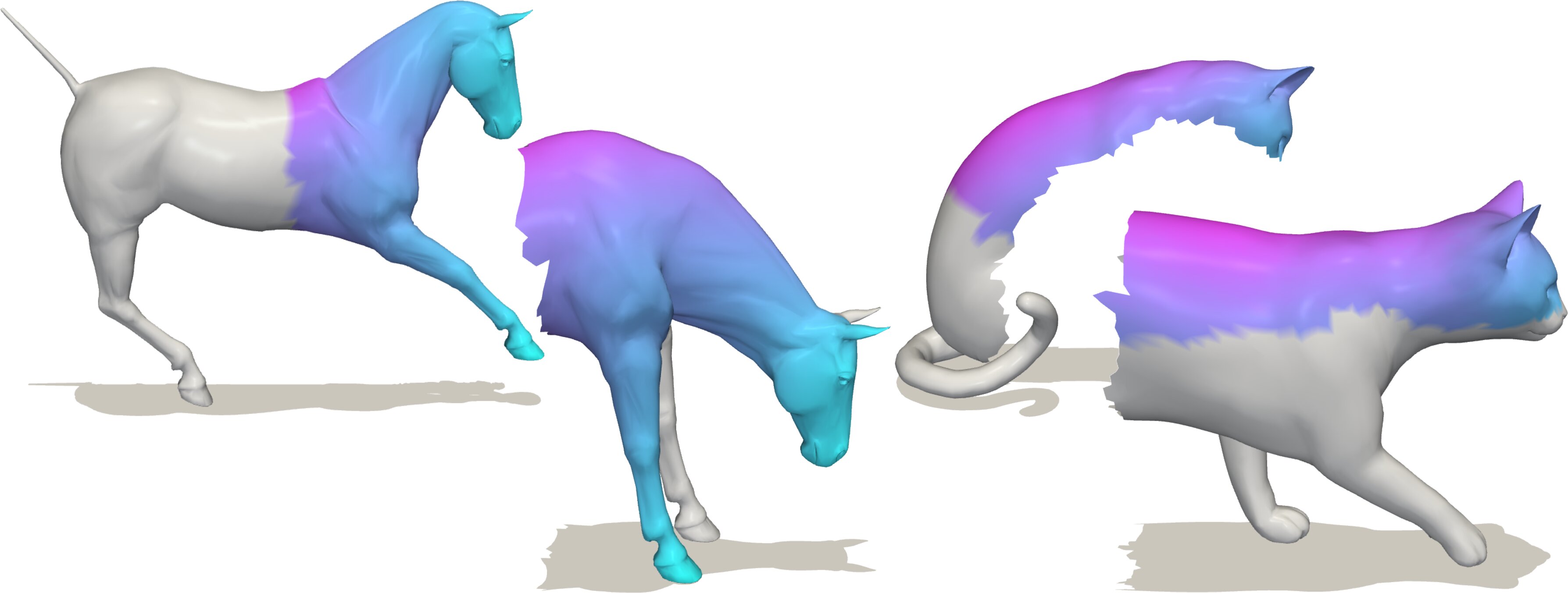}
  \end{subfigure}
  \begin{tabularx}{\textwidth}{>{\centering\arraybackslash}X
                            >{\centering\arraybackslash}X
                            >{\centering\arraybackslash}X}
\textbf{Geometric consistency} & \textbf{Datasets} & \textbf{Partial shape matching} \\
\end{tabularx}

  \caption{
    This STAR surveys modern approaches to non-rigid 3D shape correspondence, spanning spectral, combinatorial, and deformation-based methods. We further highlight emerging opportunities, such as zero-shot matching and open challenges involving partiality. 
    Image sources:~\cite{ehm2025beyond,marin2024nicp,cao_revisiting_2024}. 
  }
  \label{fig:teaser}
  \vspace{5mm}
}

\maketitle

\begin{abstract}
Estimating correspondences between deformed shape instances is a long-standing problem in computer graphics; numerous applications, from texture transfer to statistical modelling, rely on recovering an accurate correspondence map.
Many methods have thus been proposed to tackle this challenging problem from varying perspectives, depending on the downstream application.
This state-of-the-art report is geared towards researchers, practitioners, and students seeking to understand recent trends and advances in the field.
We categorise developments into three paradigms: spectral methods based on functional maps, combinatorial formulations that impose discrete constraints, and deformation-based methods that directly recover a global alignment.
Each school of thought offers different advantages and disadvantages, which we discuss throughout the report. Meanwhile, we highlight the latest developments in each area and suggest new potential research directions.
Finally, we provide an overview of emerging challenges and opportunities in this growing field, including the recent use of vision foundation models for zero-shot correspondence and the particularly challenging task of matching partial shapes. 
Project page available at \url{https://nonrigid-shape-correspondences.github.io}.


\end{abstract}

\section{Introduction} 
From character expression transfer to the study of variability in human anatomy, non-rigid shape correspondence is the cornerstone of numerous applications. 
It is an inescapable step when transferring or comparing information between 3D shapes, from texture to deformation, and segmentation to motion. 
At scale, correspondences across shape collections enable parametric shape models of various types, such as bodies, faces, hands, and organs \cite{bogo_faust_2014,loper2015smpl,romero2017mano,pavlakos2019smplx}. 
These models are indispensable for numerous downstream tasks: animation pipelines~\cite{igarashi2005rigid}, 3D reconstruction systems~\cite{loper2015smpl}, and morphometric analysis of organs~\cite{el_amrani_universal_2024} are all enabled by parametric models whose power comes from accurate pointwise correspondences. 
Developing non-rigid shape correspondence methods is, therefore, of fundamental importance and an active area of ongoing research.

But such an important task does not come without challenges. 
Data scarcity is a hindrance to development in the field, as geometric 3D acquisitions require specialised technologies, modelling is generally done by expert artists, and annotations necessitate extensive manual work, which can be too ambiguous for shape collections with large variations. 
In the rigid case, a correspondence model can be reduced to a simplified parametrisation with six degrees of freedom \cite{besl1992method}, but this does not hold for \textit{non-rigid deformations}, which are often too complex to obtain an analytical formulation \cite{vestner2017efficient}.

Even the widely studied near-isometric problem setting (e.g., humans simulated in different poses) still requires a deep understanding of geometrical structures to, for instance, distinguish intrinsic symmetries \cite{bogo_faust_2014,wang2025symmetry}. 
As structural differences grow, such as between animals of different species, so does the semantic ambiguity in the matching (how can one map the neck of a giraffe to that of a dog?) \cite{zuffi2017small}. 
Moving to real-world acquisitions, issues such as noise, holes, and partiality are common and severely degrade the performance even of the best available methods. 
Such issues continue to motivate scholars to develop robust, widely applicable shape correspondence tools.

Despite the longstanding nature of the shape correspondence problem, recent years have seen continued progress, even in classical settings such as matching two geometrically complete and clean meshes.
However, researchers have also explored new relevant directions such as extreme non-isometry, zero-shot matching, partiality, and topological changes. 
These challenges inspire new paradigms, such as the application of \textit{foundation model} features in 3D geometric contexts. 
Methods for \textit{partial-to-partial} and \textit{multi-shape matching} address more complex correspondence settings beyond the widely considered \textit{full-to-full} pairwise setting. 

Finally, the progress in non-rigid shape correspondence fields owes much to the parallel evolution of benchmarks. 
Early evaluations focused on synthetic near-isometric shape pairs with uniform meshing, while recent ones have shifted attention toward non-isometry, partiality, real-world noise, and irregular meshing, progressively closing the gap between laboratory and real-world settings.
The scope of shape correspondence is as wide as ever, with exciting application domains for interdisciplinary approaches.

\subsection{Scope of this STAR}

This state-of-the-art report is motivated by recent developments in shape correspondence, both in advancing classical foundations and introducing new paradigms.
Our primary goals are to: (i) summarise the main directions of current research in shape correspondence, and (ii) highlight open problems and future avenues of research.
Our review mainly focuses on works published from 2022 until 2025, including earlier works for background and historical context.

Our report primarily focuses on correspondence problems for 3D meshes, which remain the dominant representation in computer graphics. 
Meshes incorporate rich geometric structure into their surface representation, which has inspired numerous innovative methods that leverage the wealth of knowledge in geometry processing and numerical optimisation.
While point clouds lack such structural information, a review of shape correspondence would be incomplete without mentioning several recent influential approaches as well. Registration of neural implicit representations~\cite{park2019deepsdf,mescheder2019occupancy} is out of our scope in this paper.
Moreover, we do not consider rigid correspondence, as the complexity is vastly reduced when considering transformation models with only six degrees of freedom. 
Image-based feature matching is also beyond the scope of this report; we refer the readers to~\cite{ma2021image} for an overview.

\subsection{Related surveys}  

Previous surveys on shape correspondence have taken various perspectives.
\cite{sahillioglu_recent_2020} covers methods published between 2011 and 2019, many of which provide the foundation for approaches discussed in our report.
The more recent survey reviews correspondence methods up to 2021, with an emphasis on techniques applicable for dynamic shape acquisition and reconstruction \cite{deng_survey_2022}.
We build on both, extend the timeline, introduce novel paradigms, and cover directions that were not fully addressed, such as combinatorial methods.
\cite{van2011survey} summarises shape correspondence methods developed before 2011.
An earlier review on functional maps also offers an excellent theoretical primer~\cite{ovsjanikov_computing_2016}, but focuses on pre-2016 methods and thus excludes the deep learning approaches that dominate the recent literature.

\subsection{Structure}
This STAR is organised as follows.
In \cref{sec:fundamentals}, we introduce the mathematical foundations and problem setting of non-rigid 3D shape correspondence.
The recent body of methods is primarily covered in the next three sections:
\cref {sec:spectral} reviews spectral methods, built on the functional map framework; \cref {sec:combinatorial} covers combinatorial approaches that formulate correspondence as a discrete optimisation problem. 
\cref{sec:deformation} focuses on deformation-based methods that align shapes through spatial transformations.

Following the main body of works, we discuss recent emerging trends that expand the scope of correspondence beyond classical settings in~\cref{sec:emerging}.
Suitable datasets and benchmarks, including recently published ones, are summarised in \cref{sec:datasets}. 
Finally, we outline practical applications of shape correspondence in \cref{sec:applications}, highlight open challenges and discuss future research directions in \cref{sec:challenges}.

\section{Background \& fundamentals}
\label{sec:fundamentals}

This section provides an overview of the shape correspondence problem setting. 
We begin by discussing how surfaces can be characterised locally using \textit{feature descriptors}, which is essential for estimating correspondence between shapes.
These descriptors serve as an initial, local similarity measure but are often insufficient on their own due to ambiguity and limited discriminative power.
Additional \textit{structural priors} can be introduced to constrain the solution space and improve the faithfulness of the recovered correspondences.
These priors provide the conceptual foundation for the three methodological categories discussed in later sections.
However, no type of structural prior has emerged as a silver bullet; each has its \textit{limitations} and is best suited to specific settings.
The notation we adopt throughout this report is summarised in Table~\ref{tab:notation}.

\begin{table}[t]
\centering
\footnotesize
\begin{tabularx}{\linewidth}{lll}
\toprule
\textbf{Symbol} & \textbf{Domain} & \textbf{Description} \\
\midrule
$\mathcal{M}, \mathcal{N}$ & & 3D shapes with $n_\mathcal{M}$ and $n_\mathcal{N}$ vertices \\
$X_\mathcal{M}$ & $\mathbb{R}^{n_\mathcal{M} \times 3}$ & Vertex positions of $\mathcal{M}$ \\
$E_\mathcal{M}$ & $\subseteq \{1,\dots,n_\mathcal{M}\}^2$ & Edges of $\mathcal{M}$ \\
$T_\mathcal{M}$ & $\subseteq \{1,\dots,n_\mathcal{M}\}^3$ & Faces (triangles) of $\mathcal{M}$ \\
$f_\mathcal{M}$ & $\mathbb{R}^{n_\mathcal{M} \times d}$ & Vertex-wise features on $\mathcal{M}$ \\
$L_\mathcal{M}$ & $\mathbb{R}^{n_\mathcal{M} \times n_\mathcal{M}}$ & Discrete Laplace–Beltrami operator on $\mathcal{M}$ \\
$\phi_\mathcal{M}$ & $\mathbb{R}^{n_\mathcal{M} \times k}$ & Eigenfunctions of $L_\mathcal{M}$ (spectral basis) \\
$C_{\mathcal{M} \mathcal{N}}$ & $\mathbb{R}^{k \times k}$ & Functional map from $\mathcal{M}$ to $\mathcal{N}$ \\
$\Pi_{\mathcal{M} \mathcal{N}}$ & $\{0,1\}^{n_\mathcal{M} \times
n_\mathcal{N}}$ & Point-wise map from $\mathcal{M}$ to $\mathcal{N}$ \\
$\Energy(\cdot)$ & & Energy or cost function to be minimized \\
$\Loss(\cdot)$ & & Loss for neural network training \\
\bottomrule
\end{tabularx}
\caption{Summary of \textbf{notation} used throughout the report.}
\label{tab:notation}
\end{table}

\subsection{The shape correspondence problem}

In the context of 3D geometry processing, a shape is typically defined as a two-dimensional surface embedded in three-dimensional space. 
Such a surface is often discretised as a triangular mesh, consisting of vertices \( X_\mathcal{M} \), edges \( E_\mathcal{M} \), and faces \( T_\mathcal{M} \). 
When a surface triangulation is not given, the vertices \( X_\mathcal{M} \) can be considered as an unstructured point cloud.
Given two shapes, denoted \( \mathcal{M} \) and \( \mathcal{N} \), the goal of shape correspondence is to determine which points on \( \mathcal{M} \) match semantically or geometrically to points on \( \mathcal{N} \). 
This relationship can be expressed as a point-wise map \( \Pi_{\mathcal{M} \mathcal{N}}: \mathcal{M} \rightarrow \mathcal{N} \), assigning to each point on the source shape a corresponding point on the target shape.

Throughout the report, we use the terms \textit{correspondence} and \textit{matching} interchangeably.
We focus primarily on the dense correspondences setting, which establishes a matching for every point on the source shape to the target shape (formally, we consider $n_{\mathcal{M}} \leqslant n_{\mathcal{N}}$ with $\Pi_{\mathcal{M}\mathcal{N}}$ encoding an injective map, i.e., $\Pi_{\mathcal{M}\mathcal{N}} \mathds{1} = \mathds{1}$ and $\Pi_{\mathcal{M}\mathcal{N}}^\top \mathds{1} \leqslant \mathds{1}$, where we denote the all-ones vector as $\mathds{1}$).
Sparse, landmark-based approaches can also be useful in certain applications, specifically as they can be annotated more easily than dense maps (see \cref{sec:combinatorial}).
The deformation-based family of works also considers the registration settings, where a deformation field is recovered in the ambient domain without first matching points locally (see \cref{sec:deformation}).
When two shapes are aligned in this manner, dense correspondence maps can typically be recovered via a heuristic.

\begin{figure}[t]
  \centering

  \begin{subfigure}{0.98\linewidth}
    \centering
    \includegraphics[width=0.36\linewidth]{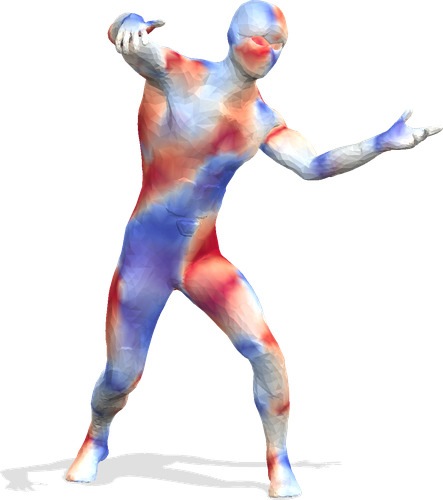}
    \includegraphics[width=0.36\linewidth]{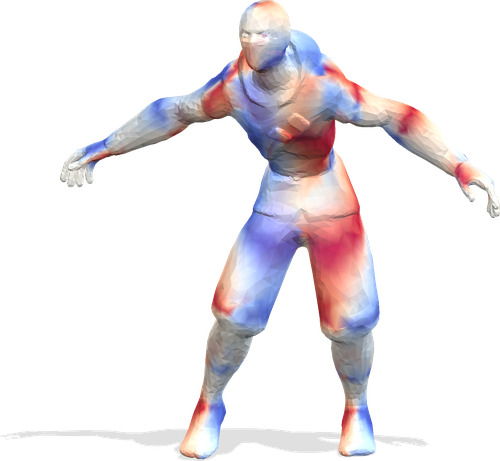}
    \caption{Feature functions on source and target shape.}
    \label{fig:02_features}
  \end{subfigure}
  \hfill
  \vspace{2mm} %

  \begin{subfigure}{0.6\linewidth}
    \centering
    \includegraphics[width=\linewidth]{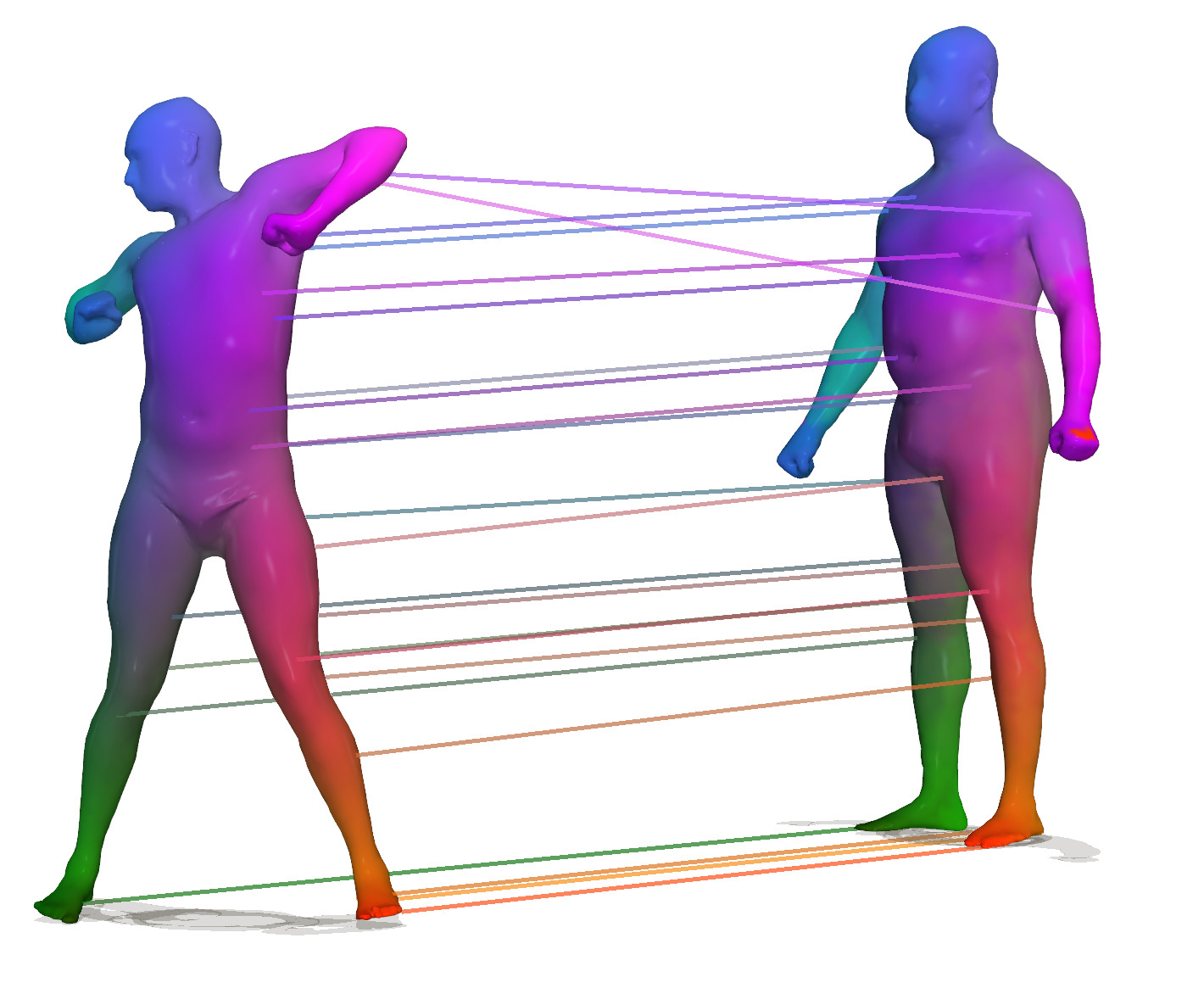}
    \caption{Visualised correspondences.}
    \label{fig:02_matches}
  \end{subfigure}

  \caption{
    \textbf{Visualising} and \textbf{matching features} between shapes. 
    \textbf{(a)} Per-vertex feature functions are computed on two non-rigidly deformed shapes visualised by colour. %
    When these features are sufficiently descriptive (as evidenced by similar regions coloured identically), a correspondence mapping can be derived.
    \textbf{(b)} Dense correspondences (visualised by transferring colours from one to the other shape and by connecting lines) computed between two non-rigidly deformed shapes, demonstrating geometric and semantic alignment.
    Image source:~\cite{gao_sigma_2023}.
    }
  \label{fig:fundamentals}
\end{figure}

\subsection{Characterising a shape's surface}

A central goal in correspondence estimation is to quantify the similarity between shapes in a tractable form. 
Given shapes \( \mathcal{M} \) and \( \mathcal{N} \), per-vertex feature functions \( f_\mathcal{M} \in \mathbb{R}^{n_\mathcal{M} \times d} \) and \( f_\mathcal{N} \in \mathbb{R}^{n_\mathcal{N} \times d} \), also referred to as descriptors, can be specified based on a heuristic or estimated in a data-driven manner. 
These descriptors are intended to encode geometric or semantic properties of the shapes, see Fig.~\ref{fig:02_features} for an illustration. 
Then, for any point on shape \( \mathcal{M} \) and a ground-truth point-wise map \( \Pi_{\mathcal{M}\mathcal{N}} \in \{0,1\}^{n_\mathcal{M} \times n_\mathcal{N}} \), the descriptor preservation principle assumes that the corresponding point on \( \mathcal{N} \) should have similar descriptors:
\begin{equation}
\label{eq:descr_preservation}
    f_\mathcal{M} \approx \Pi_{\mathcal{M}\mathcal{N}} f_\mathcal{N}.
\end{equation}

Several types of descriptors have been proposed for both structured and unstructured shape representations. 
They can be broadly categorised by their complexity and level of abstraction.

The most basic descriptors are derived directly from \emph{geometric properties} of the shapes themselves. 
These include both extrinsic properties, such as 3D vertex coordinates~\cite{tombari2010unique} or surface normals~\cite{salti2014shot}, and intrinsic properties, such as local curvature~\cite{meyer_discrete_2003,windheuser2011geometrically}.
While the resulting \textit{hand-crafted} descriptors capture local geometric information, they each have inherent limitations. 
Extrinsic descriptors such as SHOT \cite{tombari2010unique} characterise surfaces through local coordinate systems but are not isometrically invariant. 
Conversely, intrinsic descriptors such as the Heat and Wave Kernel Signatures \cite{sun2009concise,aubry2011wave} naturally inherit the Laplace-Beltrami operators' isometric invariance but are not symmetry aware, can change drastically under non-isometric deformations, and remain sensitive to noise.
Despite these limitations, these descriptors can still yield effective matches when combined with strong geometric constraints on the point-wise map~\cite{windheuser2011geometrically}.
Moreover, they are easily applicable in various domains, in contrast to the more discriminative learned descriptors.

Recent approaches increasingly rely on \emph{learning} geometric descriptors by optimising for correspondence on collections of shapes~\cite{litany2017deep}. 
These models can transform raw geometric inputs (e.g. vertex positions) or axiomatic descriptors (e.g. HKS, WKS, SHOT) to more expressive, domain-specific features \cite{sharp2022diffusionnet, saleh2020graphite}.
Learned descriptors can overcome many of the limitations of hand-crafted features and adapt to challenging non-isometric or partial matching scenarios.
In practice, however, learning sufficiently discriminative feature descriptors remains challenging.
Surface ambiguities, noise, symmetries, shape partiality, and discretisation artefacts can yield descriptors with numerous plausible matching candidates on the target shape.

While obtaining informative descriptors is essential for solving the correspondence problem, recovering a point-wise map (visualised in Fig.~\ref{fig:02_matches}) is non-trivial even from descriptors learned by state-of-the-art models.
A naive approach to obtain a correspondence map would be to match each point on shape \( \mathcal{M} \) to the most similar descriptor on shape \( \mathcal{N} \) via nearest-neighbour search:
\begin{equation}\label{eq:matching-intro}
\underset{\Pi_{\mathcal{M}\mathcal{N}} \in \{0, 1\}^{n_{\mathcal{M}} \times n_{\mathcal{N}}}}{\arg \min} \| f_\mathcal{M} - \Pi_{\mathcal{M}\mathcal{N}} f_\mathcal{N} \|_F^2. 
\end{equation}

\noindent However, this formulation imposes no constraints on the resulting map. 
Inaccuracies or ambiguities in the feature descriptors representation can lead to errancies: for example, neighbouring points on \( \mathcal{M} \) may be mapped to distant locations on \( \mathcal{N} \), yielding an undesirable \textit{geometric inconsistency}. 
To overcome this, many recent methods impose additional assumptions or structural priors to obtain more accurate correspondences.

\subsection{Methodological Overview}

Solving the matching problem in \cref{eq:matching-intro}, while straightforward, typically yields point-wise maps with errant feature pairings.
To address this, correspondence optimisation is typically augmented with prior knowledge about the structure and geometry of the shapes, constraining solutions to more plausible point-wise maps. 
We group recent developments into three main categories:

\paragraph{Spectral methods.} 
A common assumption is that the desired map is smooth and can be compactly represented in a low-frequency basis.
The resulting insight that such a smooth map can be represented in a low-frequency function space has led to the rise of the seminal \textit{functional maps} framework~\cite{ovsjanikov_functional_2012}.
Most commonly used to construct this function space are the first $k$ eigenfunctions of the Laplace-Beltrami operator (LBO), which form an orthonormal basis that captures the intrinsic geometry of the surface and remains invariant under isometric deformations~\cite{belkin2003laplacian}.
The key insight of the functional map framework is that instead of directly matching points between shapes, one can align the eigenfunctions $\phi = [\phi_1, \phi_2, \ldots, \phi_k] \in \mathbb{R}^{n \times k}$ for a pair of shapes $\mathcal{M}$ and $\mathcal{N}$ in spectral space through a low-rank linear map $C_{\mathcal{MN}} \in \mathbb{R}^{k \times k}$. 
Given eigenfunctions $\phi_{\mathcal{M}}$ and $\phi_{\mathcal{N}}$ on shapes $\mathcal{M}$ and $\mathcal{N}$, and a pointwise correspondence $\Pi_{\mathcal{NM}}$, the relationship between them is captured as:
\begin{equation}
\label{eq:intro-approx-fm}
\Pi_{\mathcal{NM}}\phi_{\mathcal{M}} \approx \phi_{\mathcal{N}}C_{\mathcal{MN}}.
\end{equation}
\noindent Intuitively, this aligns the spectral bases between shapes by transferring an indicator function from shape $\mathcal{M}$ through the point-wise map, yielding a function on $\mathcal{N}$ whose spectral coefficients are determined by the functional map $C_{\mathcal{MN}}$.

The LBO eigenfunctions are famously invariant under non-isometric deformation, and robust even under mild deviations from isometry \cite{reuter2006laplace}. 
Numerous recent works leverage the insight that shape descriptors, whether hand-crafted or learned, can be projected onto the basis functions to establish correspondences in spectral space.
The functional map $C$ naturally regularises the point-wise map due to its low dimensionality and the smoothness of the LBO eigenfunctions.
Structural priors have also been derived for $C_{\mathcal{MN}}$, encouraging desirable properties such as promoting approximately isometric mappings or bijectivity~\cite{roufosse2019unsupervised}.
The use of such geometric priors has led to unsupervised paradigms that enable state-of-the-art pipelines to learn robust and discriminative feature descriptors from collections of shapes~\cite{cao_unsupervised_2023}.

However, the low-rank spectral representation inherently compresses the recovered map, limiting the ability to represent high-frequency geometric details.
Moreover, the quality of the LBO eigenfunctions degrades in the presence of non-isometries \cite{bastian2024hybrid} and holes or partiality~\cite{ehm2025beyond}, and extracting a faithful point-wise map from a functional map remains a non-trivial step that can introduce additional error \cite{vigano_adjoint_2023}.
These trade-offs are discussed in detail in \cref{sec:spectral}.

\paragraph{Combinatorial methods.}
Solving the naive nearest-neighbour matching in \cref{eq:matching-intro} yields an initial point-wise map, but imposes no structure on the solution.
A natural extension is the Linear Assignment Problem (LAP) which considers recovering a bijective mapping (i.e., $n_{\mathcal{M}} = n_{\mathcal{N}}$, obtainable through re-meshing or subsampling) that minimises a linear dissimilarity cost $D \in \mathbb{R}^{n_{\mathcal{M}} \times n_{\mathcal{N}}}$ over the set of permutation matrices $\mathbb{P}$:
\begin{equation}\label{eq:intro-lap}
\underset{\Pi \in \mathbb{P}}{\arg \min} \; \langle D, \Pi \rangle \quad \text{where, e.g.} \quad D_{ij} = \| f_\mathcal{M}(i) - f_\mathcal{N}(j) \|^2.
\end{equation}
\noindent This formulation has been extensively studied and can be solved efficiently through the Auction algorithm~\cite{bertsekas1992auction}.
Nevertheless, it still relies entirely on feature similarity and neglects neighbourhood relationships between features.
This can lead to \textit{geometrically inconsistent} matchings where neighbouring points on $\mathcal{M}$ map to distant locations on $\mathcal{N}$.

On the other hand, the Quadratic Assignment Problem (QAP) incorporates a metric that is preserved between matching pairs of points (geodesic distances are commonly chosen) but is famously NP-hard~\cite{rendl1994quadratic}, making solutions computationally intractable for most meshes of practical size.
Many recent combinatorial shape matching methods, therefore, seek a middle ground, incorporating geodesic constraints by exploiting the surface's discrete structure while avoiding the full quadratic cost.
Recent approaches have improved scalability through convex relaxations~\cite{kushinsky2019sinkhorn,droge_kissing_2023}, proposed heuristics~\cite{vestner2017efficient, haller2022comparative, roetzer_scalable_2022}, or additional constraints like neighbourhood preservation~\cite{windheuser2011geometrically,ehm_geometrically_2024,roetzer_spidermatch_2024}.

Despite these scalability challenges, the combinatorial family of methods discussed in \cref{sec:combinatorial} has recently gained traction, as they allow enforcing that solutions have certain geometric properties by constraining the feasible set.
Beyond neighbourhood preservation, this also extends to cycle consistency across shape collections~\cite{bhatia_ccuantumm_2023,kahl2025towards}.
However, combinatorial methods require reasonably discriminative feature representations as input and are typically non-differentiable, thus serving as a complement rather than a replacement to learned methods.

\paragraph{Deformation-based methods.} This class of methods directly recovers a global deformation in the ambient embedding space without first decomposing the problem into one of matching local elements.
In general, one may seek to recover a global deformation field $\mathcal{F}: \mathbb{R}^3 \rightarrow \mathbb{R}^3$ that aligns the source shape $\mathcal{M}$ to the target shape $\mathcal{N}$:
\begin{equation}
\mathcal{F}(X_{\mathcal{M}}) \approx {\Pi_{\mathcal{MN}}}X_{\mathcal{N}}. 
\end{equation}
Pioneered by iterative closest point (ICP) \cite{besl1992method} and coherent point drift (CPD) \cite{myronenko2010point}, this class of methods remains widely applicable in settings where obtaining discriminative feature descriptors is challenging.
As the general formulation is highly ill-posed, the deformation field $\mathcal{F}$ is typically regularised with priors that induce spatial smoothness, or require that it is locally rigid.
Recent methods have emphasised more robust regularisation techniques~\cite{yao2023fast,yao2025spare}, parametrising the deformation with neural networks~\cite{li2022non,sundararaman_deformation_2024}, and the incorporation of parametric templates into the optimisation or learning problem~\cite {Feng_2023_ICCV,li2025etchgeneralizingbodyfitting}.

The deformation-based methods discussed in \cref{sec:deformation} operate in the ambient Euclidean space and are thus broadly applicable to various 3D data representations, including unstructured point clouds, without requiring manifold connectivity or consistent topology.
This extrinsic paradigm is particularly suited to settings where noise, topological changes, or incomplete data pose challenges for methods requiring clean surface geometry.
Recent works have further extended the framework with robust norms~\cite{yao2023fast}, parametric templates~\cite{Feng_2023_ICCV,li2025etchgeneralizingbodyfitting}, and learned deformation priors~\cite{li2022non,marin2024nicp}.
However, the joint recovery of correspondence and deformation is inherently ill-posed, and alternating optimisation schemes inherited from ICP~\cite{besl1992method} remain sensitive to initialisation and prone to local minima.

\subsection{Challenges across matching scenarios}

Choosing a suitable correspondence method for a given application largely depends on the characteristics of the shapes being matched. 
As shape correspondence methods are applied to an increasingly wide range of geometric inputs, no single approach has remained universally applicable. 
In practice, the assumptions underlying a given method may hold only approximately, or be entirely violated, depending on the specific matching scenario. 
Therefore, it is essential to consider the nature of the shapes, particularly their degree of similarity, partiality, and the availability of training data (with or without annotations), when deciding on the appropriate methodology. 

A key factor is the nature of the deformation. 
When the shapes are represented as water-tight meshes and near-isometric, as in cases of simulated pose variation or mild articulation, most structural and spectral assumptions of matching methods remain valid. 
In contrast, non-isometric scenarios (e.g., matching across different animal species) are more challenging.
As geometric similarity decreases, the availability of sufficiently discriminative features decreases as well, requiring stronger constraints in correspondence optimisation, such as strictly preserving local neighbourhoods.
Especially difficult non-isometric cases occur when shapes deviate from idealised, clean, and manifold conditions. 
Real-world 3D data often contains topological noise, including non-uniform sampling, small holes, or self-intersections. 
Such violations of smooth manifold structure can compromise correspondence methods that depend on local geometric information, resulting in unstable or inconsistent matches between shapes.

A further challenge arises in the case of \textit{partiality}, where one or both shapes are incomplete.
In full-to-full matching, it is generally assumed that every point of one shape has a corresponding partner on the other.
In the case of partial shape matching, this assumption is often grossly violated.
In the partial-to-full matching scenario, a mapping can be defined from the partial shape to the full one, but not the other way round, since not every region of the full shape is observed in the partial shape.
This becomes even more challenging in the partial-to-partial case, where both shapes are incomplete. 
Here, it is necessary not only to establish correspondences but also to identify the overlapping region between the two shapes.
These difficulties demand the development of more sophisticated matching strategies, which we emphasise in Section~\ref{sec:partiality} as the problem is largely unresolved.

\begin{figure*}[t]
  \centering
  \begin{subfigure}[t]{0.32\textwidth}
    \centering
    \includegraphics[width=0.80\linewidth]{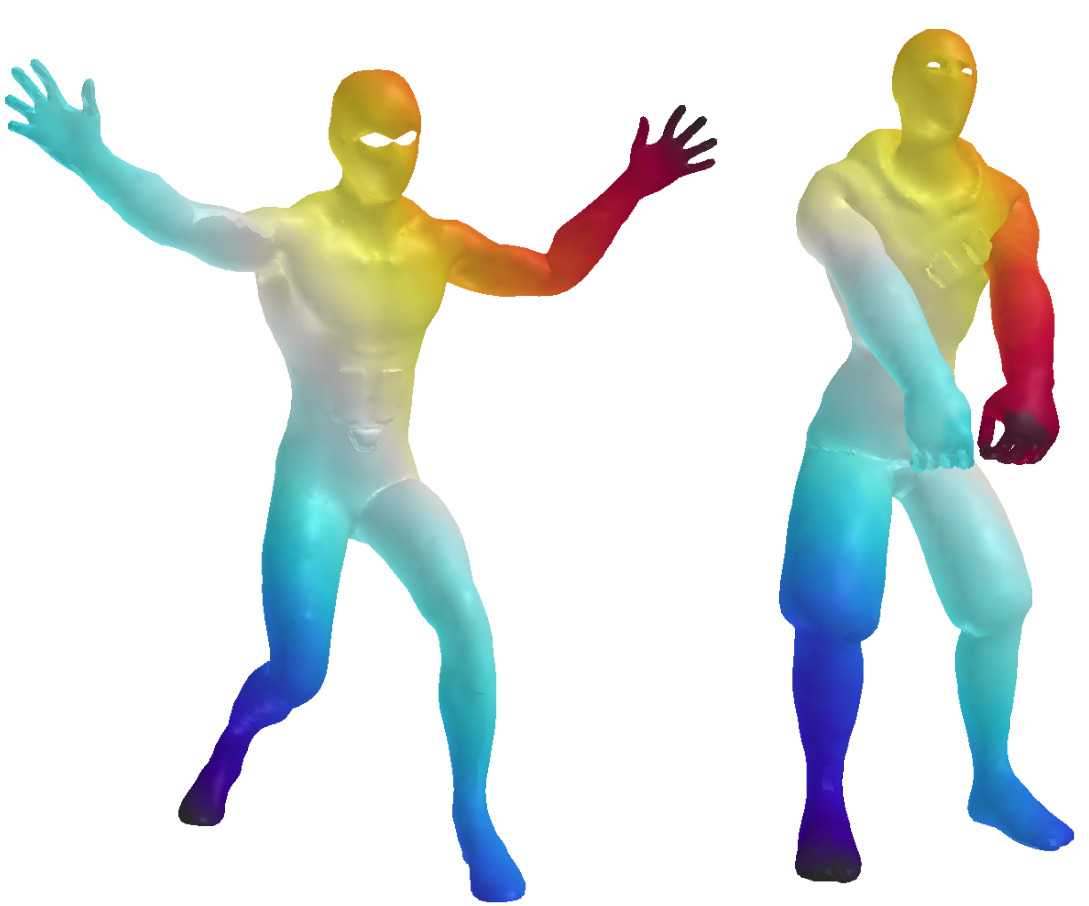}
    \vfill
    \caption{Color transfer}
    \label{fig:vis_color}
  \end{subfigure}
  \hfill
  \begin{subfigure}[t]{0.32\textwidth}
    \centering
    \includegraphics[width=0.75\linewidth]{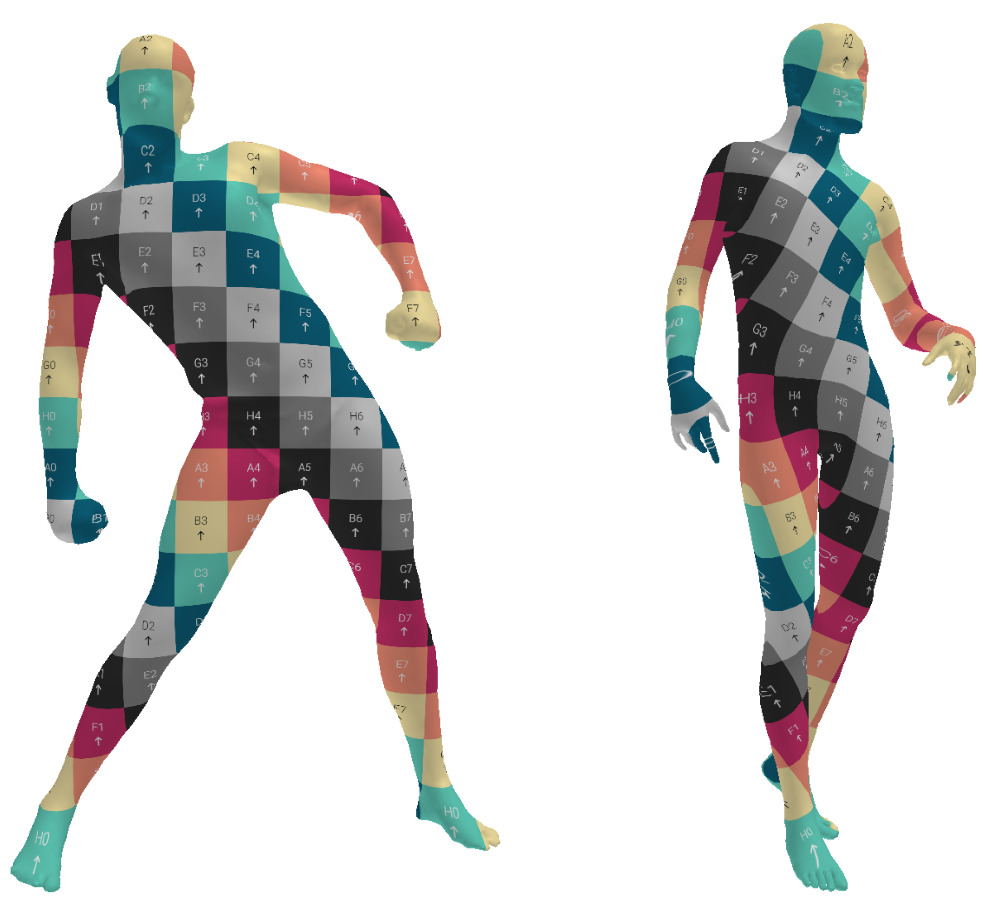}
    \vfill
    \caption{Texture transfer}
    \label{fig:vis_texture}
  \end{subfigure}
  \hfill
  \begin{subfigure}[t]{0.32\textwidth}
    \centering
    \includegraphics[width=0.60\linewidth]{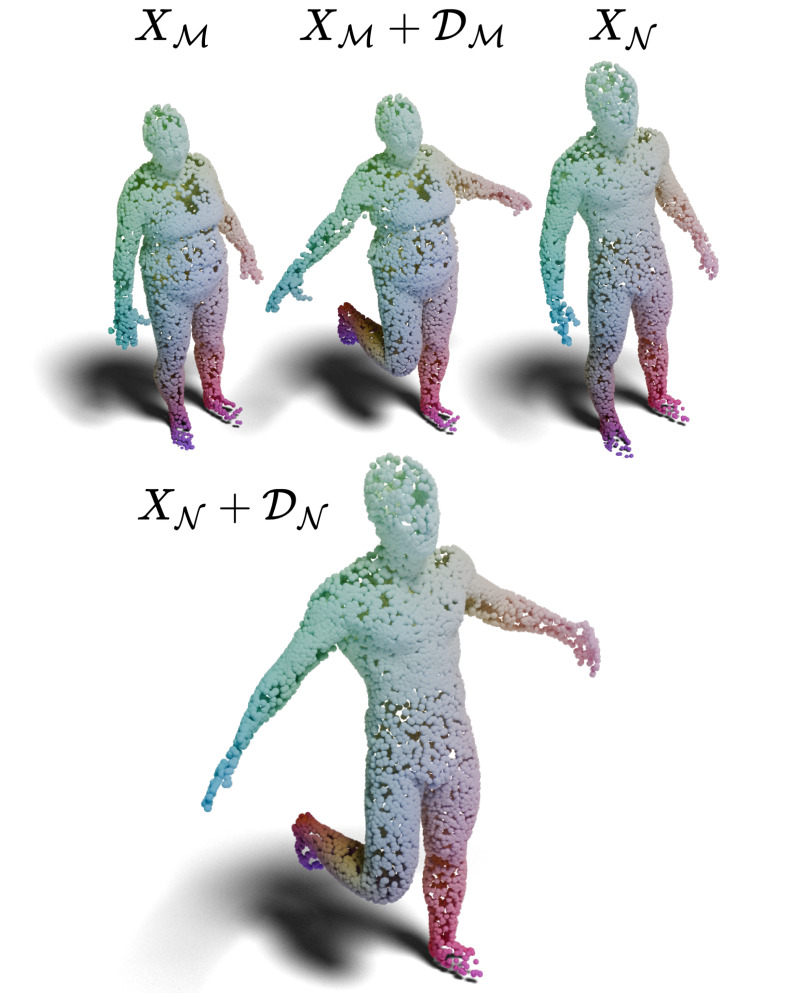}
    \caption{Deformation transfer}
    \label{fig:vis_deformation}
  \end{subfigure}

  \caption{
    Qualitative visualisation of \textbf{correspondence quality} using different transfer strategies.
    \textbf{(a)} Color transfer provides an overview of the global correspondence structure.
    \textbf{(b)} High-frequency texture transfer can reveal small local misalignments.
    \textbf{(c)} Deformation transfer can similarly visualise small topological inconsistencies caused by inaccurate correspondences. 
    Figures adapted from~\cite{zhuravlev_denoising_2025,vigano2025nam}.
    }
  \label{fig:02:correspondence_visualisation}
\end{figure*}

\subsection{Correspondence evaluation}
\label{sub_sec:eval_metrics}

Once a correspondence map $\Pi_{\mathcal{M}\mathcal{N}} \in \{0,1\}^{n_\mathcal{M} \times n_\mathcal{N}}$ between a pair of shapes is recovered, it can be evaluated through various metrics.

\paragraph{Accuracy.}
When working with standard synthetic datasets, ground-truth correspondence matrices
$\Pi_{\mathcal{M}\mathcal{N}}^*$ are generally available.
This allows one to directly evaluate the accuracy of the predicted map  $\Pi_{\mathcal{M}\mathcal{N}}$.
Let $D_\mathcal{N} \in \mathbb{R}^{n_\mathcal{N} \times n_\mathcal{N}}$ be the pairwise
geodesic distance matrix on the target shape $\mathcal{N}$, and
$\operatorname{Area}(\mathcal{N})$ the total surface area of the target shape.
Then, the vector of normalised geodesic errors over all vertices in $\mathcal{M}$
is defined as~\cite{kim2011blended}:

\begin{equation}
    e = 
    \frac{1}{\sqrt{\operatorname{Area}(\mathcal{N})}}
    \cdot 
    \operatorname{diag}\!\left(
        \Pi_{\mathcal{M}\mathcal{N}} \,
        D_\mathcal{N} \,
        \Pi_{\mathcal{M}\mathcal{N}}^{*\top}
    \right).
\end{equation}

\noindent
Here, 
$\Pi_{\mathcal{M}\mathcal{N}} \, D_\mathcal{N} \, \Pi_{\mathcal{M}\mathcal{N}}^{*\top} \in \mathbb{R}^{n_\mathcal{M} \times n_\mathcal{M}}$
contains the geodesic distances between the predicted and ground-truth correspondences for each source vertex, 
and $\operatorname{diag}(\cdot)$ extracts diagonal entries.

From the vector of normalised geodesic errors $e \in \mathbb{R}^{n_\mathcal{M}}$, 
three standard evaluation metrics are typically computed.
\textit{Mean geodesic error (MGE)} over all source vertices is defined as:
\begin{equation}
    \mathrm{MGE} = \frac{1}{n_\mathcal{M}} \sum_{i=1}^{n_\mathcal{M}} e(i).
\end{equation}

\noindent
\textit{Percentage of correct keypoints (PCK)} measures the fraction of correspondences whose normalised geodesic error is below a threshold $\tau > 0$. 
Using $\mathbf{1}[\cdot]$ as the indicator function,
\begin{equation}
    \mathrm{PCK}(\tau) = \frac{1}{n_\mathcal{M}} \sum_{i=1}^{n_\mathcal{M}} \mathbf{1}\left[ e(i) < \tau \right].
\end{equation}

\noindent
\textit{Area under the PCK curve (AUC)} captures the overall accuracy of the predicted correspondences for varying thresholds, typically up to the maximum threshold $\tau_{\max}$. This is defined as:
\begin{equation}
    \mathrm{AUC}(\tau_{\max}) = \frac{1}{\tau_{\max}} \int_0^{\tau_{\max}} \mathrm{PCK}(\tau) \, d\tau.
\end{equation}

\paragraph{Smoothness.}
Smoothness metrics assess whether neighbouring surface elements on the source and target shapes are consistently mapped under the predicted correspondence.
\textit{Geodesic distortion}~\cite{Ren2021ShapeMatching} measures how well intrinsic distances are preserved
by comparing pairwise geodesic distances under the correspondence:
\[
\sum_{i,j} \bigl( D_\mathcal{M}(i,j) - D_\mathcal{N}(\Pi_{\mathcal{M}\mathcal{N}}(i), \Pi_{\mathcal{M}\mathcal{N}}(j)) \bigr)^2 .
\]

The \textit{Dirichlet energy}~\cite{magnet2022smooth,ezuz2019reversible} can be used to evaluate the smoothness of the induced map, essentially penalising large differences between the mapped positions of neighbouring vertices.
\textit{Conformal distortion}~\cite{ezuz2019reversible,ren2020maptree} quantifies deviations from local angle preservation, indicating the extent to which small surface elements are stretched or sheared under the mapping.
Lower values of these metrics indicate smoother and more geometrically consistent correspondences, which has important ramifications for downstream applications (see \cref{sec:challenges}).

\paragraph{Coverage.}

Coverage metrics evaluate how completely the target shape is reached by the correspondence and are often used in conjunction with smoothness.
A commonly used criterion is the \emph{un-coverage rate}, defined as the fraction of target vertices (or surface area) that are not mapped to by any source vertex under the correspondence.
Low un-coverage means that the correspondence utilises the target shape more uniformly, while high
un-coverage indicates a collapsing or highly non-injective mapping.

\paragraph{Bijectivity.}
Bijectivity metrics assess whether a correspondence behaves approximately as a one-to-one mapping.
This can be evaluated by \textit{inverse consistency}, which computes correspondences in both directions,
$\Pi_{\mathcal{M}\mathcal{N}}$ and $\Pi_{\mathcal{N}\mathcal{M}}$, by applying the same model
with swapped inputs.
The forward and backward maps are then composed (i.e., $\Pi_{\mathcal{N}\mathcal{M}} \circ \Pi_{\mathcal{M}\mathcal{N}}$ and $\Pi_{\mathcal{M}\mathcal{N}} \circ \Pi_{\mathcal{N}\mathcal{M}}$), and the average geodesic distance between the composed maps and the identity map $I$ is measured.
Low values indicate approximately bijective and mutually consistent correspondences.
In addition to inverse-consistency, bijectivity can be further assessed using \textit{orientation preservation}
criteria, such as the triangle inversion rate, which measures the fraction of faces whose orientation is
reversed after mapping and indicates locally inconsistent correspondences.

\paragraph{Visualisation.}
In addition to quantitative metrics, correspondences are commonly assessed qualitatively through visualisation.
A standard approach is \textit{colour transfer} between shapes, which provides an illustration of the overall correspondence structure but may fail to reveal subtle local errors.
More diagnostic visualisations can be obtained by transferring \textit{high-frequency textures}, which are much more sensitive to small-scale errors.
Similarly, \textit{deformation transfer}~\cite{sumner2004deformation,vigano2025nam} assesses correspondence quality by propagating the deformation field defined on a source shape to the target, making local mismatches appear as visible topological artefacts.

\paragraph{Accuracy to downstream application.}
While the metrics above provide a comprehensive quantitative assessment of correspondence quality, many methods in the literature optimise primarily for geodesic accuracy or other quantifiable metrics.
In practice, however, downstream applications may have additional requirements that pure vertex-to-vertex or spectral maps may not satisfy.
For instance, deformation transfer~\cite{sumner2004deformation} and topology transfer require correspondences that are not only accurate but also smooth and bijective; without such guarantees, transferred deformations can exhibit artefacts (see \cref{fig:vis_deformation}).
Similarly, statistical shape modelling demands geometrically consistent maps across shape collections to produce meaningful modes of variation~\cite{loper2015smpl,magnet2023assessing}. 
In domain-specific settings such as medical shape analysis, purely geometric descriptors may be insufficient or prohibitively challenging to obtain. Incorporating prior knowledge from domain experts, such as anatomical landmarks or physiological constraints, can thus be essential for incorporating such methods into clinical pipelines~\cite{el_amrani_universal_2024,bongratz2023abdominal}.
These application-specific requirements highlight that correspondence accuracy, while necessary, can be insufficient on its own, motivating the development of methods that jointly account for the demands of the intended downstream application.
We revisit this gap between benchmark performance and practical applicability in \cref{sec:challenges}.

\section{Spectral methods}
\label{sec:spectral}

Spectral methods have emerged as a powerful framework, relying on intrinsic properties of the shapes that are invariant to isometric deformations. Building on this spectral representation, the functional map framework offers a compact and elegant formulation for expressing correspondences between shapes in the spectral domain, reducing the high-dimensional point-wise matching problem to the estimation of a small matrix of coefficients based on frequency decomposition. 
In this section, we first provide a theoretical background on the Laplace–Beltrami operator (LBO) and functional maps, later discussing recent classical refinement methods as well as learning-based methods.  

\subsection{Background}

In this subsection, we review the theoretical background for spectral methods. 
\paragraph{Laplace–Beltrami operator.}
The Laplace–Beltrami operator is a generalisation of the Laplace operator to Riemannian manifolds~\cite{guggenheimer2012differential}. 
It is defined as the divergence of the gradient of a function $f$:
\begin{equation}
    \label{eq:laplace}
    \Delta f = \operatorname{div}\left(\nabla f\right).
\end{equation}

\noindent
There exist many different ways to compute the LBO in the discrete setting~\cite{wardetzky2007discrete}. For triangle meshes, the most common discretisation is based on the cotangent Laplacian matrix $L_{\mathcal{M}} \in \mathbb{R}^{n_{\mathcal{M}} \times n_{\mathcal{M}}}$, computed as~\cite{pinkall1993computing}:

\begin{equation}
    \label{eq:lap_mat}
    L_{\mathcal{M}}= A_{\mathcal{M}}^{-1}W_{\mathcal{M}}.
\end{equation}

\noindent
Here $A_{\mathcal{M}}$ is the diagonal matrix of lumped area elements and $W_{\mathcal{M}}$ is the cotangent weight matrix, defined in the following form:
\begin{equation}
    \label{eq:lap_weight}
    W_{ij} = \begin{cases}
\frac{\cot(\alpha_{ij})+\cot(\beta_{ij})}{2} & \text{if } (i,j) \in \mathcal{E} \\
- \sum_{k \in N(i)} W_{ik} & \text{if } i = j \\
0 & \text{otherwise,}
\end{cases}
\end{equation}

\begin{equation}
    \label{eq:area_mat}
    A_{ij} = \begin{cases}
        \sum_{k \in N(i)} \frac{T_{k}}{3} & \text{if } i = j \\
        0 & \text{otherwise.}
    \end{cases}
\end{equation}

\begin{figure}[t]
    \centering
    \includegraphics[width=\linewidth]{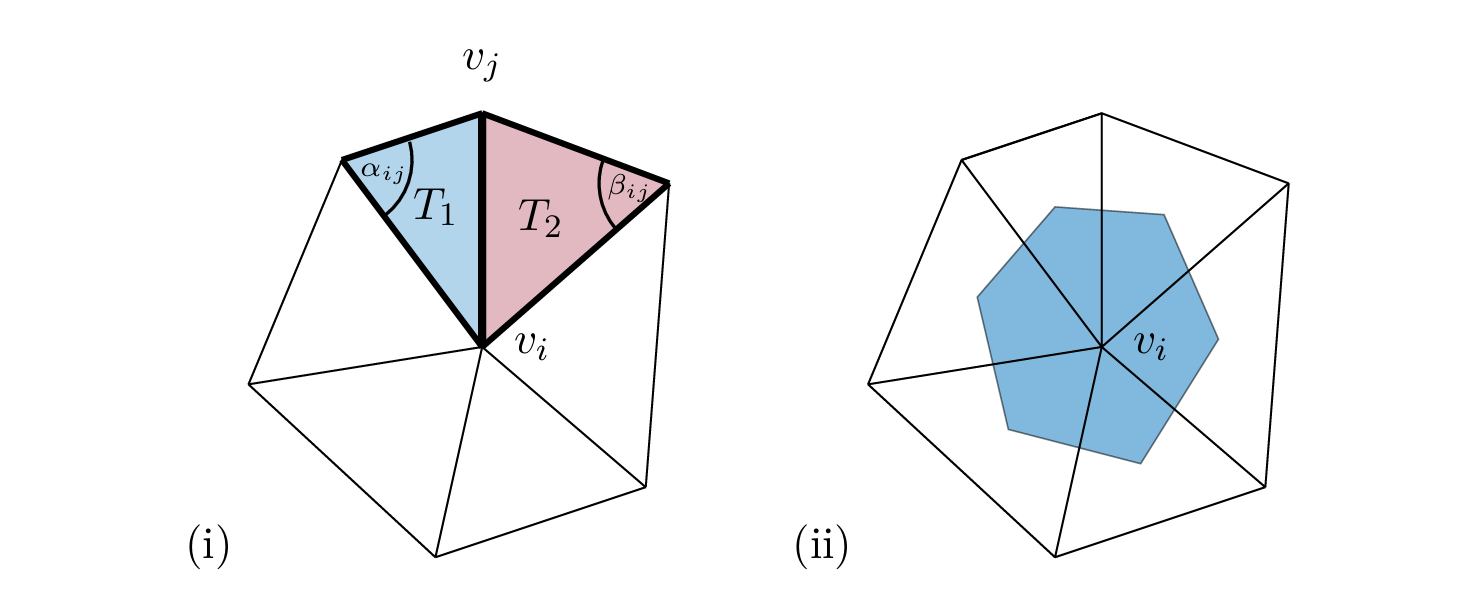}
    \caption{Illustration of the \textbf{notation} used for computing the cotangent weight matrix $W_{ij}$ and the diagonal area matrix $A_{ij}$. The cotangent weight matrix $W_{ij}$ uses the angles opposite the edge between $v_i$ and $v_j$; $T_1$ and $T_2$ denote the areas of the corresponding triangles. The diagonal area matrix $A_{ij}$ assigns each vertex one-third of the total area of its adjacent triangles.}
    \label{fig:cotangent_matrix}
\end{figure}

\noindent
Here $T_k$ is the area of triangle $k$ and $N(i)$ denotes the index of the one-ring neighbour vertices. The notations in~\cref{eq:lap_weight} and~\cref{eq:area_mat} are visualized in~\cref{fig:cotangent_matrix}. 

The eigendecomposition of the LBO plays a crucial role in spectral shape analysis. Let $\Delta_{\mathcal{M}}$ be the LBO on shape $\mathcal{M}$ and
\begin{equation}
    \label{eq:eigendecom}
    \Delta_{\mathcal{M}} \phi_{i} = \lambda_i  \phi_{i},
\end{equation}
where $\phi_{i} \in L^2(\mathcal{M})$ is the $i$-th LBO eigenfunction with corresponding eigenvalue $\lambda_i$.
In the discrete setting, the eigendecomposition can be computed via the generalised eigenvalue problem,
\begin{equation}
    W_{\mathcal{M}} \phi_{i} = \lambda_i A_{\mathcal{M}} \phi_i.
\end{equation}

\noindent Since the LBO is positive semi-definite, its eigenvalues are real and non-negative,
\begin{equation}
    0 = \lambda_0 \leqslant \lambda_1 \leqslant \ldots \lambda_{n_{\mathcal{M}}}.
\end{equation}

\noindent The eigenfunctions of the Laplace–Beltrami operator encode the intrinsic geometry of a surface, their eigenvalues correspond to the frequency, and thus provide a natural basis for representing functions defined on it as well as for frequency filters.
This spectral representation serves as the foundation for functional approaches to shape correspondence.

\paragraph{Functional maps.}
The functional map framework was first introduced by Ovsjanikov et al.~\cite{ovsjanikov_functional_2012}.
Rather than establishing correspondences directly between individual points, it represents them implicitly as linear operators that transfer functions between shapes.
Once the functional map is estimated, point-wise correspondences can be recovered, for example, by transferring delta functions.
This formulation casts the correspondence problem as solving a compact linear system that maps functions to functions, rather than points to points. Given two sets of pointwise descriptor functions $f_{\mathcal{M}} \in \mathbb{R}^{n_{\mathcal{M}} \times d}$ and $f_{\mathcal{N}} \in \mathbb{R}^{n_{\mathcal{N}} \times d}$ and the orthogonal basis sets $\phi_{\mathcal{M}} \in \mathbb{R}^{n_{\mathcal{M}} \times k}$ and $\phi_{\mathcal{N}} \in \mathbb{R}^{n_{\mathcal{N}} \times k}$, which are often chosen as the Laplace-Beltrami eigenfunctions, on each shape, the functional map $C_{\mathcal{MN}} \in \mathbb{R}^{k \times k}$ can be computed as 
\begin{equation}
    C_{\mathcal{MN}} = \arg \min_{C} \|C\phi_{\mathcal{M}}^{\dagger}f_{\mathcal{M}} - \phi_{\mathcal{N}}^{\dagger}f_{\mathcal{N}}\|_F^2,
\end{equation}
where $\phi^{\dagger}_{\bullet} = \phi^{\top}_{\bullet}A_{\bullet}$, which is the pseudo-inverse of $\phi$.

\begin{figure*}[t]
    \centering
    \includegraphics[width=.8\linewidth]{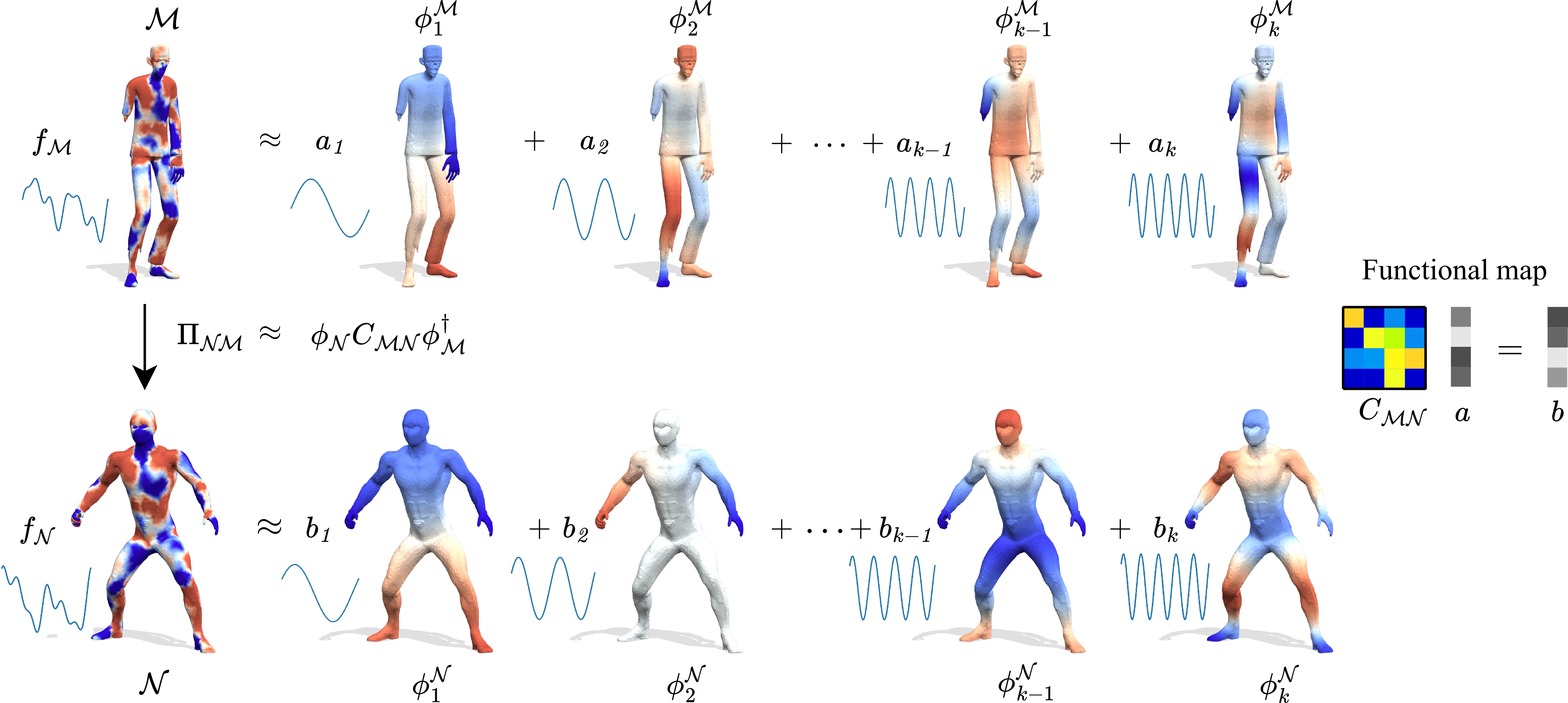}
    \caption{Visualisation of the \textbf{functional map framework} as an analogy of Fourier analysis in signal processing. Given two functions $f_{\mathcal{M}}$ and $f_{\mathcal{N}}$ defined on shapes $\mathcal{M}$ and $\mathcal{N}$, respectively, the functional map $C_{\mathcal{MN}}$ transfers $f_{\mathcal{M}}$ from $\mathcal{M}$ to $f_{\mathcal{N}}$ on $\mathcal{N}$,  each function is approximated using the first $k$ LBO eigenfunctions $\phi$ of the corresponding shape, which is similar to Fourier transform that uses sinusoidal functions with different frequencies as basis functions. In this representation, the functional map transfers the coefficients $a$ of the basis functions on $\mathcal{M}$ to the coefficients $b$ on $\mathcal{N}$, thereby encoding the point-wise correspondences $\Pi_{\mathcal{NM}}$ in a compact $k \times k$ matrix.} 
    \label{fig:functional_map}
\end{figure*}

The corresponding point-wise map $\Pi_{\mathcal{NM}}$ can be recovered from the functional map based on the following relationship:
\begin{equation}
    \label{eq:map_relation}
    \Pi_{\mathcal{NM}}\phi_{\mathcal{N}} \approx \phi_{\mathcal{M}}C_{\mathcal{NM}},
\end{equation}
The left-hand side of the equation aligns the LBO eigenfunctions in the spatial domain, while the right-hand side aligns them in the spectral domain.~\cref{fig:functional_map} visualizes the functional map framework. 

\subsection{Axiomatic methods}
Axiomatic methods assume that noisy or incomplete correspondences are obtained either based on handcrafted features~\cite{sun2009concise,bronstein2010scale,aubry2011wave} or manual annotations and aim to improve the final matching performance by iteratively refining functional maps and point-wise maps as a post-processing step. 
The key idea relies on the relationship between the two representations, as defined in~\cref{eq:map_relation}.

As the simplest axiomatic technique, ICP iteratively converts the functional map to a point-wise map via nearest neighbour search in spectral space~\cite{ovsjanikov_functional_2012}. Follow-up works~\cite{rodola2015point,rodola2017regularized,ezuz2017deblurring} iteratively optimise the functional map and point-wise map by minimising an objective function based on~\cref{eq:map_relation} with additional regularisation terms or constraints encouraging smoothness and bijectivity. 
BCICP~\cite{ren2018continuous} further promotes correspondence continuity and overall coverage by introducing bijective ICP with additional point-wise correspondence smoothing as well as outlier detection and fix. ZoomOut~\cite{melzi_zoomout_2019} gradually increases the spectral resolution of the functional map during the ICP process. 
Despite its simplicity, it achieves better performance than previous, more complex refinement methods. 

RHM~\cite{ezuz2019reversible} prompts reversible harmonic maps, thereby resulting in correspondences with lower conformal distortion. FSF~\cite{pai2021fast} theoretically analyses the relationship between functional maps and point-wise maps and proposes a point-wise map recovery method based on spectral alignment of adjoint maps. Later, DiscreteOpt~\cite{ren2021discrete} introduces the \emph{proper} functional map space and optimises the functional maps in the proper functional map space, leading to superior performance in the context of accuracy and smoothness. The \emph{proper} functional map space is the set of functional maps that arise from pointwise correspondences:
\begin{equation}\label{eq:fmap-to-pmap}
    \left\{C_{\mathcal{MN}}\; |\; \exists \Pi_{\mathcal{NM}}, \;\mathrm{s.t.}\;C_{\mathcal{MN}} = \phi_{\mathcal{N}}^{\dagger}\Pi_{\mathcal{NM}}\phi_{\mathcal{M}}\right\}.
\end{equation}
SmoothFM~\cite{magnet2022smooth} extends DiscreteOpt by introducing spatial smoothness based on minimising the Dirichlet energy, similar to RHM~\cite{ezuz2019reversible}.

Instead of using LBO eigenfunctions for functional map computation, several works~\cite{panine2022non,donati_complex_2022,hartwig_elastic_2023} propose to use different eigenfunctions to improve the matching performance in terms of non-isometry and orientation awareness.
Another line of works encourages cycle-consistency~\cite{huang2014functional,bernard2019hippi,huang2020consistent} in the case of processing a collection of shapes.

The study of axiomatic methods establishes a solid theoretical analysis of the relationship between functional maps and point-wise maps. However, in real-world applications, the axiomatic methods often fail due to the violation of the underlying assumptions (e.g., near-isometric deformation). Therefore, it is interesting to combine axiomatic methods with learning-based approaches to improve the matching accuracy. Previous learning-based approaches~\cite{halimi2019unsupervised,roufosse2019unsupervised} typically use them as a post-processing step to improve the matching accuracy. Recent learning-based methods~\cite{attaiki_understanding_2023,cao_unsupervised_2023,magnet_memory-scalable_2024} are inspired by the insights from those axiomatic methods~\cite{ren2021discrete,melzi_zoomout_2019}. It is interesting to see that future research in this area integrates and extends existing axiomatic methods within end-to-end learning-based frameworks, combining strong theoretical foundations with data-driven learning to further improve matching accuracy.

\begin{table}[t]
\centering
\small
\begin{tabular}{lcccccc}
\toprule
\textbf{Method} & \textbf{Cat.} & \textbf{MS} & \textbf{NI} & \textbf{P} & \textbf{TN} \\
\midrule
ZoomOut~\cite{melzi_zoomout_2019} & \textbf{\textcolor{blue}{A}} & \xmark & \cmark & \cmark & \xmark \\
ConsistentZoomOut~\cite{huang2020consistent} & \textbf{\textcolor{blue}{A}} & \cmark & \cmark & \cmark & \xmark \\
RHM~\cite{ezuz2019reversible} & \textbf{\textcolor{blue}{A}} & \xmark & \cmark & \xmark & \xmark \\
ComplexFMaps~\cite{donati_complex_2022} & \textbf{\textcolor{blue}{A}} & \xmark & \cmark & \xmark & \xmark \\
DiscreteOpt~\cite{ren2021discrete} & \textbf{\textcolor{blue}{A}} & \xmark & \cmark & \xmark & \xmark \\
SmoothFM~\cite{magnet2022smooth} & \textbf{\textcolor{blue}{A}} & \xmark & \cmark & \xmark & \xmark \\
\midrule
GeomFMaps~\cite{donati2020deep} & \textbf{\textcolor{orange!90!black}{S}} & \xmark & \xmark & \xmark & \xmark \\
DPFM~\cite{attaiki2021dpfm} & \textbf{\textcolor{orange!90!black}{S}} & \xmark & \xmark & \cmark & \xmark \\
SRFeat~\cite{li_srfeat_2022} & \textbf{\textcolor{orange!90!black}{S}} & \xmark & \cmark & \xmark & \xmark \\
EchoMatch~\cite{xie_echomatch_2025} & \textbf{\textcolor{orange!90!black}{S}} & \xmark & \xmark & \cmark & \xmark \\
DenoisFM~\cite{zhuravlev_denoising_2025} & \textbf{\textcolor{orange!90!black}{S}} & \xmark & \cmark & \xmark & \xmark \\
FRIDU~\cite{rimon_fridu_2025} & \textbf{\textcolor{orange!90!black}{S}} & \xmark & \xmark & \xmark & \xmark \\
DiffuMatch~\cite{pierson2025diffumatch} & \textbf{\textcolor{orange!90!black}{S}} & \xmark & \cmark & \xmark & \xmark \\
\midrule
UnsupFMNet~\cite{halimi2019unsupervised} & \textbf{\textcolor{green!50!black}{U}} & \xmark & \xmark & \xmark & \xmark \\
SURFMNet~\cite{roufosse2019unsupervised} & \textbf{\textcolor{green!50!black}{U}} & \xmark & \xmark & \xmark & \xmark \\
DPFM-unsup~\cite{attaiki2021dpfm} & \textbf{\textcolor{green!50!black}{U}} & \xmark & \xmark & \cmark & \xmark \\
UDMSM~\cite{cao2022unsupervised} & \textbf{\textcolor{green!50!black}{U}} & \cmark & \xmark & \xmark & \xmark \\
AttentiveFMaps~\cite{li_learning_2022} & \textbf{\textcolor{green!50!black}{U}} & \xmark & \cmark & \xmark & \xmark \\
ULRSSM~\cite{cao_unsupervised_2023} & \textbf{\textcolor{green!50!black}{U}} & \xmark & \cmark & \cmark & \cmark \\
G-msm~\cite{eisenberger2023g} & \textbf{\textcolor{green!50!black}{U}} & \cmark & \cmark & \xmark & \cmark \\
HybridFMaps~\cite{bastian2024hybrid} & \textbf{\textcolor{green!50!black}{U}} & \xmark & \cmark & \xmark & \cmark \\
DiffZoomOut~\cite{magnet_memory-scalable_2024} & \textbf{\textcolor{green!50!black}{U}} & \xmark & \cmark & \xmark & \cmark \\
\bottomrule
\end{tabular}

\caption{
Overview of \textbf{spectral} shape correspondence methods.
Each method is categorized by approach (\textbf{Cat.}): 
axiomatic (\textbf{\textcolor{blue}{A}}), 
supervised (\textbf{\textcolor{orange!90!black}{S}}), 
unsupervised (\textbf{\textcolor{green!50!black}{U}}).
The remaining columns indicate supported properties: 
multi-shape consistency (\textbf{MS}), 
non-isometric deformations (\textbf{NI}), 
partiality (\textbf{P}), 
topological noise (\textbf{TN}).
}
\label{tab:spectral_methods}
\end{table}

\subsection{Supervised methods}
Instead of relying on handcrafted features~\cite{sun2009concise,bronstein2010scale,aubry2011wave,salti2014shot} for finding correspondences, learning-based methods attempt to use neural networks to learn the feature descriptors directly from training data. FMNet~\cite{litany2017deep} was among the first supervised methods to learn a non-linear transformation of SHOT descriptors~\cite{salti2014shot} based on ground-truth correspondences.
Instead of directly obtaining soft correspondences based on feature similarity, learning-based methods first compute a functional map based on extracted features on each shape $f_{\mathcal{M}} \in \mathbb{R}^{n_{\mathcal{M}} \times c}, f_{\mathcal{N}} \in \mathbb{R}^{n_{\mathcal{N}} \times c}$ by solving the linear least square problem:
\begin{equation}
    \label{eq:fmap}  C_{\mathcal{MN}}=\mathrm{argmin}_{C}~ \Energy_{\mathrm{data}}\left(C\right)+\lambda \Energy_{\mathrm{reg}}\left(C\right).
\end{equation}
Here, minimizing $\Energy_{\mathrm{data}}=\left\|C\phi_{\mathcal{M}}^{\dagger}f_{\mathcal{M}}-\phi_{\mathcal{N}}^{\dagger}f_{\mathcal{N}}\right\|^{2}_{F}$ enforces descriptor preservation, while minimizing the regularization term $\Energy_{\mathrm{reg}}$ imposes some form of structural properties~\cite{ovsjanikov_functional_2012}. After obtaining the functional map, the soft correspondences can be computed as:
\begin{equation}
    \label{eq:soft_corr}
    \Pi_{\mathcal{NM}}^{\mathrm{soft}} = \operatorname{softmax}\left(\phi_{\mathcal{N}}C_{\mathcal{MN}}\phi_{\mathcal{M}}^{\dagger}\right).
\end{equation}
The loss is measured as the geodesic distance error between the predicted soft correspondences and the ground truth. 
Instead of using geodesic distance, which is computationally expensive, GeomFMaps~\cite{donati2020deep} directly compute the loss in the spectral domain based on the difference between the predicted functional map and the ground-truth functional map. 
DPFM~\cite{attaiki2021dpfm} extends this idea to partial shape matching and, for the first time, enables partial-to-partial shape matching by introducing an additional overlap prediction module. 
Afterwards, SRFeat~\cite{li_srfeat_2022} introduces additional contrastive loss to improve the local discriminative power of learned features and to overcome the limitation that functional maps mainly focus on low-frequency areas. 
EchoMatch~\cite{xie_echomatch_2025} improves the partial-to-partial matching performance by introducing the concept of correspondence reflection. Correspondence reflection indicates that the valid correspondences for partial shapes should be bijective. 
Another line of research focuses on learning semantically meaningful basis functions to replace the standard LBO eigenfunctions~\cite{marin2020correspondence,huang_multiway_2022} and thereby improve matching performance. 
Recently, NAM~\cite{vigano2025nam} extended the concept of functional maps to non-linear neural adjoint maps based on multi-layer perceptions. 

Despite the appealing matching performance of supervised learning methods, they still face challenges in generalisation. 
Due to the lack of large annotated training data with ground-truth correspondences across different object categories, supervised methods cannot generalise well to unseen object categories and thus limit their practical applications. 
Further research in this area should focus on improving generalisation to unseen object categories and on creating large-scale training datasets~\cite{ehm2025beyond,liu_stable-score_2025}.    

\subsection{Unsupervised methods}
Due to the lack of large, well-annotated training datasets with ground-truth correspondences, many learning-based methods focus on developing unsupervised training strategies. 
Starting from UnsupFMNet~\cite{halimi2019unsupervised}, the method proposes an unsupervised loss to train FMNet~\cite{litany2017deep} based on geodesic distance preservation:
\begin{equation}
    \mathcal{L}_{\mathrm{geo}} = \frac{1}{n_{\mathcal{M}}}\|D_{\mathcal{M}} - \Pi_{\mathcal{MN}} D_{\mathcal{N}} \Pi_{\mathcal{MN}}^{\top} \|_{{F}}^{2}, 
\end{equation}
where $D_{\mathcal{M}} \in \mathbb{R}^{n_{\mathcal{M}} \times n_{\mathcal{M}}}$ and $D_{\mathcal{N}} \in \mathbb{R}^{n_{\mathcal{N}} \times n_{\mathcal{N}}}$ are the geodesic distance matrices for shape $\mathcal{M}$ and shape $\mathcal{N}$, respectively. Similar to~\cite{litany2017deep}, the soft correspondences are obtained based on~\cref{eq:soft_corr}. Later work~\cite{aygun2020unsupervised} replaces geodesic distance matrices with heat kernels to be more efficient and robust. Instead of directly regularising the soft correspondences, follow-up works attempt to regularise the more compact functional maps. SURFMNet~\cite{roufosse2019unsupervised} is one of the first works in this direction. It exploits the structure of the functional map to enable unsupervised learning. Specifically, it uses Laplacian commutativity as the $\Energy_{\mathrm{reg}}$ in~\cref{eq:fmap}, which can be expressed as
\begin{equation}
    \label{eq:l_lap}
    \left\|C_{\mathcal{MN}}\Lambda_{\mathcal{M}} - \Lambda_{\mathcal{N}}C_{\mathcal{MN}}\right\|_{F}^{2},
\end{equation}
where $\Lambda_{\mathcal{M}}$ and $\Lambda_{\mathcal{N}}$ are the diagonal eigenvalue matrix of shape $\mathcal{M}$ and $\mathcal{N}$, respectively. As for the unsupervised loss terms, SURFMNet uses bijectivity and orthogonality regularisation to promote the matching to be bijective and area-preserving~\cite{ovsjanikov_functional_2012}:
\begin{equation}
    \label{eq:l_orth}
    \mathcal{L}_{\mathrm{orth}} = \left\|C_{\mathcal{MN}}^{\top}C_{\mathcal{MN}}-I\right\|^{2}_{F}+\left\|C_{\mathcal{NM}}^{\top}C_{\mathcal{NM}}-I\right\|^{2}_{F},
\end{equation}
\begin{equation}
    \label{eq:l_bij}
    \mathcal{L}_{\mathrm{bij}}=\left\|C_{\mathcal{MN}}C_{\mathcal{NM}}-I\right\|^{2}_{F}+\left\|C_{\mathcal{NM}}C_{\mathcal{MN}}-I\right\|^{2}_{F}.
\end{equation}
Later work~\cite{sharma2020weakly} systematically analyses each component, and DPFM~\cite{attaiki2021dpfm} extends the idea to partial-to-full shape matching. 

Nevertheless, these unsupervised methods still rely on off-the-shelf refinement methods~\cite{melzi_zoomout_2019,ovsjanikov_functional_2012} to obtain better matching results. AttentiveFMaps~\cite{li_learning_2022} proposes an attention mechanism to aggregate predicted functional maps with different resolutions and significantly improves the matching performance for non-isometric shapes. ULRSSM~\cite{cao_unsupervised_2023} utilises the concept of proper functional map space~\cite{ren2021discrete,attaiki_understanding_2023,sun_spatially_2023,cao_revisiting_2024} to establish a relationship between the soft correspondences based on feature similarity and the functional maps from~\cref{eq:fmap} and leads to state-of-the-art performance at the time of publishing for near-isometric, non-isometric as well as topologically noisy shapes. 
HybridFMaps~\cite{bastian2024hybrid} extends ULRSSM by incorporating the elastic basis~\cite{hartwig_elastic_2023} to further improve the matching performance, especially for non-isometric deformations.
DiffZoomOut~\cite{magnet_memory-scalable_2024} proposes a differentiable version of ZoomOut~\cite{melzi_zoomout_2019}, leading to more efficient inference, especially for high-resolution shapes. 
Another line of work combines functional maps with spatial alignment, leading to spatially smooth matching \cite {eisenberger2020deep,attaiki_ncp_2022,cao_spectral_2024}. 

Recent unsupervised methods~\cite{cao_unsupervised_2023,cao2024synchronous} have shown significant progress in shape matching for near-isometric, non-isometric shapes, even outperforming some supervised approaches~\cite{donati2020deep}. However, for partial shape matching (particularly the partial-to-partial setting), unsupervised methods struggle to achieve satisfactory results due to the absence of adequate regularization~\cite{rodola2017partial,ehm2025beyond}.
Additionally, if significant topological noise arises from self-intersections or poor reconstructions, unsupervised methods are likely to fail. 
These limitations severely hinder their applicability in real-world scenarios. 
Future work in this area should focus on improving matching performance for partial and topologically noisy shapes. 

\paragraph{Pointwise map recovery.}
An essential step in any spectral pipeline is the recovery of a pointwise map from the functional map. 
The most straightforward way to achieve this is via nearest-neighbour search in the spectral embedding space~\cite{ovsjanikov_functional_2012}.
However, this treats each vertex independently, and the resulting maps thus lack guarantees on smoothness or neighbourhood preservation~\cite{magnet2022smooth,vigano_adjoint_2023,ren2021discrete}, which limits their direct applicability in downstream tasks such as deformation transfer~\cite{sumner2004deformation,vigano2025nam} or statistical shape modelling~\cite{greenspan_s3m_2023}.
While the refinement methods discussed above mitigate these artefacts~\cite{melzi_zoomout_2019,magnet2022smooth}, they operate as post-processing and cannot fully recover information lost due to the low-rank approximation.
In \cref{sec:combinatorial}, we discuss combinatorial methods that partially address this limitation by encoding neighbourhood constraints directly into the matching formulation.

\subsection{Functional map estimation using image diffusion}
\label{subsec:fm_diffusion}

Diffusion models have emerged as the dominant paradigm in image synthesis, enabling high-quality generation through iterative denoising \cite{sohl2015deep,ho2020denoising,karras2022elucidating,po2024state}.
They operate by gradually corrupting data with Gaussian noise in a forward process and then learning to reverse this process step by step.
A popular approach is Denoising Diffusion Probabilistic Models~\cite{ho2020denoising}, where a generative model is trained by recovering content from a noise distribution, potentially conditioned on additional data.

Several works have observed that functional maps \cite{ovsjanikov_functional_2012} provide a suitable domain for image diffusion.  
Given the truncated Laplacian eigenbases $\phi_\mathcal{M}, \phi_\mathcal{N} \in \mathbb{R}^{n \times k}$ of shapes $\mathcal{M}$ and $\mathcal{N}$, the resulting functional map $C_{\mathcal{M} \mathcal{N}} \in \mathbb{R}^{k \times k}$ is a compact matrix, typically with $k \leqslant 200$.  
This compactness makes it feasible to repurpose architectures originally designed for image generation to predict functional maps.

\begin{figure}[t]
    \centering
    \includegraphics[width=0.95\linewidth]{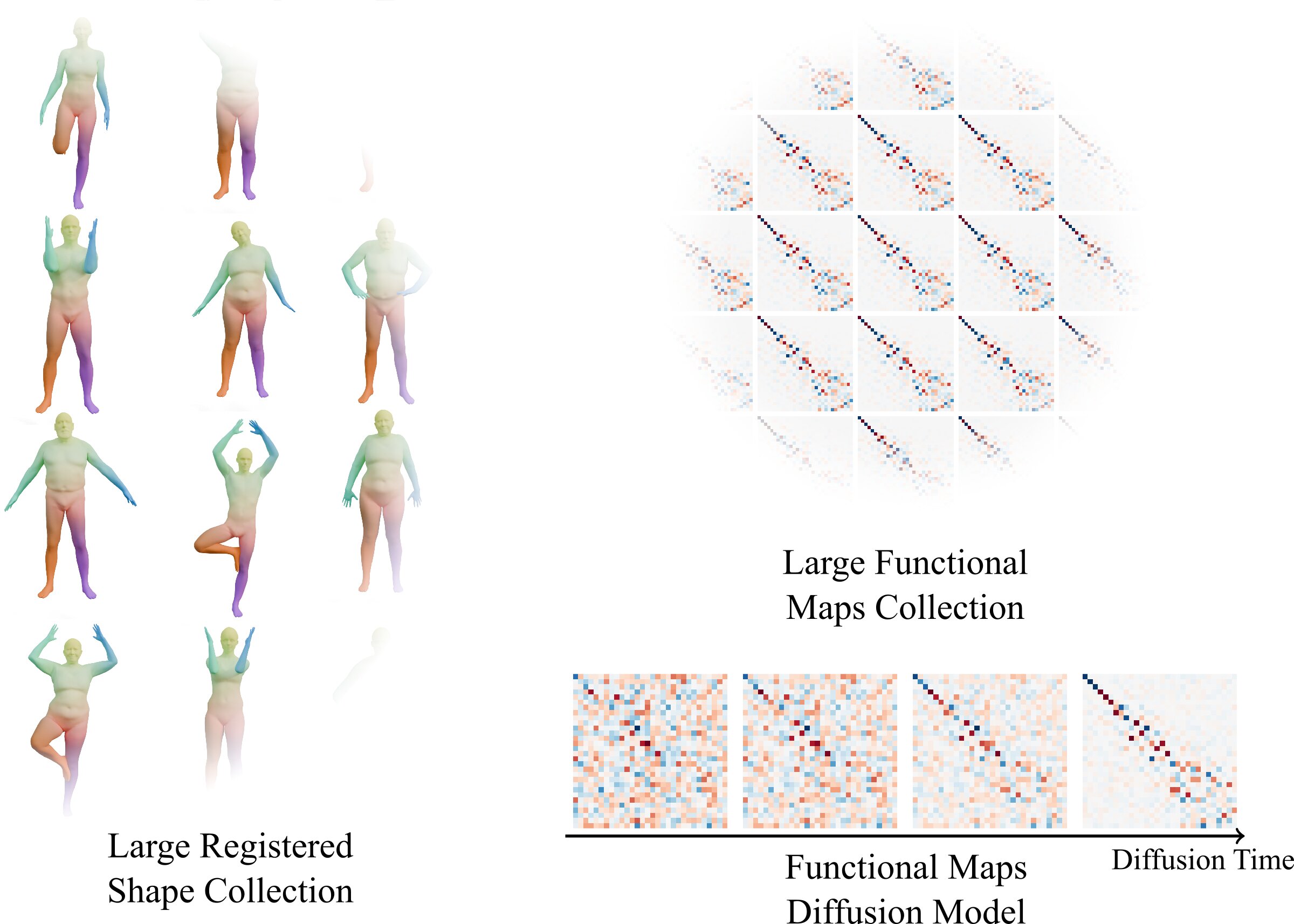}
    \caption{Functional maps provide a suitable domain for \textbf{2D diffusion models} due to their compact matrix representation. Such spectral diffusion models, trained on large collections of registered shapes, introduce a learned prior for functional map solvers. Adapted from \cite{pierson2025diffumatch}.}
    \label{fig:6_2_classification}
\end{figure}

DenoisFM~\cite{zhuravlev_denoising_2025} trains a diffusion model to directly generate a functional map between a pair of shapes, addressing sign ambiguity of Laplacian eigenfunctions using an unsupervised sign-selection strategy based on learned surface features \cite{aubry2011wave}.  
FRIDU~\cite{rimon_fridu_2025} presents an approach for functional map refinement, where a diffusion model learns to generate refined maps conditioned on noisy initial maps, guiding learning objectives that encourage plausible output correspondences.
DiffuMatch~\cite{pierson2025diffumatch} is trained on the absolute values of functional maps $|C_{\mathcal{M} \mathcal{N}}|$ to overcome the sign ambiguity of Laplacian eigenbasis, replacing existing axiomatic constraints such as orthogonality \cref{eq:l_orth} and Laplacian commutativity \cref{eq:laplace} with the learned regularizer.
These early works demonstrate that diffusion models can serve as a substitute for an explicit functional map solver.

\section{Combinatorial methods}
\label{sec:combinatorial}

While the features obtained through neural network architectures based on functional maps are remarkably descriptive, the differentiability requirement on the matching algorithm often yields undesirable matches on disparate but geometrically similar regions of the shape.
Another line of work seeks to exploit the manifold structure and geometric coherence of feature matching through powerful combinatorial optimisation techniques.

\subsection{Background}
Combinatorial methods provide powerful tools for shape matching by formulating correspondence problems as discrete optimisation tasks over finite (but potentially very large) solution spaces.
As such, combinatorial methods are particularly well-suited for handling discrete shape representations such as meshes and point clouds~\cite{schrijver2003combinatorial,korte2011combinatorial} and usually seek an optimal assignment between discrete elements of shapes, such as vertex-to-vertex correspondences.

Formally, in the shape matching context, combinatorial methods aim to find an optimal solution $x^*\in\mathcal{F}\subset\{0,1\}^n$ 
(which usually resembles an assignment between surface elements) within a set of feasible solutions $\mathcal{F}$ while minimizing some (matching cost) function $\Energy(x) :\;\mathcal{F} \rightarrow \mathbb{R}$ such that 
\begin{equation} \label{eq:combinatial-general}
    x^*=\underset{x \in\mathcal{F}}{\min} \;\Energy(x).
\end{equation}
Consequently, the advantage of combinatorial formalisms is that they allow (i) to enforce that the solution is guaranteed to have certain properties by constraining the feasible set $\mathcal{F}$ adequately, and (ii) to add soft constraints and similarity measures by designing appropriate cost functions $\Energy(x)$.
In addition to the binary variables $x\in\mathcal{F}$ used in \eqref{eq:combinatial-general}, some methods also introduce additional continuous variables to model deformation alongside matching~\cite{bernard2020mina, gao_sigma_2023}.

Among the most prominent examples of combinatorial matching methods are the linear assignment problem (LAP) and the quadratic assignment problem (QAP).
The LAP was introduced to the shape matching community in~\cite{belongie2001matching}.
Yet, the LAP does not take into account relations between surface elements, such as their geodesic distances, and is therefore not suitable for shape matching when geometric relations must be preserved.
On the contrary, the QAP allows for incorporating such relations, but due to the quadratic objective function and resulting NP-hardness~\cite{rendl1994quadratic} it does not scale to practically relevant resolutions.
Thus, shape matching approaches based on QAP-formalisms rely on finding solutions via approximations, including convex relaxations~\cite{schellewald2005probabilistic,dym2017relax,bernard2018ds,kushinsky2019sinkhorn,droge_kissing_2023}, and heuristics~\cite{dubrovina2011approximately, le2017alternating,vestner2017efficient, haller2022comparative}.
To obtain more scalable solutions, other combinatorial shape matching methods exploit the discrete representation of 3D shapes and incorporate neighborhood relations using linear constraints to restrict the feasible set.
Among these, one of the most prominent examples is the work by Windheuser et al.~\cite{windheuser2011geometrically} which sparked many follow-up approaches building on similar concepts. 

In the following, we summarise recent combinatorial methods. 
We further categorise approaches into landmark-based formalism, surface-to-surface formalisms, multi-shape matching approaches as well as methods based upon quantum computing. 
We summarise all methods in \cref{tab:combinatorial}.

\begin{table}[t]
    \centering
    \setlength{\tabcolsep}{3pt}
    \begin{tabular}{l@{}cccccc}
    \toprule
    \small
     \textbf{Method} & \textbf{LB} & \textbf{S2S} & \textbf{MS} & \textbf{Q} & \textbf{LO} & \textbf{RF} \\ 
     \midrule
     Maron et al.~\cite{maron2016point} & \cmark & \cmark & \xmark & \xmark & \xmark & \cmark \\
     MINA~\cite{bernard2020mina} & \cmark & \xmark & \xmark & \xmark & \xmark & \xmark \\
     SIGMA~\cite{gao_sigma_2023} & \cmark & \xmark & \xmark & \xmark & \xmark & \xmark  \\
     Ren et al.~\cite{ren2021discrete} & \xmark & \cmark & \xmark & \xmark & \xmark & \xmark  \\
     Pai et al.~\cite{pai2021fast} & \xmark & \cmark & \xmark & \xmark & \xmark & \xmark \\
     SM-Comb~\cite{roetzer_scalable_2022} & \xmark & \cmark & \xmark & \xmark &\cmark & \xmark  \\
     DiscoMatch~\cite{roetzer2024discomatch} & \xmark & \cmark & \xmark & \xmark &\cmark & \xmark  \\
     SpiderMatch~\cite{roetzer_spidermatch_2024} & \xmark & \cmark & \xmark & \xmark &\cmark & \xmark \\
     Roetzer et al.~\cite{roetzer2025higherorder} & \xmark & \cmark & \xmark & \xmark &\cmark & \xmark \\
     GeCo~\cite{roetzer2025geco} & \xmark & \cmark & \xmark & \xmark &\cmark & \xmark \\
     SuPaMatch~\cite{amrani2025highres} & \xmark & \cmark & \xmark & \xmark &\cmark & \xmark \\
     Gao et al.~\cite{gao2021isometric} & \xmark & \xmark & \cmark & \xmark & \xmark & \xmark \\
     Kahl et al.~\cite{kahl2025towards} & \xmark & \xmark  & \cmark & \xmark & \xmark & \xmark \\
     Bhatia et al.~\cite{bhatia_ccuantumm_2023} & \xmark & \xmark & \cmark & \cmark  & \xmark & \xmark \\
     Benkner et al.~\cite{benkner2020adiabatic} & \xmark & \xmark & \xmark & \cmark & \xmark & \xmark \\
     Seelbach et al.~\cite{seelbach2021qmatch} & \xmark & \xmark & \xmark & \cmark & \xmark & \xmark \\
     Meli et al.~\cite{meli_qucoop_2025} & \xmark & \xmark & \xmark & \cmark & \xmark & \xmark  \\
     \bottomrule
    \end{tabular}
    \caption{Overview over \textbf{combinatorial} matching methods and classification whether methods are \textbf{(LB)} landmark-based, are \textbf{(S2S)} Surface-to-Surface methods, are applicable to \textbf{MS} multiple shapes, are \textbf{(Q)} Quantum, have a \textbf{(LO)} linear objective $f(x)$, or have a \textbf{(RF)} relax the feasible set $\mathcal{F}$.
    }
    \label{tab:combinatorial}
\end{table}

\subsection{Landmark-based approaches}
\begin{figure}
    \centering
    \begin{tabular}{cc}
        \small
         \includegraphics[height=3cm]{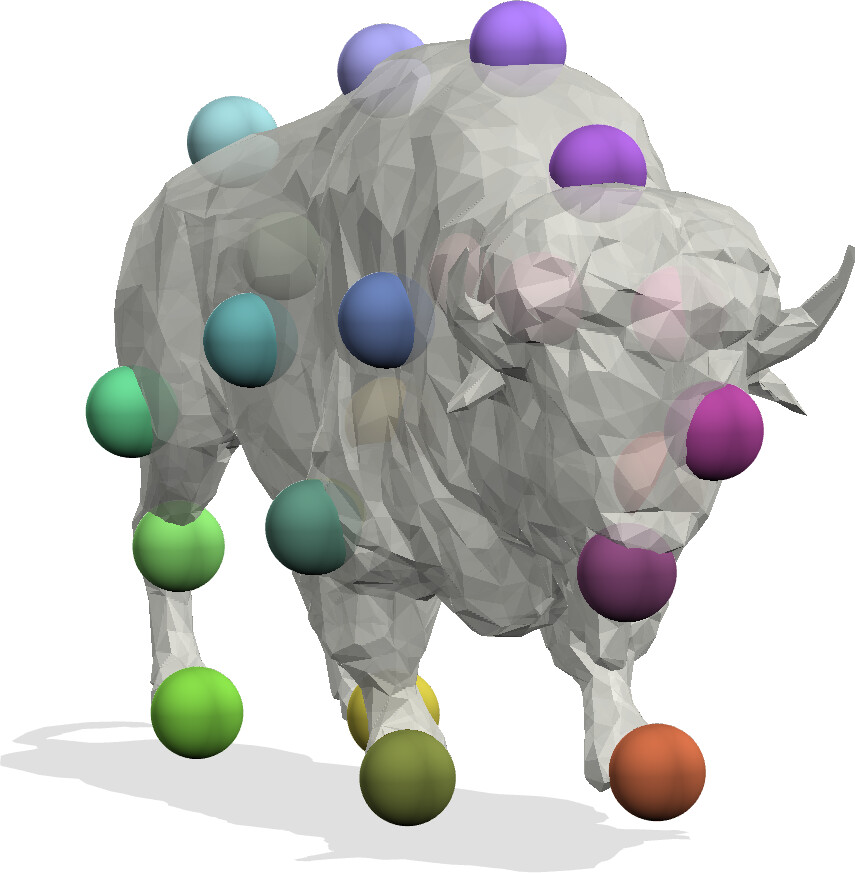}&
         \includegraphics[height=3cm]{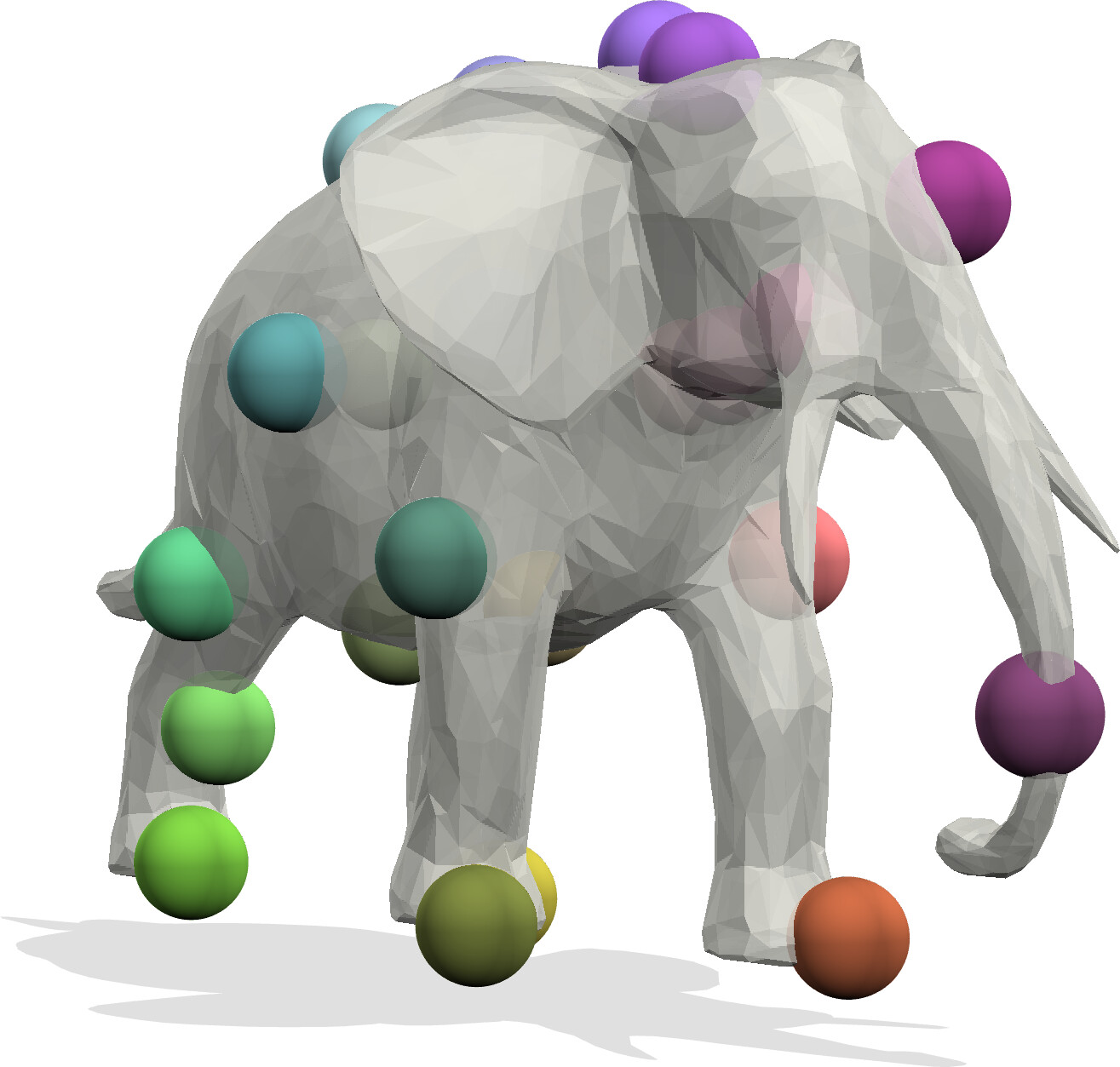}
    \end{tabular} 
    \caption{Visualisation of correspondence between \textbf{landmarks} (in colour) lying on the surface (grey) of two 3D shapes. Image source: \cite{gao_sigma_2023}}
    \label{fig:key-points}
\end{figure}

Landmark-based approaches, sometimes also referred to as \emph{sparse} shape matching approaches, focus on a small subset of vertices on each shape, 
denoted by $\Tilde{X}_\mathcal{M}\subset X_\mathcal{M}$ and $\Tilde{X}_\mathcal{N}\subset X_\mathcal{N}$, see \cref{fig:key-points}. 
These vertices, often referred to as \emph{landmarks} or sometimes also as \emph{keypoints}, are intended to capture 
salient or distinctive locations on the respective shapes $\mathcal{M}$ and $\mathcal{N}$.
Generally, such landmarks are not available and are often defined by human annotators.
Landmark-based formalisms usually aim to find correspondences only among the landmarks with the underlying assumption that correspondences between shapes can be inferred or approximated effectively from the relationships between their landmarks.
Consequently, using landmarks often allows for efficient solvability while still enabling the recovery of dense correspondences via post-processing.
Yet, landmark-based approaches assume consistent landmark positions across shapes and thus the solution quality of these methods is heavily based on their position.

Among the landmark-based approaches, one line of work seeks to uncover synergies between shape correspondences and shape reconstruction.
\def\assignmentMatrix{\Pi}
In~\cite{bernard2020mina}, Bernard et al.~propose a mixed integer programming framework which optimises for a $m\times m$ permutation matrix $\assignmentMatrix$ between landmarks $\Tilde{X}_\mathcal{M}\subset X_\mathcal{M}$, $\Tilde{X}_\mathcal{N}\subset X_\mathcal{N}$, and  a set of reconstructed landmark vertex-positions $\Tilde{X}_\mathcal{M}^\text{rec}$.
While $\Tilde{X}_\mathcal{M}$ represents the initial positions of the selected landmarks on shape $\mathcal{M}$, $\Tilde{X}_\mathcal{M}^\text{rec}$ denotes their optimized coordinates in the target space that best align with the landmarks $\Tilde{X}_\mathcal{M}$.
On a high-level the goal is to minimise
\begin{equation}
    \begin{aligned}
    \underset{
        \assignmentMatrix \in \mathbb{P}, \;
        \Tilde{X}_\mathcal{M}^\text{rec}\in \mathbb{R}^{m\times r}
    }
    {\min} 
    ||\Tilde{X}_\mathcal{M}^\text{rec} - 
    \assignmentMatrix 
    \Tilde{X}_\mathcal{N}||  + 
    \Energy_\text{rec}(\Tilde{X}_\mathcal{M}^\text{rec}, \mathcal{M})
    \end{aligned}
\end{equation}
Here, $\Energy_\text{rec}(\Tilde{X}_\mathcal{M}^\text{rec}, \mathcal{M})$ is a reconstruction function that aims for a smooth reconstruction of shape $\mathcal{M}$ using the optimised landmark positions $\Tilde{X}_\mathcal{M}^\text{rec}$.

Building on the ideas presented in~\cite{bernard2020mina},
Gao et al.~\cite{gao_sigma_2023} propose a similar mixed integer programming formalism
and introduce the 
\emph{projected Laplace-Beltrami} (LBO) operator for the reconstruction, which reads
\begin{equation}
    L_\mathcal{M}^\text{proj} \coloneqq {M}^T \Delta^\text{stiff}_\mathcal{M} {M}.
\end{equation}
Here, $\Delta^\text{stiff}_\mathcal{M}$ is the stiffness matrix component of the Laplace-Beltrami operator $L_\mathcal{M}$,
${M}\coloneqq I -
\Tilde{X}_\mathcal{M}
(
\Tilde{X}_\mathcal{M}^T  
\Tilde{X}_\mathcal{M}
)^{-1} \Tilde{X}_\mathcal{M}^T$
is a projection matrix
and $ \Tilde{X}_\mathcal{M}\coloneqq( {X}_\mathcal{M}, \boldsymbol{1})$.
As shown in \cite{gao_sigma_2023}, the projected LBO $L_\mathcal{M}^\text{proj}$ is scale-invariant and invariant under rigid body transformations.
Furthermore, it retains more high-frequency geometric details of the shapes compared to the Laplace-Beltrami operator.
This helps to increase reconstruction quality and, in turn, leads to better matching results.

While landmark-based approaches offer improved scalability due to smaller optimisation problems and, with that, greater mathematical modelling flexibility, many argue that full surface-to-surface maps must be considered to compute faithful matchings between shapes.

\subsection{Surface-to-surface approaches}
Surface-to-surface means finding a dense matching for all vertices $X_\mathcal{M}$ of a shape $\mathcal{M}$, rather than only for a subset of vertices, as in landmark-based approaches.
To find such a matching, some lines of work build on the \emph{Procrustes matching} problem where the goal is to simultaneously find correspondences between two sets of points $X_\mathcal{M}, X_\mathcal{N} \in \mathbb{R}^{n\times d}$ (in form of a assignment matrix $\Pi_{\mathcal{M}\mathcal{N}}$), as well as a rigid alignment $T \in SE(d)$. 
This is formulated as the following optimization problem
\begin{equation}
    \underset{
        \Pi_{\mathcal{M}\mathcal{N}} \in\mathbb{P}, T \in SE(d),
                  }{\min} ||\bar{X}_\mathcal{M} T - \Pi_{\mathcal{M}\mathcal{N}} \bar{X}_\mathcal{N} ||_{F}^2,
\end{equation}
with $\bar{X}_\mathcal{M} := [X_\mathcal{M}, \mathds{1}_{n}] \in \mathbb{R}^{n \times (d + 1)}$ denoting the homogeneous coordinates of shape $\mathcal{M}$, where $\mathds{1}_{n} \in \mathbb{R}^{n}$ is the all-ones vector. 
The homogeneous coordinates $\bar{X}_\mathcal{N}$ are defined analogously.
An iterative approximation to this formulation was popularised by the seminal iterative-closest point (ICP) algorithm~\cite{besl1992method}, see also \cref{sec:deformation}.
Other works~\cite{maron2016point} propose convex relaxations for the Procrustes matching problem.

Ren et al.~\cite{ren2021discrete} 
propose a discrete optimisation approach which computes surface maps between two shapes $\mathcal{M}$ and $\mathcal{N}$ by constraining a functional map $C_{\mathcal{M}\mathcal{N}}$ such that it can be related to a point-wise map $\Pi_{\mathcal{M}\mathcal{N}}$.
These are then also called \emph{proper functional maps}, see~\cref{eq:fmap-to-pmap}.
As such, proper functional maps have been further explored 
in \cite{pai2021fast} using convex relaxations of point-wise maps, as well as in
deep shape matching methods as a loss function~\cite{cao_unsupervised_2023, attaiki_understanding_2023}.

In contrast, Windheuser et al.~\cite{windheuser2011geometrically} enforce consistent neighbourhood relations of surface elements by utilising linear constraints in an integer linear programming formalism based upon a so-called \emph{product space}.
While mathematically elegant, this formalism is hard to solve in practice and thus Windheuser et al.~\cite{windheuser2011geometrically} propose to solve it using linear programming (LP) relaxations and iterative variable fixations. 
Yet, even with this approximate scheme, it only scales to shapes with around $50-100$ triangles on modern hardware and with modern LP solvers; see~\cite{roetzer2025geco} for runtime comparisons.
This inspired follow-up works~\cite{roetzer_scalable_2022, roetzer2024discomatch} to develop approximate solvers for this formalism.
Both solvers~\cite{roetzer_scalable_2022, roetzer2024discomatch}
build on dual decomposition schemes presented in~\cite{lange2021efficient,abbas2022fastdog} and run on CPU~\cite{roetzer_scalable_2022} and GPU~\cite{roetzer2024discomatch} respectively.
These solvers show significant improvements in scalability, i.e.~\cite{roetzer_scalable_2022} scales to shapes with up to a $1000$ triangles and \cite{roetzer2024discomatch} to $750$ triangles.
Yet, due to runtime and memory limitations, scalability is nevertheless limited. 

\begin{figure}
    \centering
    \includegraphics[width=0.5\linewidth]{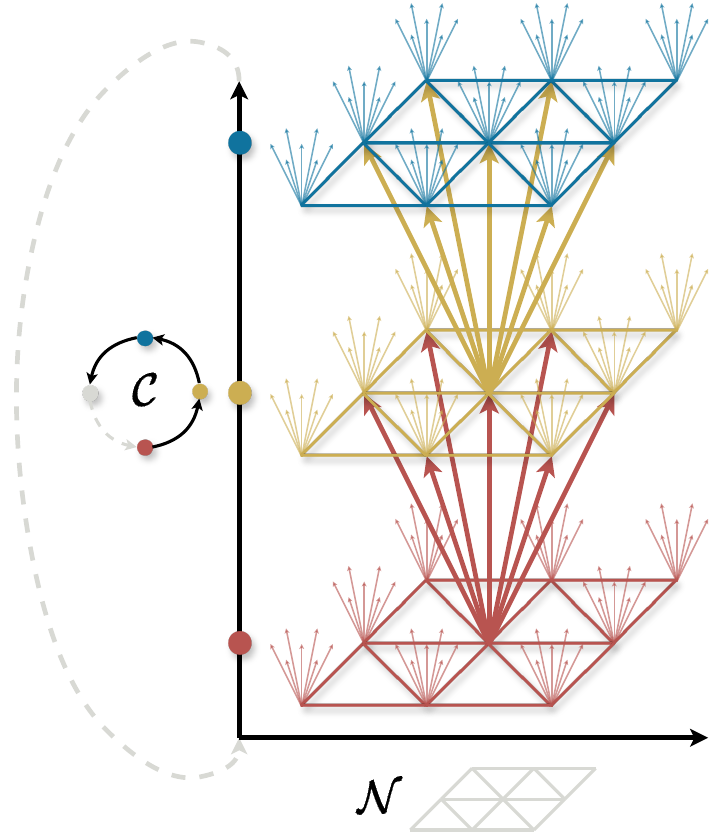}
    \caption{Visualisation of the (in-layers-structured) \textbf{product graph} $\mathcal{P}$ between a cyclic chain graph $\mathcal{C}$ and the graph of a triangle mesh of a 3D shape $\mathcal{N}$. Correspondences between $\mathcal{C}$ and $\mathcal{N}$ can be computed by finding the shortest cyclic path that visits all layers. Image source: \cite{roetzer_spidermatch_2024}.}
    \label{fig:prodgraph}
\end{figure}

This motivated follow-up works to develop new formalisms based on \emph{product graphs}. Product graphs are denoted 
\begin{equation}
    \mathcal{P}=\mathcal{G}_1\boxtimes\mathcal{G}_2,
\end{equation}
where the \emph{strong product} 
between two graphs $\mathcal{G}_1$ and $\mathcal{G}_2$, see~\cite{hammack2011handbook} for more details and \cref{fig:prodgraph} for a visualization.
In the past, these product graphs were successfully used for dynamic time-warping~\cite{sakoe1978dtw},
i.e.~finding point-to-point correspondence between two contours with start and end,
matching a shape to an image~\cite{schoenemann2009combinatorial}, 2D-to-2D shape matching~\cite{schmidt2009planar}, and 2D-to-3D shape matching~\cite{lahner2016efficient}.
To obtain an efficient-to-solve formalism for 3D-3D shape matching, SpiderMatch~\cite{roetzer_spidermatch_2024} changes the representation of shape $\mathcal{M}$, i.e.~the \emph{source} shape:
it uses a single cyclic, self-intersecting curve $\mathcal{C}$ on the surface of the shape $\mathcal{M}$ to represent the respective shape.
This allows casting the 3D shape matching problem as finding a minimum cost cyclic path in a product graph $\mathcal{P}$ with additional constraints.
As shown in~\cite{roetzer_spidermatch_2024}, this problem is globally optimal and efficiently solvable in practice, scaling to shape resolutions of up to $3500$ triangles.

In~\cite{roetzer2025higherorder}, authors utilise the curve representation proposed in SpiderMatch~\cite{roetzer_spidermatch_2024}, drop the additional constraints, and show that good shape matching results can be computed when using minimum ratio cycles with local rigidity regularisers~\cite{sorkine2007rigid,roetzer_conjugate_2023}, achieving global optimality in practice.

Contrary to these methods, \cite{roetzer2025geco} argues that representing a 3D shape with a single cyclic curve $\mathcal{C}$ and matching this curve to the target shape $\mathcal{N}$ does not lead to a faithful surface-to-surface matching.
Therefore, \cite{roetzer2025geco} proposes to use multiple cyclic curves $\mathcal{C}_1,\dots,\mathcal{C}_n$ to represent each
triangle of the source shape $\mathcal{M}$.
This allows formulating 3D shape matching via multiple product graphs $\mathcal{P}_1,\dots,\mathcal{P}_n$ as finding multiple shortest cyclic paths which have to adhere to additional coupling constraints. 
In turn, this leads to a faithful surface-to-surface matching that is globally optimal in practice and scales to shapes with up to $3000$ triangles.
Building on this formalism, \cite{amrani2025highres} propose a two-stage approach in which they first transfer patches from source to target shape 
and subsequently find dense surface-to-surface matchings for each pair of corresponding patches.
This allows scaling to shapes with up to $10$k triangles without the need for coarse-to-fine strategies.
This has the advantage of fewer discretisation artifacts which usually occur in approaches relying on coarse-to-fine strategies.

While \cite{roetzer2025geco} has shown that multiple coupled problems lead to high-quality matching results, these coupled problems only involve two 3D shapes.
As a result, their formalism is not directly applicable to problems that involve collections of shapes, which we discuss next.

\subsection{Multi-shape matching approaches}
Multi-shape matching approaches simultaneously optimise maps between a collection of shapes.
Assuming bijective mappings between shapes, \emph{cycle consistency} emerges as a fundamental property that should be fulfilled in most cases~\cite{schmidt2007intrinsic,huang2013consistent,cosmo2017consistent}.
This means that the matching from a vertex on one shape via multiple other shapes back to the first shape must end in the same vertex again.
For non-bijective mappings between shapes, the definition of cycle consistency is more intricate~\cite{bernard2019synchronisation}.
In particular, considering the bijective case, for a collection of $N$-many shapes $\mathcal{M}_1,\dots,\mathcal{M}_N$, it must hold that
\begin{equation} \label{eq:cycle-consistency}
    \forall i,j,k \in \{1,\dots,N\}:\; \Pi_{\mathcal{M}_i\mathcal{M}_j} \Pi_{\mathcal{M}_j\mathcal{M}_k} = \Pi_{\mathcal{M}_i\mathcal{M}_k}.
\end{equation}
\cite{bhatia_ccuantumm_2023} enforce cycle consistency constraints by considering triplets of shapes and by proposing a quantum-hybrid approach building on ideas of the famous alpha-expansion algorithm~\cite{boykov2002fast}.
\cite{eisenberger2023g} learns maps between multiple shapes, which are used as edge weights for all pairs $i,j \in \{1,\dots,N\}$ in the shape graph, to then propagate maps along shortest paths in the shape graph to enforce cycle consistency.
\cite{kahl2025towards} propose an approximate algorithm for cycle-consistent multi-graph matching, which exploits parallelism, performs local graph matching and maintains cycle consistency at all times.

To avoid enforcing the cycle consistency constraints \eqref{eq:cycle-consistency} explicitly, other works revert to shape-to-universe matching where cycle consistency is enforced implicitly~\cite{pachauri2013solving}.
Among these, \cite{gao2021isometric} introduces a multi-shape matching formalism for isometric deformed shapes and further proposes an algorithm based upon convex relaxations.
Later, the shape-to-universe matching scheme was also employed in unsupervised learning approaches~\cite{cao2022unsupervised}, see also \cref{tab:spectral_methods}.

Multi-shape matching approaches and many other combinatorial methods are most often hard to solve and as a consequence, on classical hardware, their scalability is limited. 
Motivated by that, other lines of work aim for formalisms which can be employed on quantum hardware
which can find global optima naturally in some configurations, 
such that formalisms might become much more scalable with advances in quantum computing.

\subsection{Quantum annealing}

Quantum annealers can efficiently solve quadratic unconstrained binary optimisation problems in the form of
\begin{equation}
 \min_{x \in \{0,1\}^n} x^\top Q x + q^\top x. \label{eq:qubo}
\end{equation}
Here $Q\in\mathbb{R}^{n\times n}$ is a symmetric (possibly dense) cost matrix containing pairwise costs and $q \in \mathbb{R}^{n}$ a linear bias. 
Current available systems, for example, the D-Wave line~\cite{dwave_website}, contain several thousand qubits (\emph{quantum bits}).
In theory, the number of qubits corresponds directly to $n$ in \eqref{eq:qubo}, however, due to hardware limitations in the connectivity of the annealer, this only holds for certain forms of $Q$ and in reality, many qubits are blocked when representing a dense $Q$ matrix~\cite{seelbach2025compensating}. 
The natural use of binary variables makes quantum annealers primed for solving combinatorial versions like \cref{eq:combinatial-general}, but the unconstrained setting cannot model many of the desired properties of the solution, such as bijectivity.
The first approach encoded the permutation constraint in $Q$, but this occupies most of the qubits on fulfilling these constraints and could only correspondence between up to 4 points~\cite{benkner2020adiabatic}. 
However, by solving the problem iteratively in subproblems based on a version of $\alpha$-expansion, quantum annealers can be scaled to several thousand vertices~\cite{seelbach2021qmatch} and even solve the hard multi-shape case with a linear runtime scaling in the number of shapes~\cite{bhatia_ccuantumm_2023}.
The permutation set can also be parametrised by the binary variables directly, which leads to an efficient approximation problem without the need for an iterative approach~\cite{meli_qucoop_2025}.
For an introduction to quantum annealing in the context of computer vision and more details on correspondence estimation approaches relying on quantum computational paradigms, interested readers can refer to the recent survey by Kuete Meli et al.~\cite{meli2025quantum}.

\section{Deformation-based methods}
\label{sec:deformation}

\begin{figure}
    \centering
    \includegraphics[width=\linewidth]{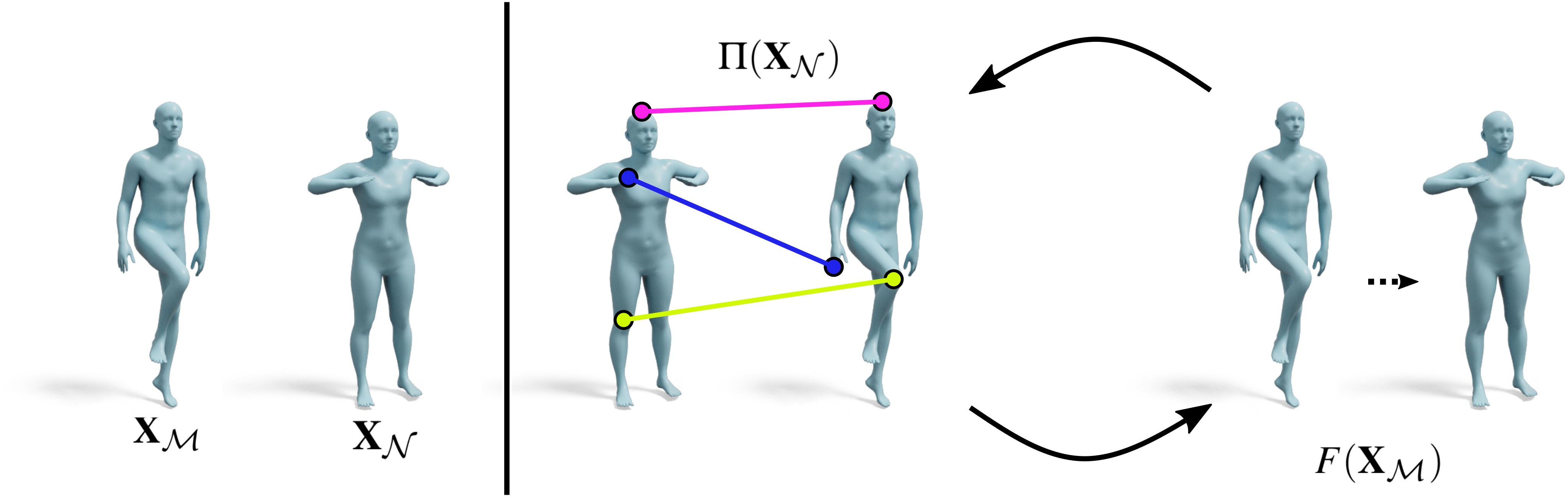}
    \caption{\textbf{Deformation methods} recover the correspondence between shapes by finding a deformation that aligns them spatially.}
    \label{fig:deform_2}
\end{figure}

\begin{table*}[ht]
    \footnotesize
    \centering
    \begin{tabular}{l@{\hspace{0.3em}}c@{\hspace{0.5em}}c@{\hspace{0.5em}}c@{\hspace{0.5em}}c@{\hspace{0.5em}}c}
        \toprule
         \textbf{Method} & \textbf{Category} & \textbf{Correspondence} & \textbf{Embedding Space} & \textbf{Deformation Model} & \textbf{Training} \\ 
         \midrule
        Revisited-CPD \cite{fan2022coherent} &  \textbf{O} & Soft & Euclidean & Coherent & \NONE \\ 
        Geo-CPD \cite{hirose2022geodesic} &  \textbf{O} & Soft & Manifold & Coherent & \NONE \\
        Tajdari et al. \cite{tajdari2022feature} &  \textbf{O} & Hard & Euclidean & Affine & \NONE \\
        Yao et al. \cite{yao2023fast} &  \textbf{O} & Hard & Euclidean & Locally-affine & \NONE \\
        Zhao et al. \cite{zhao2024correspondence} &  \textbf{O} & Soft & Euclidean & Kernel-based & \NONE \\
        Merrouche et al. \cite{merrouche2025matching} &  \textbf{O} & Hard & Neural & ARAP & \NONE \\
        SPARE \cite{yao2025spare} &  \textbf{O} & Hard & Euclidean & ARAP & \NONE \\
        \midrule
        Jiang et al. \cite{jiang2023non} &  \textbf{H} & Neural & Neural & ARAP & \SUP \\
        Chen et al. \cite{chen2023rethinking} &  \textbf{H} & Global & Neural & Parametric & \SUP \\ %
        Qin et al. \cite{qin2023deep} &  \textbf{H} & Neural & Neural & Parametric & \SUP \\
        ArtEq \cite{Feng_2023_ICCV} &  \textbf{T} & Neural & Semantic & Parametric & \SUP \\ 
        Guo et al. \cite{guo2024diffusion} &  \textbf{H} & Hard & Neural & Coherent & \SUP \\ %
        ETCH \cite{li2025etchgeneralizingbodyfitting} &  \textbf{T} & Hard & Manifold & Parametric & \SUP \\
        Jung et al. \cite{jung2025variable} &  \textbf{T} & Hard & Euclidean & Kernel-based & \NONE \\
        \midrule
        NDP \cite{li2022non} &  \textbf{L} & Neural & Neural & Neural Field & \SUP \\
        IFS \cite{avidan_implicit_2022} &  \textbf{L} & Neural & Implicit & Neural Field & \UNSUP  \\
        LVD \cite{corona2022learned} &  \textbf{H} & Neural & Neural & Neural Field & \SUP \\ %
        Sundararaman \cite{sundararaman_deformation_2024} &  \textbf{L} & Neural & Spectral & Jacobian-based & \SUP\\
        Merrouche \cite{Merrouche_2023_BMVC} &  \textbf{L} & Neural & Neural & Hierarchical & \UNSUP \\
        Lin et al. \cite{lin2023leveraging} &  \textbf{L} & Neural & Manifold & Neural Field & \UNSUP \\
        Feng et al. \cite{feng2023differentiable} &  \textbf{L} & Neural & Neural & Parametric & \SUP \\
        NICP \cite{marin2024nicp} &  \textbf{H} & Neural & Neural & Neural Field & \SUP \\
         \bottomrule
    \end{tabular}
    
    \caption{Overview of \textbf{deformation-based} matching methods. Methods are categorized by their approach (\textbf{O}ptimisation, \textbf{T}emplate, \textbf{L}earning, or \textbf{H}ybrid), correspondence type (Hard, Soft, Neural, Global), feature space, deformation model, and training requirements (\NONE one, \SUP upervised, or \UNSUP nsupervised).
    }
    \label{tab:deformation_methods}
\end{table*}

So far, the methods that we considered retrieve the correspondence without modifying the input shapes, mainly by relying on intrinsic properties of the geometry. 
In this section, we instead present methods that recover correspondence by computing a deformation field that aligns a source surface with a target surface in the ambient 3D coordinate space.
Unlike \textit{intrinsic} methods, which rely on surface-aware metrics like geodesic distances, this \textit{extrinsic} paradigm fundamentally considers Euclidean metrics and is thus broadly applicable to various 3D data representations, including unstructured point clouds, without requiring manifold connectivity or consistent topology between the shapes being registered.
It is worth mentioning that while the majority of such works define the problem in terms of point sets, they are commonly applied to mesh vertices.
Surface properties are often used both for regularisation and as a domain for point sampling. 

In revisiting the literature of recent years, we categorise deformation works into three main branches: optimisation-, template-, and learning-based. 
Drawing a clear distinction among them is impossible, as many methods are hybrid, and our categorisation follows what we believe are the main principles behind the paper.
We additionally emphasise the type of optimisation that is conducted, feature embedding space, correspondence type, and deformation model (see ~\cref{tab:deformation_methods}).
We begin with background on seminal techniques that continue to serve as inspiration.

\subsection{Background}
While intrinsic methods come with a number of advantages, such as invariance of the shape's embedding in the ambient Euclidean domain and strong structural priors, their theoretical premises are also a source of drawbacks. 
Noise, topological changes, as well as incomplete or cluttered shapes, often introduce enough instability to make purely intrinsic methods inapplicable. 
On the contrary, we have already seen that considering the extrinsic information of shapes can play an important role, for example, in distinguishing symmetries, and provide semantical nuances difficult to catch on a pure surface level.
Along this line, several methods rely on the assumption that the correspondence is equivalent to a certain, generally unknown, deformation. For the sake of simplicity, we formalise the problem in an idealised bijective setting. 
Considering two sets of points $X_{\mathcal{M}}$ and $X_{\mathcal{N}}$, and assuming there exist a permutation $\Pi: X_{\mathcal{N}} \rightarrow \widehat{X}_{\mathcal{N}} \in \mathbb{R}^{n_{\mathcal{M}} \times 3}$ that reorganizes their vertices under the desired semanticity, the idea is to recover an equivalent deformation $\mathcal{F}:X_{\mathcal{M}} \rightarrow \widehat{X}_{\mathcal{M}} \in \mathbb{R}^{n_{\mathcal{M}} \times 3}$ such that:
\begin{equation}
    \mathcal{F}(\mathbf{X}_{\mathcal{M}}) \approx {\Pi}(\mathbf{X}_{\mathcal{N}}).
    \label{eq:deformation}
\end{equation}
It is important to outline a few observations, which will help us anticipate the challenges addressed in the literature. First, perfect alignment can only occur when we have access to a perfect deformation model, and the two shapes share the same discretisation; this generally holds only for the synthetic settings. Second, given the permutation ${\Pi}$ or the deformation $\mathcal{F}$, the other can be recovered in a closed form \cite{besl1992method}; however, the standard case (and the one we will focus on) assumes that both are unknown, an ill-posed problem with many plausible solutions $\mathcal{F}$ and $\Pi$. 
Finally, it is often the case that only subsets of the points can be correctly aligned; clutter, noise, and outliers often require careful consideration and ad-hoc techniques. 
For these reasons, the problem is classically phrased as an energy minimisation.

\paragraph{Energy-based formulation.}
Solving \cref{eq:deformation} via optimisation requires defining an energy, which aims to align the shapes, but also recovers a plausible deformation. 
Hence, the energy is generally composed of data ($\Energy_\text{align}$) and regularization ($\Energy_{\text{reg}}$) terms, with a weighting factor $\alpha$:
\begin{align}
\Energy = \Energy_{\text{align}} + \alpha \Energy_{\text{reg}}
\end{align}

\paragraph{Seminal registration methods.} %
A survey on non-rigid registration would be incomplete without discussing two seminal works, which still serve as inspiration for many recent methods: Iterative Closest Point (ICP) ~\cite{besl1992method} and Coherent Point Drift (CPD) \cite{myronenko2010point}.
Since both the correspondence and the deformation are unknown, ICP proposes iterating between the two, fixing one and optimising the other. Starting from an initial configuration of the two shapes (i.e., $\widetilde{\mathcal{F}}(X_{\mathcal{M}}) = X_{\mathcal{M}}$), ICP obtains an estimated correspondence $\widetilde{\Pi}$ by Euclidean nearest-neighbour pairing:
\begin{equation}
    \widetilde{\Pi} = \arg\min\limits_{\Pi} \|\widetilde{\mathcal{F}}(X_{\mathcal{M}}) - \Pi(X_{\mathcal{N}})\|_2^2.
    \label{eq:corr}
\end{equation}
Next, the registration is updated $\widetilde{F}$ by minimizing the point-to-point distance with a least-squares objective:
\begin{equation}
    \widetilde{\mathcal{F}} = \arg\min\limits_{\mathcal{F}} \sum\limits_{i = 1}^{m}\|\mathcal{F}(X_{\mathcal{M}_i}) - \widetilde{\Pi}(X_{\mathcal{N}})_i\|_2^2.
    \label{eq:def}
\end{equation}
The two steps are iterated until convergence. 
Notice that to avoid trivial solutions, $\mathcal{F}$ should be restricted to a specific family, which originally was the class of rigid transformations~\cite{besl1992method}. 
Over the past several decades, numerous non-rigid variations have been proposed. 
The first non-rigid extension was proposed by Amberg et al. \cite{amberg2007optimal}, which revises the transformation in terms of local affine transformations, ensuring that changes are minimal between nearby vertices. 
By fixing the amount of desired stiffness, the authors derive a cost function that still admits a closed-form to recover an optimal iterate. 

As opposed to the discrete, nearest-neighbour-based approaches pioneered by ICP, probabilistic methods generate soft matches where each point in the source set $X_{\mathcal{M}}$ has a certain probability of corresponding to every point in the target set $X_{\mathcal{N}}$. 
The seminal Coherent Point Drift (CPD) algorithm by Myronenko and Song \cite{myronenko2010point} reformulates the geometric alignment problem through statistical density estimation. 
CPD treats the source points as centroids of a Gaussian Mixture Model (GMM), while the target points are considered as observations generated from this GMM. 
To handle outliers and noise, CPD augments the mixture model with a uniform distribution component, yielding the probability density:
\begin{equation}
p(\mathbf{x}) = w \cdot \frac{1}{n_{\mathcal{N}}} + (1-w) \sum_{i=1}^{n_{\mathcal{M}}} \frac{1}{n_{\mathcal{M}}} \mathcal{N}(\mathbf{x} | \mathcal{F}(x_i), \sigma^2 I)
\end{equation}
where $w \in [0,1]$ represents the assumed proportion of outliers, $\mathcal{F}$ denotes the transformation function, and $\sigma^2$ is the isotropic variance of the Gaussian components.
The registration problem can then be approximated through maximum likelihood estimation (MLE).

Since the two streams of work rely on different definitions of data fitting, the resulting regularisation techniques also differ. For ICP, the deformation is often chosen within a specific class, such as $\mathrm{SE}(3)$~\cite{besl1992method} or local Affine transformations \cite{amberg2007optimal}, which naturally restrict the valid solutions. 
CPD instead admits free displacements and thus relies on Motion Coherence Theory (MCT), penalising derivatives of all orders of the displacement field and ensuring neighbouring points are deformed consistently (this results in neighbouring points moving \textit{coherently}).

\subsection{Revisiting classical methods}
\begin{figure*}
    \centering
    \includegraphics[width=\linewidth]{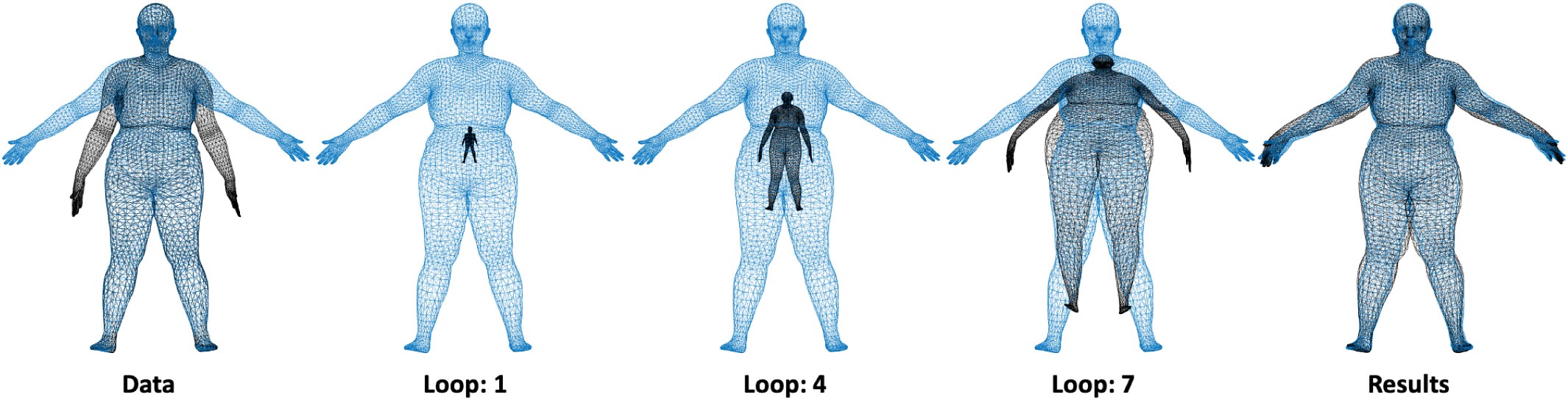}
    \caption{\textbf{An example of deformation} from GeoCPD \cite{hirose2022geodesic}. The source shape (black) is deformed toward the target one (blue), toward multiple steps till final convergence.}
    \label{fig:deform_1}
\end{figure*}

\paragraph{Optimisation-based.}

The standard point-to-plane metric as used in ICP only considers the normals of the static target surface. 
The recent SPARE \cite{yao2025spare} method thus proposes a symmetrised point-to-plane distance metric. 
This formulation incorporates normal information from both the deforming source and the target surfaces, providing a more accurate, second-order approximation of the local geometry, which can accelerate convergence and improve final accuracy \cite{yao2025spare}.
Alternatively, the problem of aligning distribution can be rephrased as a fuzzy clustering problem \cite{zhao2024correspondence}.
Beyond penalising smoothness through motion coherence or stiffness terms, geometrically derived features can provide additional constraints \cite{tajdari2022feature,hirose2022geodesic}.

To address noise, outliers, and acquisition artefacts, optimisation can be made more resilient through robust norms such as the $L_p$ norm with $p<2$ (e.g., $L_1$) for both alignment and regularisation terms \cite{yao2020quasi,zhao2024correspondence}.
Unlike the $L_2$ norm, which quadratically penalises large errors and is thus highly sensitive to outliers, these norms apply a less severe penalty, making the overall cost function more resilient to gross errors.
However, many robust norms are non-smooth and non-differentiable, complicating optimisation and inhibiting standard gradient-based solvers.
Recent work by Yao et al. \cite{yao2023fast} addresses this through the Majorisation-Minimisation (MM) algorithm with Welsch's function \cite{holland1977robust}, enabling closed-form solutions that prove computationally superior to non-smooth proximal methods (such as ADMM \cite{boyd2011distributed}) while maintaining robustness.

Despite the use of the Fast Gaussian Transform, the speed of the CPD algorithm remains a crucial drawback \cite{golyanik2016extended}. 
Subsequent works have significantly advanced the regularization framework: Hirose proposed an accelerated version using downsampling \cite{Osamu2021Acceleration} and later introduced a Bayesian formulation that generalizes both CPD and ICP by rephrasing coherence as a distribution prior \cite{Osamu2021Bayesian}, while also extending the regularization to use geodesic distances rather than Euclidean ones to better preserve the intrinsic geometry of the surface during deformation \cite{hirose2022geodesic}. 
Other extensions have adapted the coherence constraint to non-Euclidean settings \cite{fan2022coherent} and incorporated topological preservation through joint optimisation of explicit and implicit representations \cite{merrouche2025matching}.

\paragraph{Use of learned priors.}

Learned features can be used to complement non-rigid deformation-based approaches.
For instance, Jiang et al. learn deep features via functional maps (see \cref{sec:spectral}), which induces an initial set of point-to-point correspondences $\hat{\Pi}$ \cite{jiang2023non}. 
This permutation estimate can then be filtered based on geometric consistency and used to recover the deformation function $\mathcal{F}$. 
Guo et al. \cite{guo2024diffusion} address cloth deformation by learning a diffusion-based prior from real-world clothing deformations. 
Learned methods can be used to prune correspondences in a manner more suitable for non-rigid deformations \cite{qin2023deep}.
As with some functional map-based methods \cite{attaiki2021dpfm}, transformer-based modules can be useful in guiding overlap prediction to handle deformations between partial shapes \cite{mei2023overlap}.
Auxiliary tasks can also yield learned feature representations useful for registration \cite{chen2023rethinking}.

\subsection{Template fitting}

For certain classes, we might have access to an archetypal shape that describes the structure, and often comes with information on how shapes of that class are expected to deform. 
Such a model is referred to as a template, and when it is available, it provides a strong regularisation. 
Such a template is generally crafted by an artist or domain experts \cite{gao2023automated, dale1999cortical}.
The template can also be enriched with a learned parametric deformation space by aligning it to a collection of shapes. A seminal work in this sense is that of Blanz and Vetter \cite{blanz2023morphable}, which aligned a few hundred human faces to learn PCA, both to model the subject's geometry and texture. 
More recent shape models are the SMPL models for humans \cite{loper2015smpl, pavlakos2019smplx}, SMAL for the animals \cite{zuffi2017small}, and MANO for the hands \cite{romero2017mano}, but learning such morphable shapes is also popular in medicine \cite{MacFra_Unraveling_MICCAI2025, magnet2023assessing, posner2025statistical}. 

While generally such a procedure requires a large amount of curated data \cite{achenbach2017fast, loper2015smpl, pons2015dyna, zuffi2024varen}, it is worth mentioning that recent approaches also leverage data in the wild, such as images and videos from the web \cite{li2024learning}.

\paragraph {Fitting.} To take advantage of the available template, its structural properties can be used during the optimisation. For example, humans can be represented as piece-wise rigid deformations, as done in the Stitched Puppet formulation \cite{zuffi2015stitched}. 
Recent methods make use of local $\mathrm{SO}(3)$-equivariance features to represent the local rigidity \cite{li2025etchgeneralizingbodyfitting,Feng_2023_ICCV}, though in practice, such equivariance assumes little or no noise. 
Since templates provide a smooth and complete geometry, they can be integrated with intrinsic methods, enhancing their robustness \cite{marin2020farm}. 
It is likely that a template cannot catch all the nuances of a specific class. Hence, it is often the case that to align it to the data, the local details are delegated to free-form offsets. 
Examples are for skin details \cite{marin2019high}, cloths \cite{ma2020learning}, and soft-tissues \cite{pons2015dyna}. 
Jung et al. take an approach of jointly optimising a shared template with respect to a collection of input shapes \cite{jung2025variable}.
The template fitting can also be enriched by visual semantic information present on the surface, such as textures \cite{antic2024close, lazova2019360}.
Recent developments in capturing technologies have also enabled the study of fine-grained template deformation, such as that in fabrics. 
Modelling wrinkles can be achieved by physics modelling \cite{yu2025phydeformer} or learned-priors \cite{guo2024diffusion}.

\subsection{Learning-based methods}
With the advent of deep learning, many methods leverage deep neural networks trained to learn a deformation model from a collection of data.  
3DCoded \cite{groueix_3d-coded_2018} popularised this line of work, training an autoencoder to deform a template into all the shapes of the training set. 
In case ground truth registration is available, such a network can be trained in a supervised way using a vertex-to-vertex loss. 
On the contrary, it can also be trained without supervision, relying on Chamfer Distance. 
In the follow-up \cite{deprelle2019learning}, the same authors also investigated the possibility of learning primitive structures together with the deformation module. 
A revision of the 3DCoded formulation has also been proposed with Transformer architecture \cite{trappolini_shape_2021}, where the network relies on an attention mechanism that weights points by their area.

The deformation can be represented globally, for instance, as a set of hierarchical MLPs \cite{li2022non}.
Merrouche et al. \cite{Merrouche_2023_BMVC} suggest training both an association and a deformation network relying on hierarchical features.
Temporal network modules such as GRU can be used to represent the deformation incrementally, enhancing robustness \cite{feng2023differentiable}.
A more nuanced geometric view is achieved through Neural Jacobian Fields (NJF) \cite{aigerman_neural_2022}, which first learns a matrix field in the ambient space and then derives the Jacobian for each triangle via an intrinsic projection. 
A final global deformation can then be recovered by solving a Poisson equation based on the locally learned Jacobians.
Sundararaman et al. build upon NJF by abandoning the need for a global shape encoder, representing shapes' vertices with their local Jacobians, learning a coarse-to-fine deformation entirely based on local properties \cite{sundararaman_deformation_2024}.

Particularly convenient is relying on implicit representations, since they provide spatial and differentiable information useful for the fitting. 
Additionally, they offer a flexible representation that is significantly more robust to noise and partial information. Sundararaman et al. propose representing shapes as neural fields and utilising implicit representations to regularise deformation \cite{avidan_implicit_2022}. 
On a similar principle, they propose to learn the deformation as local fields \cite{sundararaman_reduced_2022}. 
In IPNET  \cite{bhatnagar2020combining}, the authors combine a deformable template with a network to predict the occupancy of local body parts, which is then used to guide the SMPL registration. 
Wang et al. \cite{wang2021locally} propose PTF, a similar framework that relies on piece-wise implicit representations. LoopReg \cite{bhatnagar2020loopreg} also proposes a weakly-supervised schema for shape registration. 

While these works rely on implicit representations to represent the shapes, a more recent trend is to leverage neural fields to also represent deformations as a velocity field defined all over $\mathbb{R}^3$, as proposed in Learned Vertex Descent \cite{corona2022learned}. 
The motivation is that they provide a richer representation for the deformation, on which further supervision can be provided. While the original paper utilises the neural field primarily during training, Marin et al. \cite{marin2024nicp} demonstrate that this formulation can also be beneficial at inference time. 
In practice, they propose Neural ICP (NICP), a procedure to iteratively refine the neural deformation field by retrieving its induced correspondence. Neural deformation fields have also been defined intrinsically, in the clothing deformation case \cite{lin2023leveraging}. Lastly, in the majority of previous works, the correspondence step is delegated to features coming from neural backbones.

\section{Challenges and emerging trends}
\label{sec:emerging}

In this section, we explore several opportunities that have recently emerged in shape correspondence research.
These include leveraging foundation models, preserving local neighbourhoods during matching, and addressing partial shape scenarios.
Such directions extend the capabilities of conventional methods, enabling them to handle more challenging and previously inaccessible settings.

\subsection{Feature aggregation from foundation models}

\newcommand{\nonisometricWidth}{0.265\columnwidth}
\newcommand{\nonisometricHeight}{2.65cm}
\begin{figure}[t]
    \centering

    \begin{tabular}{cc}
        \small

         \includegraphics[height=\nonisometricHeight,width=\nonisometricWidth]{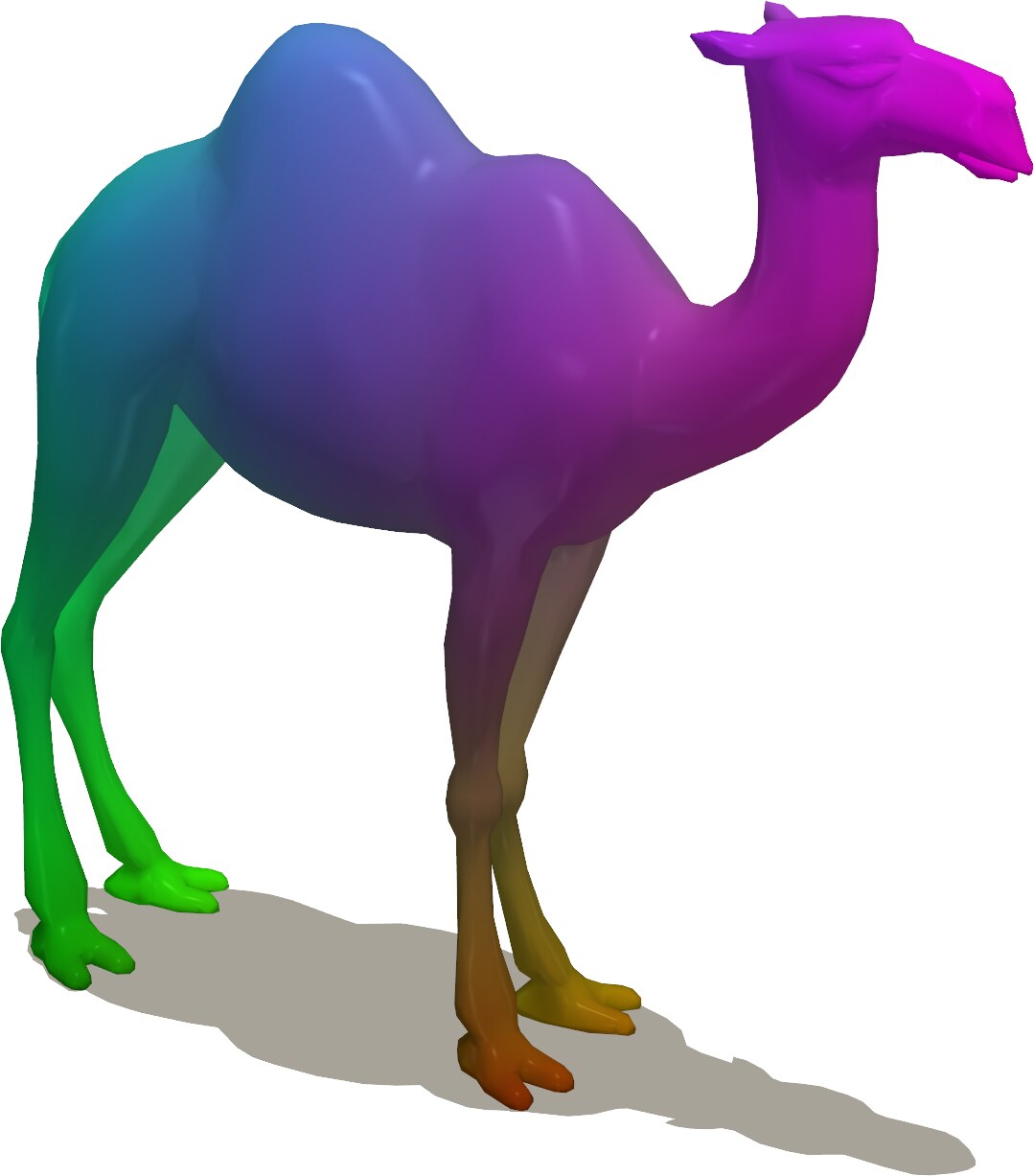}&
         \includegraphics[height=\nonisometricHeight,width=0.26\columnwidth]{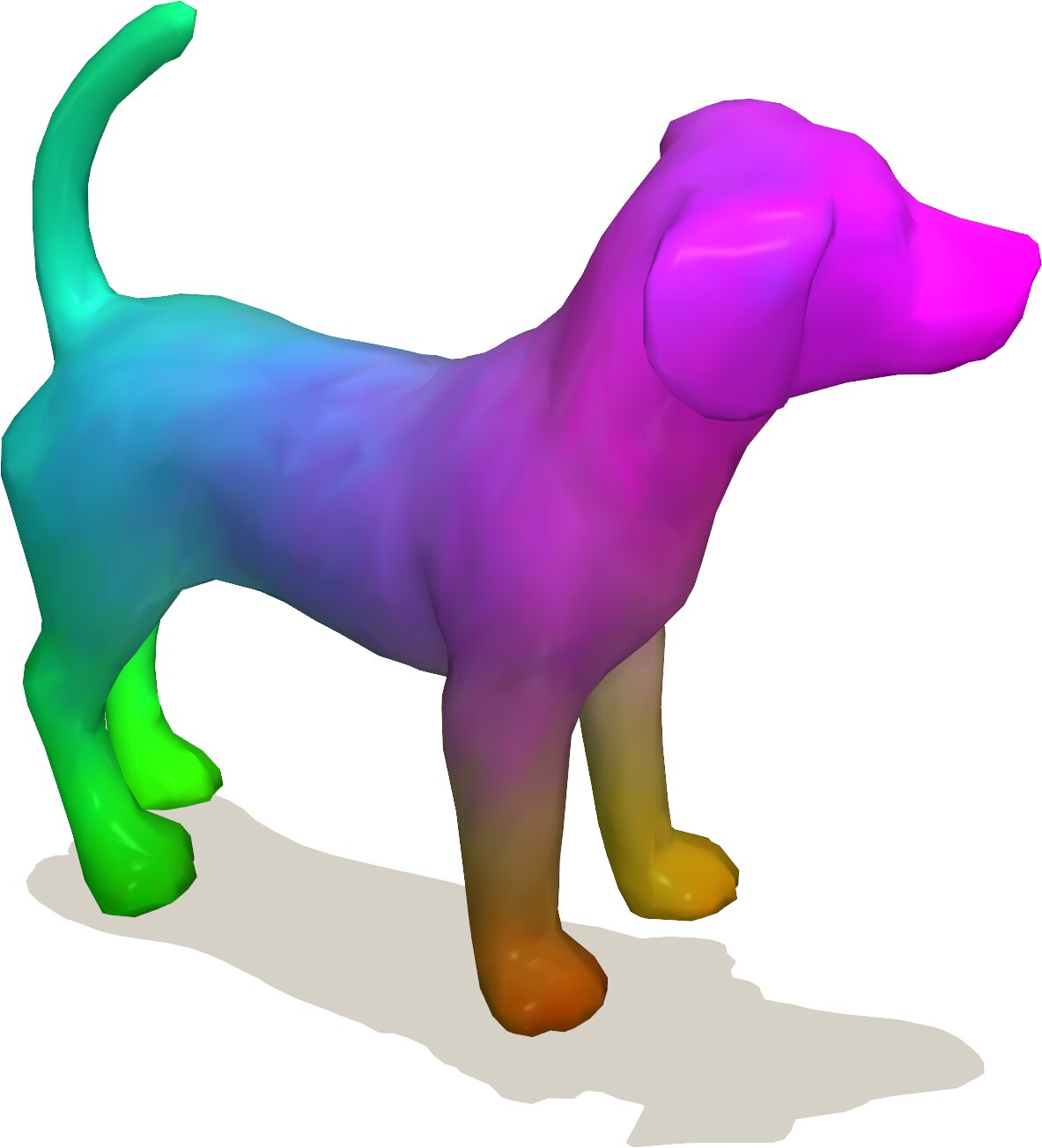}\\

         \includegraphics[height=\nonisometricHeight,width=\nonisometricWidth]{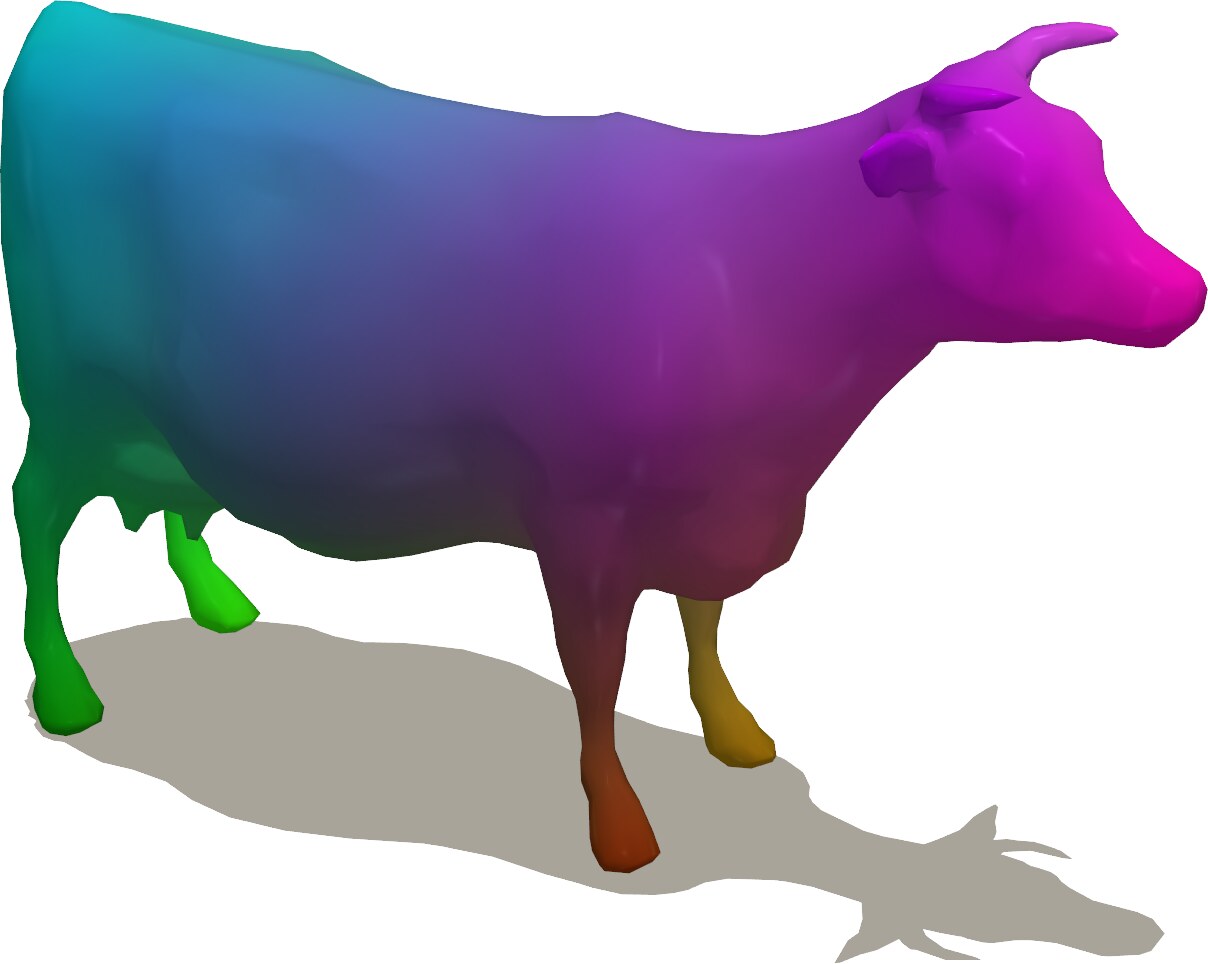}&
         \includegraphics[height=\nonisometricHeight,width=\nonisometricWidth]{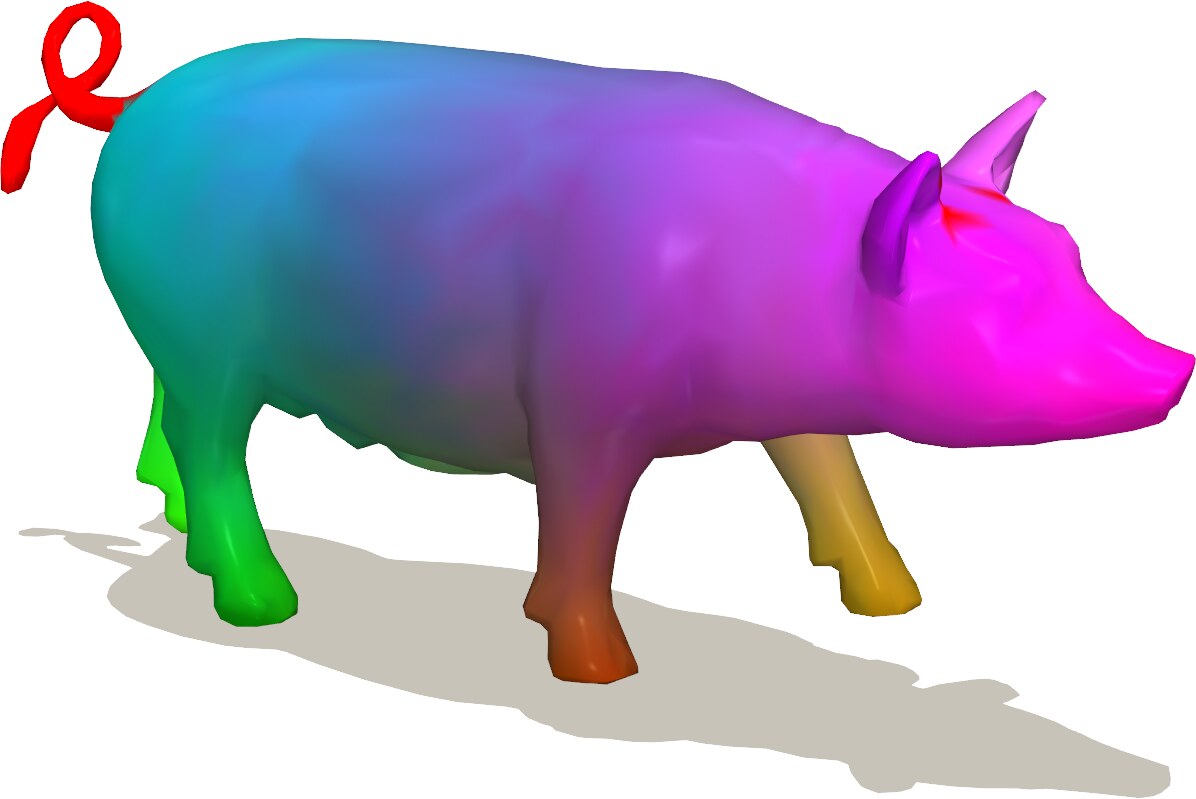}\\
         \textbf{Source} & \textbf{Target} \\
    \end{tabular}
    
    \caption{\textbf{Zero-shot matching} involves establishing correspondences between structurally distinct shape categories (such as different animal species), which often contain unique regions present on only one of the shapes (shown in red). In the absence of training data, feature descriptors from 2D vision models are particularly effective for this task. Shapes are taken from SHREC’20~\cite{dyke2020shrec} and BeCoS~\cite{ehm2025beyond} datasets.}
    \label{fig:6_2}
\end{figure}

Axiomatic and learned feature descriptors are fundamental to many correspondence estimation pipelines.
However, their key limitations lie in the former's limited expressiveness and the latter's poor generalisation to out-of-distribution shapes.
As an alternative, an increasingly popular direction is to utilise features from 2D foundation vision models trained on large-scale image datasets, such as self-supervised transformers \cite{caron2021emerging,oquab2023dinov2}, image diffusion models \cite{rombach2022high}, or segmentation models \cite{kirillov2023segment}. 
These distilled descriptors have the potential to generalise zero-shot across shape classes, without category-specific training data.

In a typical pipeline for adapting 2D vision features to 3D shape correspondence, a mesh $\mathcal{M}$ is projected to image space from $n$ viewpoints using a projection operator $\text{Proj}$:
\begin{equation}
    \text{Proj}_{\mathcal{C}_j}\left( \mathcal{M}\right)= I_j \in \mathbb{R} ^{H \times W},    
\end{equation}
\noindent where $H, W$ denote the height and width of the rendered image, with $\mathcal{C}_j$ representing the $j^{\text {th }}$ camera producing the image $I_j$.
Together with a set of additional information $\mathcal{G}$ such as depth, normal maps, or a text prompt, the rendered image $I_j$ is processed by a 2D vision model $\Psi$ to obtain $d$-dimensional features $f _j^{2D}$,
\begin{equation}
    \Psi(I_j, \mathcal{G})=  f _j^{2D} \in \mathbb{R} ^{H \times W \times d}.
\end{equation}
\noindent Known camera parameters are then leveraged to unproject features from image space to mesh vertices $f_j^{3D}$. 
These are aggregated across viewpoints to obtain shape descriptors $f_\mathcal{M}$, 
\begin{equation}
\begin{aligned}
    f _j^{2D} &\xrightarrow{\text{Proj}^{-1}} f _j^{3 D}, \quad & f_\mathcal{M} &=\frac{1}{n} \sum_{j=1}^n f _j^{3 D}.
\end{aligned}
\end{equation}

\noindent The latter can be passed to a downstream matching pipeline for computing correspondences, or combined with other axiomatic or learned descriptors.

Among the earliest works to leverage foundation models for 3D shape correspondence, Abdelreheem et al.~\cite{abdelreheem2023zero} use vision-language models~\cite{li2023blip,brown2020language} to infer shape categories and semantic regions, which are localised by a multiview object detector~\cite{liu2024grounding} and segmented using Segment Anything~\cite{kirillov2023segment}. 
The resulting masks are aggregated onto the mesh to guide the initialisation of the functional map with WKS descriptors~\cite{aubry2011wave}, and are then refined via BCICP~\cite{ren2018continuous}. 
In contrast, Diff3F renders textureless shape views and uses Stable Diffusion~\cite{rombach2022high} to synthesise textured images, extracting intermediate features and fusing them with DINO~\cite{oquab2023dinov2}. 
These are unprojected onto the mesh and used in a standard functional map pipeline~\cite{ovsjanikov_functional_2012} to compute dense correspondences.

Recent methods extend the early use of 2D vision features by integrating them into more advanced correspondence pipelines.
Morreale et al.~\cite{morreale2024neural} employ DINO to co-align shapes by estimating their upright orientation from multiview renderings and to provide semantic descriptors. 
These features are then integrated into Neural Surface Maps \cite{morreale2021neural}, a framework that computes continuous correspondences between surfaces by encoding them with dedicated neural functions. 
DenseMatcher \cite{zhu_densematcher_2024} proposes to fuse features from Stable Diffusion and DINO with HKS~\cite{tombari2010unique} descriptors and refine them using a trainable DiffusionNet \cite{sharp2022diffusionnet}, supervised by a semantic distance loss that encourages feature distances to align with semantic part similarity. 
In \cite{uzolas2025surface}, the problem of disambiguating symmetric parts on a shape is addressed by training an autoencoder network to refine Diff3F \cite{dutt_diffusion_2024} features through an objective that encourages the preservation of geodesic distances in the learned embedding space.

Several works integrate vision features into deformation-based correspondence pipelines.
In SRIF \cite{sun_srif_2024}, multiview renderings of the source and target meshes are processed by a diffusion-based image morphing model \cite{zhang2024diffmorpher}, which generates smooth transitional frames between the two shapes.
A continuous deformation field is then estimated via a normalising flow \cite{yang2019pointflow}, gradually warping the source mesh to match the target while following the reconstructed trajectory. 
In DV-Matcher \cite{chen_dv-matcher_2025}, DINO features are combined with positional encodings and processed by a Siamese network with a dual-path attention mechanism, trained to predict the deformation between source and target shapes. 
Stable-SCore \cite{liu_stable-score_2025} utilises vision features to obtain 2D correspondences between rendered views of meshes, which are then used to guide a mesh deformation model using Neural Jacobian Fields (NJF)~\cite{aigerman_neural_2022}.

We summarise the discussed methods in Table~\ref{tab:foundational}.
These approaches demonstrate the strong potential of 2D vision models for matching structurally distinct shape categories, such as animal species.
However, their limitation lies in the lack of fine-grained geometric awareness, where axiomatic or learned descriptors continue to perform strongly. 
Moreover, certain shapes may lack sufficiently discriminative textures (e.g., anatomical organs) or may not be sufficiently represented in the training distributions of these models, making them inapplicable.
Future research could investigate hybrid strategies that integrate foundation model features with other types of descriptors to leverage their complementary strengths, or explore features coming from 3D foundation models \cite{du2025hierarchical}.

\begin{table}[t]
\centering
\small

\begin{tabular}{lcc}

    \toprule
    \small
    \textbf{Method} & \textbf{FM} & \textbf{CP} \\
    \midrule
    Abdelreheem et al.~\cite{abdelreheem2023zero} & SAM + DINO & \textbf{\textcolor{red}{SP}} \\
    Diff3F~\cite{dutt_diffusion_2024} & SD + DINO & \textbf{\textcolor{red}{SP}} \\
    DenseMatcher~\cite{zhu_densematcher_2024} & SD + DINO & \textbf{\textcolor{red}{SP}} \\
    Echomatch~\cite{xie_echomatch_2025} & DINO & \textbf{\textcolor{red}{SP}} \\
    SRIF~\cite{sun_srif_2024} & DiffMorpher & \textbf{\textcolor{green!50!black}{D}} \\
    DV-Matcher~\cite{chen_dv-matcher_2025} & DINO & \textbf{\textcolor{green!50!black}{D}} \\
    Stable-SCore~\cite{liu_stable-score_2025} & SD + DINO & \textbf{\textcolor{green!50!black}{D}} \\
    Morreale et al.~\cite{morreale2024neural} & DINO & \textbf{\textcolor{blue!70!black}{O}} \\
    Uzolas et al.~\cite{uzolas2025surface} & SD + DINO & \textbf{\textcolor{blue!70!black}{O}} \\
    \bottomrule
\end{tabular}
\caption{Overview of methods using \textbf{foundation models} for shape correspondence. They are categorized by the foundation model used for descriptor extraction \textbf{(FM)}: DINO~\cite{caron2021emerging,oquab2023dinov2}, Stable Diffusion~\cite{rombach2022high}, SAM~\cite{kirillov2023segment}, DiffMorpher~\cite{zhang2024diffmorpher}, as well as the downstream correspondence pipeline \textbf{(CP)}: Spectral (\textbf{\textcolor{red}{SP}}), Deformation-based (\textbf{\textcolor{green!50!black}{D}}), or Other (\textbf{\textcolor{blue!70!black}{O}}).
}
\label{tab:foundational}
\end{table}

\subsection{Neighbourhood-preservation}
\label{sec:neigh}
Neighbourhood preservation is an important aspect in shape matching as it leads to smoother correspondences (see~\cref{fig:geometric-consistency}).
Furthermore, many downstream applications rely on the fact that matchings preserve neighbourhoods, including texture transfer, shape reconstruction, and many others.
In the literature, neighbourhood preservation is also referred to as \emph{geometric consistency}, a term which was introduced by Windheuser et al.~\cite{windheuser2011geometrically}.

\newcommand{\geoConsFigWidth}{0.3\columnwidth}
\newcommand{\geoConsFigHeight}{3cm}
\begin{figure}[t]
    \centering
    \begin{tabular}{ccc}
        \small
         \includegraphics[height=\geoConsFigHeight,width=\geoConsFigWidth]{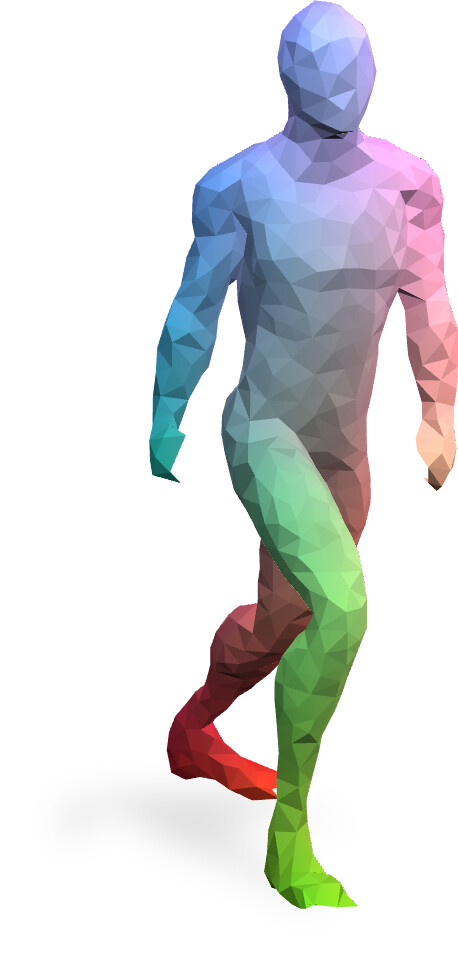}&
         \includegraphics[height=\geoConsFigHeight,width=\geoConsFigWidth]{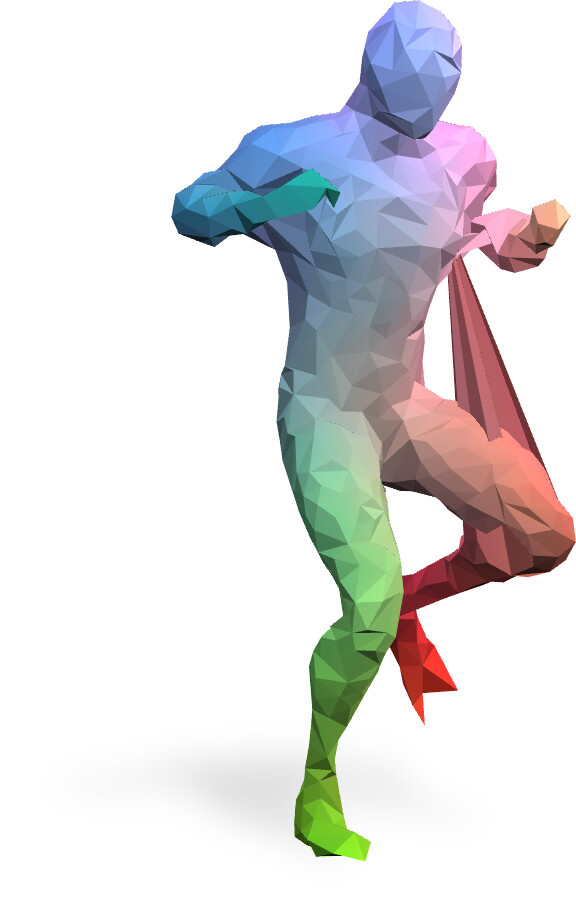}&
         \includegraphics[height=\geoConsFigHeight,width=\geoConsFigWidth]{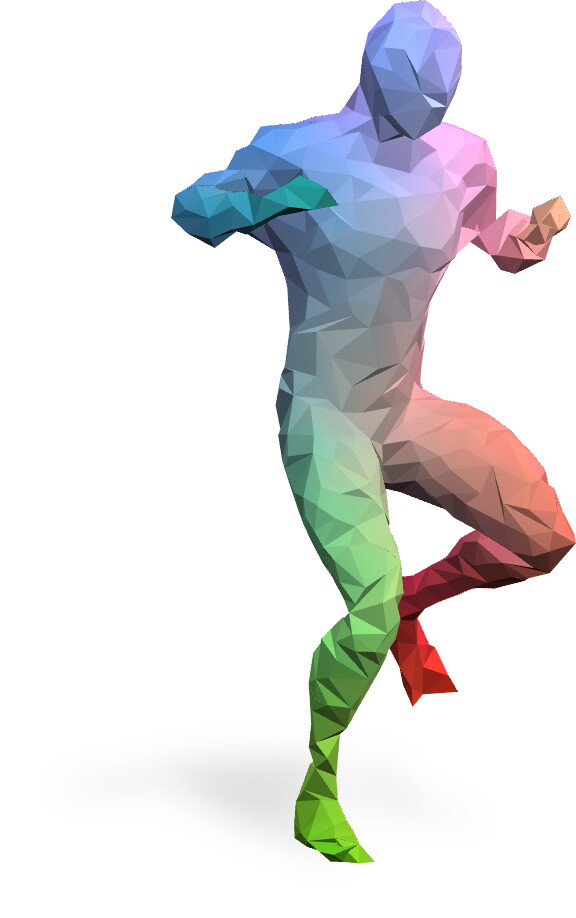}\\
         Source& \xmark\hspace{0.2cm} Neighbourhood- & \cmark\hspace{0.2cm} Neighbourhood-\\
         & preservation & preservation
    \end{tabular}
    
    \caption{Visualisation of the importance of \textbf{neighbourhood preservation} for correspondence computation by transferring triangulation from source to target shape via computing matchings. 
    Matchings that violate neighbourhood preservation can lead to arbitrary distortions of triangles.
    Image source:~\cite{roetzer2025geco}.}
    \label{fig:geometric-consistency}
\end{figure}

Despite its importance, neighbourhood preservation is often overlooked. 
The resulting optimisation problems are usually harder to solve, and, as a consequence, scalability is limited, as neighbourhood-preserving approaches typically do not scale to shapes with thousands of triangles.
For example, the LAP (not neighbourhood preserving) can be solved in polynomial time while the QAP (incorporates neighbourhood relations via costs) is generally NP-hard~\cite{rendl1994quadratic}.

Approaches that do incorporate neighbourhood preservation often use combinatorial optimisation techniques.
Thus, we pick up many of the aforementioned combinatorial methods in this section.
Nevertheless, there are other lines of work which build on other ideas.

In the following, we categorise neighbourhood-preserving methods into local and global approaches; see also \cref{tab:neighbourhood}.
Local approaches rely on initialisation in the form of sparse landmark correspondences or noisy input maps, whereas global approaches consider neighbourhood relations directly during optimisation of shape correspondences.

\begin{table}[t]
    \centering
    \small
    \setlength{\tabcolsep}{8pt}
    \begin{tabular}{l@{}ccccc}
        \toprule
        \textbf{Method} & \textbf{IF} & \textbf{S} & \textbf{TF} & \textbf{GNP} & \textbf{P}\\
        \midrule
        Takayama \cite{takayama2022compatible} & \xmark &\xmark & \xmark & \cmark & \xmark\\
        Schmidt et al.~\cite{schmidt2023surface} & \xmark &\xmark & \xmark & \cmark & \xmark\\
        Rorberg et al.~\cite{rorberg2023bpm} & \xmark &\cmark & \xmark & \xmark & \xmark\\
        Born et al.~\cite{born2021surface} & \xmark &\cmark & \xmark & \xmark & \xmark\\
        Magnet et al.~\cite{magnet2022smooth} & \xmark &\cmark  &(\cmark) & \xmark & \cmark\\
        Xia et al.~\cite{xia2024locality} & \xmark &\cmark &(\cmark) & \xmark & \xmark\\
        SM-Comb~\cite{roetzer_scalable_2022} & \cmark & \xmark & \xmark & \cmark & (\cmark)\\
        DiscoMatch~\cite{roetzer2024discomatch} & \cmark & \xmark & \xmark & \cmark & \xmark\\
        SpiderMatch~\cite{roetzer_spidermatch_2024} & \cmark & (\cmark) & \xmark & \xmark & (\cmark)\\
        Roetzer et al.~\cite{roetzer2025higherorder} & \cmark & \cmark & \xmark & \xmark & (\cmark) \\
        GeCo3D~\cite{roetzer2025geco} & \cmark & \cmark & \xmark & \cmark & \cmark\\
        SuPaMatch~\cite{amrani2025highres} & \cmark & \cmark & \xmark & \cmark & \cmark\\
        GC-PSM~\cite{ehm_geometrically_2024}& \cmark & \xmark & \xmark & \cmark & \cmark\\
        GC-PPSM~\cite{ehm_partial--partial_2024}& \cmark & \xmark & \xmark & \cmark & \cmark\\
        Partial-GeCo~\cite{ehm2026partialgeco} & \cmark & \cmark & \xmark & \cmark & \cmark\\
        \bottomrule
    \end{tabular}
    \caption{Overview of \textbf{neighbourhood preserving} approaches and classification if methods are \textbf{(IF)} initialisation-free, are \textbf{(S)} scalable, are \textbf{(TF)} topologically flexible, are \textbf{(GNP)} globally neighbourhood preserving, or can solve \textbf{(P)} shape matching problems where at least one shape is partial.
    The scalability of SpiderMatch~\cite{roetzer_spidermatch_2024} depends on the input representation; see \cite{roetzer2025geco} for more details.
    Approaches~\cite{magnet2022smooth,xia2024locality} do not show experiments on topologically noisy shapes but would, in theory, support such inputs.
    The formalism solved by SM-Comb~\cite{roetzer_scalable_2022} is not designed for partial-to-full or partial-to-partial shape matching, yet the authors show that it is applicable when closing shapes using triangle fans.
    Further, the approaches~\cite{roetzer_spidermatch_2024, roetzer2025higherorder} do not show partial-to-full experiments, but their formalisms naturally support this case.
    }
    \label{tab:neighbourhood}
\end{table}

\subsubsection{Local optimisation}
Several lines perform local optimisation, or \emph{refinement}, to obtain neighbourhood-preserving maps, or rather to convert existing maps into neighbourhood-preserving ones.

In recent years, Takayama~\cite{takayama2022compatible} and Schmidt et al.~\cite{schmidt2023surface} propose using piece-wise linear transformations (which can be seen as piece-wise plane embeddings) to refine sparse landmark correspondences into dense, neighbourhood-preserving matchings by building on so-called intrinsic triangulations, see~\cite{sharp2019navigating}.
Other works use piece-wise Möbius transformations to obtain smoother maps~\cite{rorberg2023bpm}, inspired by ~\cite{lipman2009mobius}.

For shape pairs with genus greater than zero, \cite{born2021surface} propose to use homology transfer to obtain high-quality maps. 
Here, the homology bases and their respective transfer functions act as low-pass filters, making the resulting maps smooth by filtering out local noisy/inconsistent matches.
In~\cite{magnet2022smooth}, authors propose to use Dirichlet energy minimisation in a functional map framework to obtain smooth maps.
Further, \cite{xia2024locality} finds inliers of given noisy input correspondences by exploiting neighbourhood information, and subsequently, by utilising these inliers, they refine the outliers.

These refinement methods, by definition, rely on the quality of the input map and, therefore, matchings could in theory be arbitrarily bad.
In contrast, global approaches are not dependent on initialisation.

\subsubsection{Global approaches}\label{sec:glob-nei-nei}

Global neighbourhood preserving approaches refer to 
approaches that are initialisation-free, explicitly consider neighbourhood relations during optimisation, and provide (some form of) optimality certificates.
Among the most important approaches in this category is the integer linear programming formalism proposed in~\cite{windheuser2011geometrically} in which neighbourhoods of surface elements are preserved as a hard constraint and, as already mentioned, which shaped the term ``geometric consistency''.
Yet, this formalism is hard to solve and does not scale to higher shape resolutions.
Motivated by this, \cite{roetzer_scalable_2022, roetzer2024discomatch} proposed approximate solvers which yield improved scalability.

Other lines of work aim for novel formalisms.
Among these, SpiderMatch~\cite{roetzer_spidermatch_2024} (see \cref{sec:combinatorial}) achieves geometric consistency by representing a 3D shape as a cyclic, self-intersecting curve and by preserving intersections in their matching formalism.
Building on this idea, \cite{roetzer2025geco} encodes each triangle of a 3D shape with a single cyclic curve, leading to independent matching problems for each curve. 
Geometric consistency is then enforced by coupling the problems at opposite edges such that the constraints always lead to a faithful surface-to-surface matching and, with that, to better geometric consistency compared to \cite{roetzer_spidermatch_2024}.
In addition, this formalism is faster to solve and, in practice, can be solved to global optimality.
Motivated by the improved scalability, follow-up work~\cite{amrani2025highres} builds on the idea of an alternative shape representation and proposes a two-stage formalism: first, transfer patches from source to target shape and second, find a dense surface-to-surface matching for each pair of corresponding patches.
Yet, for both approaches, there are no theoretical guarantees of polynomial runtime, and thus it remains an open research question whether there exists a neighbourhood-preserving shape-matching approach that is globally optimal in polynomial time.

In contrast, \cite{ehm_geometrically_2024,ehm_partial--partial_2024} propose extensions of the formalism by Windheuser et al.~\cite{windheuser2011geometrically} to partial-to-full~\cite{ehm_geometrically_2024} and partial-to-partial shape matching~\cite{ehm_partial--partial_2024}.
For partial-to-full shape matching \cite{ehm_geometrically_2024} relaxes the hard neighbourhood constraints proposed in~\cite{windheuser2011geometrically}.
For the more challenging partial-to-partial shape-matching task, \cite{ehm_partial--partial_2024} proposes solving for the mean matching cost rather than the sum of matching costs.
This effectively allows for solving for correspondences and unknown overlap at the same time.
Yet, it is harder to solve due to the resulting non-linear objective function and requires solving multiple integer linear programs.
To address the challenges posed by the non-linear objective, \cite{ehm2026partialgeco} adapts \cite{roetzer2025geco} to the partial-to-partial setting and introduces an integer linear programming formalism that explicitly enforces geometric consistency.
By integrating a predicted overlap region directly into the objective function, the resulting approach scales significantly better than the previous geometrically consistent method for partial-to-partial shape matching.
In Section~\ref{sec:partiality}, we discuss additional approaches to partial shape matching.

\subsection{Neural correspondence optimization}

The machine learning and computer vision communities alike have become increasingly interested in approximating complex optimisation problems with neural networks~\cite{wang2025vggt}.
Deep learning provides a flexible mechanism for integrating intricate, data-driven priors to approximate what would otherwise be solved iteratively, such as the correspondence between two shapes.
As we saw in~\cref{subsec:fm_diffusion}, the optimisation to recover a functional map for a given set of spectral correspondences can be approximated with a diffusion model trained on a large collection of such maps \cite{zhuravlev_denoising_2025,pierson2025diffumatch}.
\cite{wu2024diff} recently observed that the analogous can be achieved for doubly stochastic ``soft'' correspondence matrices.
This raises the question of whether more complex costs and objectives, such as neighbourhood constraints, can be incorporated into such a formulation, paying the cost of computationally intensive numerical optimisation upfront during training time. 

\subsection{Partial shape matching}
\label{sec:partiality}

As mentioned in previous chapters, the problem of matching two complete shapes has been explored in many works.
However, partial shape matching, in which at least one shape is incomplete, has so far received less attention.
In the context of partial shape matching, we distinguish between two cases: partial-to-full shape matching, in which only one of the shapes is partially observed, and partial-to-partial shape matching, in which both shapes are incomplete.

\begin{figure}[t]
    \centering
    \includegraphics[width=1\linewidth]{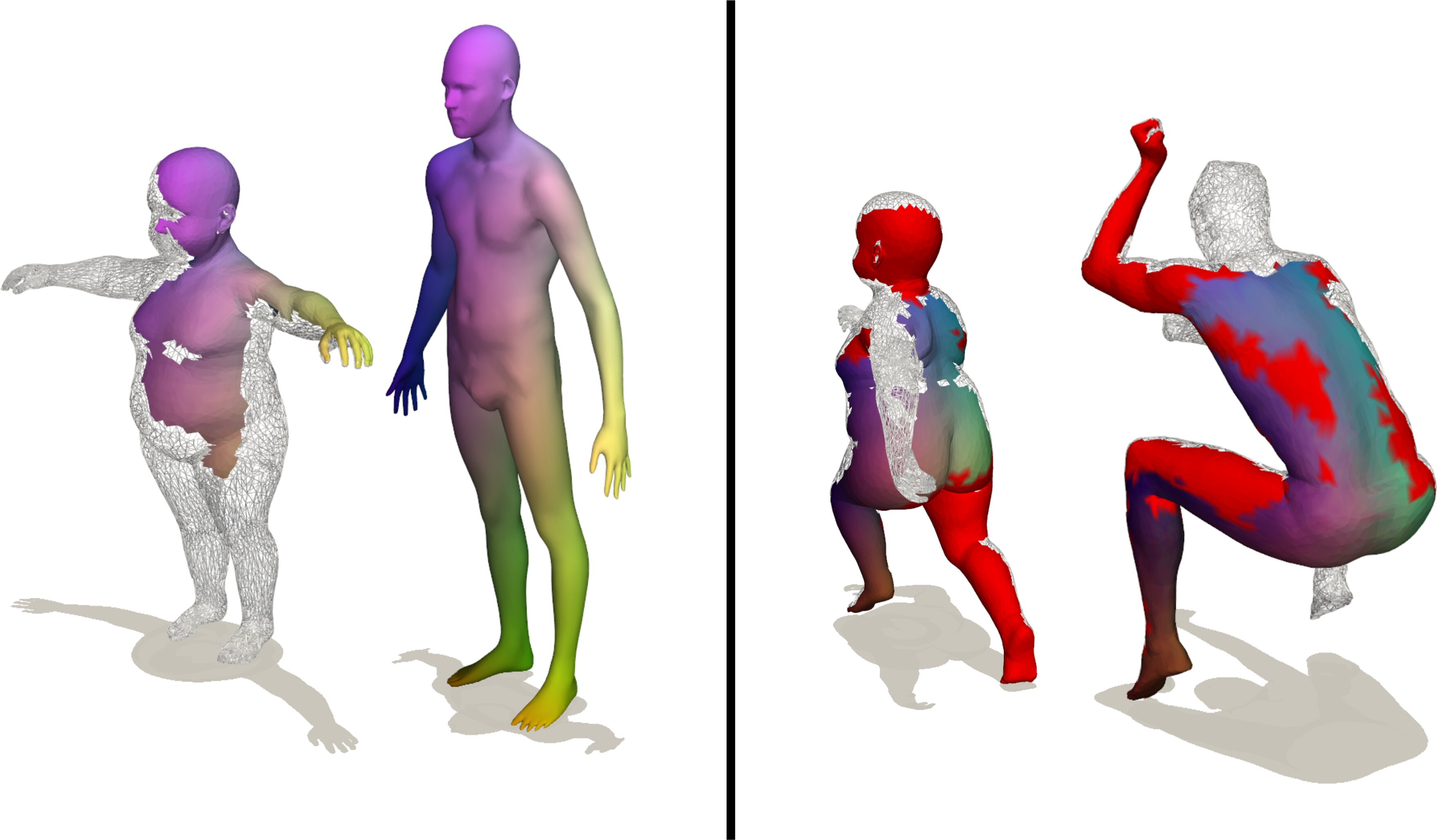}
    \caption{\textbf{Partial shape matching settings}: In the partial-to-full setting (left), a partial shape is matched to a template shape. In comparison, in the partial-to-partial setting, both shapes are only partially observed (right). The grey grid indicates that this part is missing, where red indicates that this part is observed but not in the overlapping region.
    Shapes are taken from FAUST~\cite{bogo_faust_2014}, KIDS~\cite{rodola2014dense} and BeCoS~\cite{ehm2025beyond} datasets.}
    \label{fig:partial_settings}
\end{figure}

\subsubsection{Partial-to-full shape matching} \label{sub:partial-to-full}
In partial-to-full shape matching, we explore the task of a partially observed shape $\mathcal{N}$ being matched on a full template shape $\mathcal{M}$ (see Fig.~\ref{fig:partial_settings} left).

\begin{table}[t]
    \centering
    \small
    \begin{tabular}{l@{}ccc}
    \toprule
        \textbf{Method} & \textbf{Setting} & \textbf{Framework} & \textbf{Type} \\
        \midrule
        DPFM~\cite{attaiki2021dpfm} & P2F,P2P & PFM & \textcolor{orange!90!black}{S},\textbf{\textcolor{green!50!black}{U}}\\
        Bensaid et al.~\cite{bensaid_partial_2023} & P2F, P2P & PFM & \textcolor{orange!90!black}{S} \\
        Bensaid et al.~\cite{bensaid_multi-spectral_2024} & P2F & Spectral & \textbf{\textcolor{red!70!black}{A}} \\
        Bracha et al.~\cite{bracha_unsupervised_2024} & P2F & Dist & \textbf{\textcolor{green!50!black}{U}} \\
        Wormhole\cite{bracha_wormhole_2024} & P2F & Dist & \textbf{\textcolor{green!50!black}{U}} \\
        ULRSSM~\cite{cao_unsupervised_2023} & P2F & PFM & \textbf{\textcolor{green!50!black}{U}} \\
        Magnet et al.~\cite{magnet2022smooth}& P2F & Neigh & \textbf{\textcolor{red!70!black}{A}} \\
        GC-PSM~\cite{ehm_geometrically_2024} & P2F & Neigh & \textbf{\textcolor{red!70!black}{A}} \\
        GeCo3D~\cite{roetzer2025geco} & P2F & Neigh & \textbf{\textcolor{red!70!black}{A}}\\
        El Amrani et al.~\cite{amrani2025highres} & P2F & Neigh & \textbf{\textcolor{red!70!black}{A}}\\ EchoMatch~\cite{xie_echomatch_2025} & P2P & PFM & \textcolor{orange!90!black}{S} \\
        GC-PPSM~\cite{ehm_geometrically_2024} & P2P & Neigh & \textbf{\textcolor{red!70!black}{A}} \\
        Partial-GeCo~\cite{ehm2026partialgeco} & P2P & Neigh & \textbf{\textcolor{red!70!black}{A}}\\
    \bottomrule
    \end{tabular}
    \caption{Categorization of Partial Shape Matching Methods: We categorize partial shape matching methods in their setting (partial-to-full (\textbf{P2F}) or partial-to-partial (\textbf{P2P})), their underlying framework (\textbf{P}artial \textbf{F}unctional \textbf{M}aps, \textbf{Spectral} alignment, \textbf{Dist}ance-based or \textbf{Neigh}bourhood-preserving) and their optimisation type (axiomatic (\textbf{\textcolor{red!70!black}{A}}), 
supervised (\textbf{\textcolor{orange!90!black}{S}}), unsupervised (\textbf{\textcolor{green!50!black}{U}})).
}
\label{tab:partial}
\end{table}

\paragraph{Partial functional maps.}
In one of the first works to address partial-to-full shape matching, the authors observe a partial setting in the functional map context~\cite{rodola2017partial} for isometric shape matching.
For reference, for isometric shapes we observe a close to diagonal structure of the functional map $C_{\mathcal{MN}} = \phi_\mathcal{N}^{\dagger}\Pi_{\mathcal{NM}}\phi_\mathcal{M}$ with an optimal point-wise map $\Pi_{\mathcal{NM}}$.
In the partial-to-full scenario, this is not the case, as some eigenfunctions of the full shape do not have a corresponding eigenfunction in the partial shape.
Therefore, instead of a full diagonal, we observe a slanted diagonal structure with rank $r$ of the partial functional map $C_{\mathcal{MN}}$.
The rank $r$ can be estimated by comparing the surface areas of the two shapes~\cite{weyl1911asymptotische}.

Similar to the full-to-full case, the partial functional map has several key properties.
One is semi-orthogonality, which ensures that the map preserves volume when transferring functions from the partial shape to the full one, but not necessarily in the opposite direction. The other is bijectivity, which requires that mapping a function from the partial shape to the full shape and then back recovers the original function.
While supervised methods rely on the ground-truth functional map to guide training~\cite{attaiki2021dpfm}, these structural properties provide a foundation for unsupervised learning~\cite{attaiki2021dpfm, cao_unsupervised_2023}, including the orthogonality losses

\begin{equation}
\left\lVert C_{\mathcal{M}\mathcal{N}} C_{\mathcal{M}\mathcal{N}}^{\top} - I_r \right\rVert_F^2 
\text{ and } 
\left\lVert C_{\mathcal{N}\mathcal{M}}^{\top} C_{\mathcal{N}\mathcal{M}} - I_r \right\rVert_F^2,
\end{equation}
\noindent and the bijectivity loss,
\begin{equation}
\left\|C_\mathcal{MN}C_\mathcal{NM} - I_r\right\|^2_F.
\end{equation}
Here, $I_r$ denotes the identity matrix with only the first $r$ elements being 1.
DPFM~\cite{attaiki2021dpfm} introduces a cross-attention refinement module that enables feature communication across shapes.
Bensaid et al.~\cite{bensaid_partial_2023} propose learning a piecewise-smooth function over the surface: sparse landmark correspondences between a full shape and its partial counterpart are first established via feature similarity, and a neural representation is then optimised to fit these landmarks, enabling smooth interpolation between matched features that serve as anchors.

\paragraph{Spectral alignment methods.}
Rampini et al.~\cite{rampini2019correspondence} align spectral representations derived from the Laplace–Beltrami operator (LBO).
Building on this idea, Bensaid et al.~\cite{bensaid_multi-spectral_2024} combine the standard LBO with its scale-invariant variant (SI-LBO), leveraging the complementary information from both spectra to better capture semantically meaningful curved regions.

\paragraph{Distance-based methods.}
Another line of work shows that using functional maps in the partial shape matching setting introduces unavoidable errors~\cite{bracha_unsupervised_2024}.
Therefore, the method primarily operates in the spatial domain rather than the spectral domain.
The main loss preserves the geodesic distance (surface distance) between points on both shapes.
Bracha et al.~\cite{bracha_wormhole_2024} tackle the problem that the geodesic distance of a partial shape of two points does not necessarily have the same length as the same two points on a full shape, due to possible holes.
Therefore, reliable point pairs are identified and are later used in the loss formulation.

\paragraph{Neighbourhood-preserving methods.}
\cite{magnet2022smooth,ehm_geometrically_2024,amrani2025highres,roetzer2025geco} use integer neighbourhood-preserving techniques to find solutions for the partial-to-full shape matching problem, see also \cref{sec:glob-nei-nei}.

\subsubsection{Partial-to-partial shape matching} \label{sub:partial-to-partial}
The more challenging problem in partial shape matching is the partial-to-partial setting, where we have to solve two problems in parallel: finding the overlapping region and determining correspondences in it (see Figure~\ref{fig:partial_settings} right). Due to its complexity, it remains an underexplored problem and warrants further consideration.

So far, only a few supervised learning-based methods~\cite{attaiki2021dpfm, bensaid_partial_2023,xie_echomatch_2025} exist that tackle the partial-to-partial shape matching problem.
Two methods~\cite{xie_echomatch_2025, attaiki2021dpfm} mainly operate similarly to the partial-to-full or full-to-full setting, but add an overlap prediction module to determine the overlapping region.
For two partial shapes $\mathcal{M}$ and $\mathcal{N}$ define a overlap prediction vector $p_\mathcal{M} \in [0,1]^{n_\mathcal{M}}$ and $p_\mathcal{N} \in [0,1]^{n_\mathcal{N}}$, indicating the probability that a vertex is in the overlapping region or not.
These predictions are compared with a binary cross entropy loss (BCE) with the ground truth overlap $\text{gt}_\mathcal{M}$ and $\text{gt}_\mathcal{N}$:

\begin{equation}
    \Loss_{\text{overlap}} = 
    \frac{1}{2} 
    \Bigl[ 
      \text{BCE}\!\left(p_\mathcal{M},\, \text{gt}_\mathcal{M} \right)
      + 
      \text{BCE}\!\left(p_\mathcal{N},\, \text{gt}_\mathcal{N} \right)
    \Bigr].
\end{equation}

DPFM~\cite{attaiki2021dpfm} uses the predicted features directly in an MLP to predict the overlap predictions $p_\mathcal{M}$ and $p_\mathcal{N}$.
In contrast, EchoMatch~\cite{xie_echomatch_2025} introduces a correspondence reflection module that leverages cycle consistency of the point-wise maps, \( P_\mathcal{M} = \Pi_{\mathcal{M}\mathcal{N}} \Pi_{\mathcal{N}\mathcal{M}} \), to verify whether a point returns to its original neighbourhood.  
It is further demonstrated that image-based DINOv2 features are effective for handling partial shapes.  
A different strategy is proposed in~\cite{bensaid_partial_2023}, where anchor correspondences are first established to enable smooth interpolation.  
For the partial-to-partial setting, two additional area-based losses are introduced: one to enforce similarity of overlapping regions across shapes, and another to encourage large overlap.
\cite{ehm_partial--partial_2024, ehm2026partialgeco} propose an axiomatic neighbourhood-preserving solution for partial-to-partial shape matching, as mentioned in described ~\ref{sec:neigh}.

\paragraph{Metric adaptations.}

Metrics must be properly adapted for the partial correspondence settings.
In the partial-to-full (P2F) case, where the source shape $\mathcal{M}$ is a partial observation and the target shape $\mathcal{N}$ is complete, the metrics are computed identically to the full-to-full case.
In contrast, the partial-to-partial (P2P) setting assumes that both the predicted and ground-truth correspondences are defined only on the overlapping regions of the source and target shapes. 
Let $\Omega^* \subseteq \mathcal{M}$ denote the ground-truth overlap region, and $\Omega \subseteq \mathcal{M}$ the predicted one. 
To ensure consistency across partial and full scenarios, all geodesic distances are normalised by the square root of the area of the full target shape, $\sqrt{\text{Area}(\mathcal{N}_{\text{full}})}$.
Vertices in the ground-truth overlap that were not matched ($i \in \Omega^* \setminus \Omega$), as well as vertices falsely matched outside the overlap ($i \in \Omega \setminus \Omega^*$), are penalised with infinite error: $e(i) = \infty$. 
When evaluating the mean geodesic error, PCK, or AUC, only correspondences with finite error ($e(i) < \infty$) are considered. 
This ensures that the metrics measure accuracy within the overlap area, while ignoring unmatched or undefined regions. 
Additional evaluation metrics for the P2P setting are discussed in~\cite{ehm2025beyond,ehm_partial--partial_2024}.

While previously discussed mesh-based approaches rely on explicit surface connectivity, the next section focuses on matching pairs of point clouds, where such connectivity information is unavailable. 
Structural priors can therefore only approximate connectivity.

\subsection{Point cloud matching}

Several works have focused specifically on the point cloud setting.
Such methods are generally also applicable to structured mesh representations but cannot leverage as strong a structural prior; thus, methods like those in \cref{sec:spectral} and \ref{sec:combinatorial} are preferred.
Nevertheless, unstructured point cloud acquisitions are ubiquitous in real-world settings, and thus these methods remain of fundamental importance.

Point cloud correspondence methods generally rely on strong feature representations for matching because there are few structural priors to exploit.
Many require ground-truth supervision; those that don't typically regularise the resulting deformation using the chamfer loss, which requires that the shapes already have a reasonable extrinsic alignment, or using piecewise ARAP deformations.
Corrnet3d~\cite{zeng2021corrnet3d} predicts high-dimensional point-wise features for both point clouds, which are used to regress the permutation matrix between them. 
This permutation matrix is used to deform the point clouds.
DPC~\cite{lang2021dpc} uses point-wise features for self-construction and cross-construction, where self-construction encourages smooth correspondence maps and cross-construction unique matches.
In~\cite{netto2022robust}, the authors use self- and cross-attention modules to predict a soft correspondence matrix, where, in comparison to Corrnet3d and DCP, they also adopt a \textit{dustbin} paradigm to discard correspondences.
In~\cite{he2023hierarchical}, the authors use \textit{multi-receptive-field point encoders} with different receptive radii to compute hierarchical features.
SE-ORNet~\cite{deng2023se} aligns the orientation of shapes to tackle the symmetry problem in human and animal shapes.
In~\cite{deng2024unsupervised}, the authors propose learning a set of templates with explicit structures that can later be used for better shape correspondences.
Spectral methods can indeed also be applied in point cloud settings~\cite{cao_self-supervised_2023} due to the development of point cloud Laplacians~\cite{sharp2020laplacian}, yielding feature encoders that can operate on both mesh and point clouds.
The same holds for the Gaussian splatting representation~\cite{zhou2025gaussLBO}.
Alternatively, graphs are also well-suited for such multi-modal  encoders~\cite{saleh_bending_2022}.

While these methods mostly handle the task of matching complete, noisy point clouds, some also target partiality.
Lepard~\cite{li2022lepard} uses a Transformer architecture with self and cross-attention layers. 
While primarily designed for the rigid case, design a deformation module which predicts a warp function for every point in a dense point cloud for non-rigid settings. 
Training also proceeds in a supervised manner. 
Later, a rotation invariant method~\cite{yu2023rotation} was introduced to prevent instabilities for unseen rotations. 
They introduce a rotation-invariant attention-based encoder-decoder architecture for local geometry and a global transformer with rotation-invariant cross-frame position awareness. 
Similar to Lepard, they predict a per-point flow to also tackle the deformable case.
DiffReg~\cite{wu2024diff} builds upon these techniques, but proposes to recover a correspondence map through a diffusion process in the space of double-stochastic matrices.

\section{Datasets}
\label{sec:datasets}

\begin{figure*}
 \includegraphics[width=\linewidth]{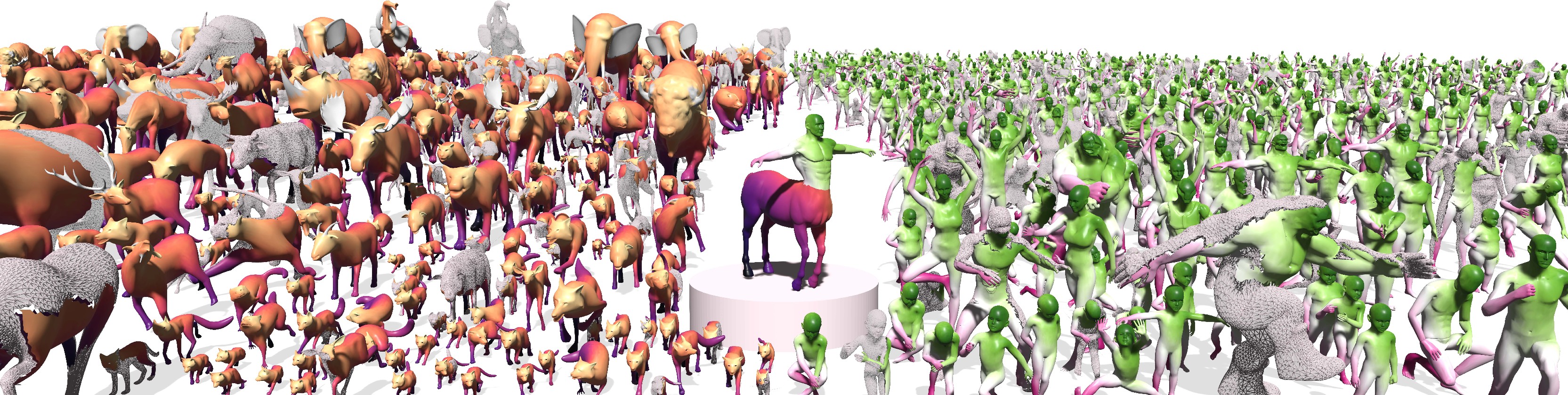}
 \centering
  \caption{\textbf{The BeCoS dataset}~\cite{ehm2025beyond} connects seven existing 3D shape matching datasets~\cite{bogo_faust_2014,zuffi2017small,anguelov_scape_2005,bronstein2008numerical,li20214dcomplete,rodola2014dense,dyke2020shrec}, enabling the propagation of dense correspondences across a wide range of categories, including humanoids, cats, elephants, and others. Furthermore, it provides annotations for partial correspondences, supporting both partial-to-full (\textbf{P2F}) and partial-to-partial (\textbf{P2P}) matching settings. Image source: \cite{ehm2025beyond}.
  }
  \label{fig:becos_teaser}
\end{figure*}

\begin{table}[!t]
\centering
\footnotesize
    \begin{tabular}{cl@{}cccc@{}}
        \toprule
        \textbf{Setting} & \textbf{Dataset}  & \textbf{NS} & \textbf{\# US} & \textbf{TN} & \textbf{SC}\\
        \midrule
        \multirow{13}{*}{\textbf{F2F}}    &FAUST~\cite{bogo_faust_2014}                                         & \xmark & 100 & \xmark & \Strichmaxerl \\
                                &SCAPE~\cite{anguelov_scape_2005}                                     & \xmark & 71  &  \xmark  & \Strichmaxerl \\
                                &SMAL~\cite{zuffi2017small}                                           & \cmark & 49 &   \xmark & \Cat \\
                                &DT4D~\cite{magnet2022smooth, li20214dcomplete}                       & \xmark & 2198 &  \xmark & \Strichmaxerl / \Cat \\
                                &TOSCA~\cite{bronstein2008numerical}                                  & \xmark & 82 & \xmark & \Strichmaxerl / \Cat \\
                                &KIDS~\cite{rodola2014dense}                                          & \xmark  & 32 & \xmark & \Strichmaxerl \\
                                &SHREC'20~\cite{dyke2020shrec}                                        & \cmark & 11 & \xmark & \Cat \\
                                &SHREC'19~\cite{melzi2019shrec}                                       & \xmark & 44 & \xmark & \Strichmaxerl \\
                                &BeCoS (F2F)~\cite{ehm2025beyond}                                   & \cmark & 2543 & \xmark &  \Strichmaxerl / \Cat \\
                                &SURREAL~\cite{varol_learning_2017}                                   & \xmark & 230K & \xmark & \Strichmaxerl\\
                                &TOPKIDS~\cite{laehner2016topkids,rodola2014dense}                    & \xmark & 32 & \cmark & \Strichmaxerl \\
                                &SCAPET~\cite{yusuf2023scapet,anguelov_scape_2005}                    & \xmark & 71 & \cmark & \Strichmaxerl\\
                                &ExtFAUST~\cite{merrouche2025matching,bogo_faust_2014}                & \xmark & 100 & \cmark & \Strichmaxerl \\
                                
    \midrule
    \multirow{7}{*}{\textbf{P2F}}        & SHREC'16~\cite{bronstein2008numerical,cosmo2016shrec}               & \xmark & 76 & \xmark &  \Strichmaxerl / \Cat \\
                                & PFAUST~\cite{bracha_partial_2023,bogo_faust_2014}                   & \xmark & 10 & \xmark & \Strichmaxerl \\
                                & FARM partial~\cite{kirgo2021wavelet}                                & \xmark & 5 & \xmark & \Strichmaxerl \\
                                & PFARM~\cite{attaiki2021dpfm,kirgo2021wavelet}                       & \xmark & 28 & \xmark & \Strichmaxerl \\
                                & SHREC'20~\cite{dyke2020shrec}                                       & \cmark & 14 & \xmark & \Cat \\
                                & BeCoS (P2F)~\cite{ehm2025beyond}                                  & \cmark & 2543 & \xmark & \Strichmaxerl / \Cat \\
    \addlinespace
    \midrule
    \multirow{3}{*}{\textbf{P2P}}        & CP2P~\cite{attaiki2021dpfm,cosmo2016shrec}                          & \xmark & 76 & \xmark & \Strichmaxerl / \Cat  \\
                                & PSMAL~\cite{ehm_partial--partial_2024,zuffi2017small}               & \cmark & 43 & \xmark &  \Cat \\
                                & BeCoS (P2P)~\cite{ehm2025beyond}                                  & \cmark & 2543 & \xmark & \Strichmaxerl /  \Cat \\                    
    \bottomrule
    \end{tabular}
    
\caption{An overview of \textbf{3D deformable shape datasets}.
        The datasets are categorized according to their shape matching setting: full-to-full (\textbf{F2F}), partial-to-full (\textbf{P2F}), and partial-to-partial (\textbf{P2P}).
        For each dataset, we report several characteristics, including non-isometry (\textbf{NS}), number of unique shapes (\textbf{\#US}), topological noise (\textbf{TN}), and type of shape categories (\textbf{SC}), indicating whether the dataset contains humanoid shapes (\Strichmaxerl) or animal shapes (\Cat).
        } 
\label{tab:datasets}
\end{table}

While large-scale 3D shape datasets such as ShapeNet~\cite{chang_shapenet_2015} and Objaverse~\cite{objaverse} have enabled progress in general shape understanding, they typically lack the dense ground-truth correspondences required for evaluating deformable shape matching.
In line with prior work in this domain, we instead focus on datasets that contain deformable objects annotated with consistent correspondences across shape instances. 

\paragraph{Near-isometric.}
Early 3D shape matching datasets such as the CAESER dataset \cite{robinette1999caeser} provided thousands of human body scans annotated with hand-placed landmarks, offering an important benchmark but with limited ground-truth correspondences. To address this limitation, the SCAPE dataset \cite{anguelov_scape_2005} introduced 71 registered meshes of a single subject in a single pose, providing dense correspondences but restricting variability since all shapes originate from one individual. The TOSCA dataset \cite{bronstein2008numerical} expanded diversity by including multiple object categories (e.g., humans, cats and dogs), yet its synthetic nature with fixed topology introduced a domain gap between real-world scans and artist-crafted meshes. Subsequent datasets such as FAUST~\cite{bogo_faust_2014} improved diversity with 10 individuals captured in multiple poses, using registrations to a common template to establish dense correspondences, while DFAUST extended FAUST to dynamic sequences, yielding over 40,000 aligned meshes, though its focus on adults limited generalisation to children. To address this gap, the KIDS dataset \cite{rodola2014dense} introduced eight high-resolution synthetic models (80K vertices) of a child in near-isometric poses.

Other datasets emerged from parametric models such as SMPL~\cite{loper2015smpl}, its variants~\cite{romero2017mano,pavlakos2019smplx}, and the COMA model~\cite{ranjan2018coma}, enabling the creation of very large collections such as SURREAL \cite{varol_learning_2017}, which contains 230,000 meshes. However, these datasets suffer from biases introduced by repeated use of identical triangulations across thousands of meshes. More recent efforts, such as DT4D~\cite{li20214dcomplete}, extend diversity by providing animatable shapes from multiple categories, such as different animal species, with consistent triangulations across meshes of the same class. Despite these advances, nearly all of these datasets remain constrained to the near-isometric case and, therefore, fail to fully capture the complexities of real-world shape variability. This highlights the need for datasets that provide reliable correspondences under non-isometric deformations.

\paragraph{Beyond isometry.}
A more general and challenging setting for 3D shape matching arises in the non-isometric regime, where datasets need to provide correspondences across fundamentally different object categories, such as establishing mappings between a cat and a horse. Constructing such datasets is particularly difficult because non-isometric correspondences are inherently ill-defined. While one may attempt to align overall body structures, it is unclear how to meaningfully associate, for example, the trunk of an elephant with the body of a cat. Early efforts in this direction include the SMAL dataset \cite{zuffi2017small}, which introduced a parametric model for multiple animal categories trained on 41 scans. While this dataset offers cross-category representations, it relies on fitting a single parametric model across all meshes, which limits its expressiveness and its ability to capture the fine geometric details unique to each class of shapes. Another step forward is SHREC’20 \cite{dyke2020shrec}, which explicitly defines correspondences across different animal categories, such as elephants and cats. However, it only provides sparse correspondences, thereby limiting its utility for dense shape analysis. 
More recently, Magnet et al.~\cite{magnet2022smooth} addressed this gap by manually establishing dense correspondences across most categories in the DT4D dataset~\cite{li20214dcomplete}, thereby creating the first dataset to offer dense non-isometric mappings.
Building on this idea, BeCoS~\cite{ehm2025beyond} further expanded the scope by leveraging seven existing isometric datasets. Instead of introducing new shapes, BeCoS builds upon seven existing isometric datasets and manually generates dense correspondences between a set of selected template meshes, effectively enabling cross-category mappings. This extension expands deformable shape analysis to the non-isometric setting, resulting in a large, interconnected collection of shapes, as illustrated in Fig.~\ref{fig:becos_teaser}.

\paragraph{Challenging connectivity and topological noise.}
Another line of research in 3D shape matching has focused on datasets designed to introduce specific challenges, particularly related to mesh connectivity and topological noise. Variations in mesh connectivity can arise from anisotropy, where parts of the same surface are represented with coarse triangulation, while others use much finer discretisation, making correspondence estimation substantially more difficult. To evaluate generalisation to such scenarios, SHREC’19~\cite{melzi2019shrec} introduced human scans with altered mesh connectivity. Similarly, Donati et al.~\cite{donati_deep_2022} created a challenging setting by remeshing both the SCAPE~\cite{anguelov_scape_2005} and FAUST~\cite{bogo_faust_2014} datasets anisotropically. 

A related challenge is topological noise, which includes the presence of holes, self-intersections, or other surface artefacts. SHREC’10~\cite{bronstein2010shrec} addressed this by extending TOSCA with artificially injected topological noise, though it inherited TOSCA’s broader limitations. Building on this idea, TOPKIDS~\cite{laehner2016topkids} extended the KIDS dataset \cite{rodola2014dense} by synthetically introducing topological perturbations, while TACO~\cite{pedico_taco_2024} extended TOSCA using a non-connectivity-preserving remeshing strategy. In TOCA~\cite{pedico_taco_2024}, mesh connectivities were intentionally altered, even among shapes of the same class. Leveraging the 3D alignment between original and remeshed shapes as well as TOSCA’s ground-truth correspondences, the authors generated new ground-truth correspondences across the perturbed surfaces. 
Using a different approach, SCAPET~\cite{yusuf2023scapet} and ExtFAUST~\cite{basset2021extFaust,Merrouche_2023_BMVC} build upon existing datasets, SCAPE~\cite{anguelov_scape_2005} and FAUST~\cite{bogo_faust_2014}, respectively. These methods modify the topology of the shapes by either adding shortcut tunnels or by merging nearby surface regions, effectively connecting previously separate parts of the mesh.
Together, these datasets emphasise the importance of evaluating shape matching methods under realistic noise and connectivity variations, moving beyond the simplified settings of earlier benchmarks.

\paragraph{Partial 3D shapes datasets.}
Beyond the full-to-full (F2F) setting discussed so far, the availability of datasets for the other two key settings, partial-to-full (P2F) and partial-to-partial (P2P), as described in Sec.~\ref{sub:partial-to-full}, is equally important.
The first benchmark in this direction was SHREC’16~\cite{cosmo2016shrec}, which extended the TOSCA dataset \cite{bronstein2008numerical} to include partial shapes. This was later followed by PFAUST~\cite{bracha_partial_2023} and PFARM~\cite{attaiki2021dpfm}, based respectively on the FAUST~\cite{bogo_faust_2014} and FARM~\cite{kirgo2021wavelet} datasets, while SHREC’20~\cite{dyke2020shrec} also included a subset of partial-to-full correspondence pairs. 

In contrast, P2P shape matching is considerably more challenging, as it requires establishing correspondences between two partial shapes with unknown overlap. The first benchmark for this setting, CP2P~\cite{attaiki2021dpfm}, was introduced by extending the SHREC’16 CUTS dataset \cite{cosmo2016shrec}. More recently, PSMAL~\cite{ehm_partial--partial_2024} extended the SMAL dataset~\cite{zuffi2017small} to incorporate partial-to-partial correspondences across animal categories. However, in both P2F and P2P settings, partiality has traditionally been simulated by artificially removing mesh parts or introducing holes, which does not accurately reflect the challenges of real-world partial observations. To address this limitation, BeCoS~\cite{ehm2025beyond} introduces a more realistic framework for generating partial shapes through ray-casting, enabling the creation of P2F, P2P, and F2F correspondence pairs within a unified framework.

\paragraph{Performance of 3D shape matching methods.}
Over the past few years, performance on well-established benchmarks has largely saturated, as indicated by the marginal improvements achieved by recent methods (e.g.~the mean geodesic error (×100) on FAUST~\cite{bogo_faust_2014} has remained at approximately 1.6~\cite{cao2024synchronous, bastian2024hybrid}). This stagnation highlights the need for more challenging benchmarks, preferably based on real-world shapes obtained from 3D scans, in order to further drive methodological progress. In addition, the reliability of ground-truth annotations in several commonly used datasets, including DT4D~\cite{li20214dcomplete}, BeCoS~\cite{ehm2025beyond}, and SHREC’19~\cite{melzi_shrec19_2019}, may contain errors, suggesting that improvements in numerical accuracy do not necessarily reflect genuine advances in correspondence quality~\cite{li_learning_2022}. These challenges are even more pronounced in the partial shape setting, where benchmarks and dedicated methods remain scarce, despite this scenario being the most representative of real-world data. 
At the time of writing, state-of-the-art results in full-to-full, partial-to-full and partial-to-partial matching scenarios, were obtained by~\cite{bastian2024hybrid}, ~\cite{cao2024synchronous} and~\cite{xie_echomatch_2025}, respectively.

\section{Practical applications}
\label{sec:applications}

\begin{figure*}[t]
  \centering
\begin{subfigure}{0.49\linewidth}
    \centering
    \setlength{\fboxsep}{0pt}
    \includegraphics[width=0.9\linewidth]{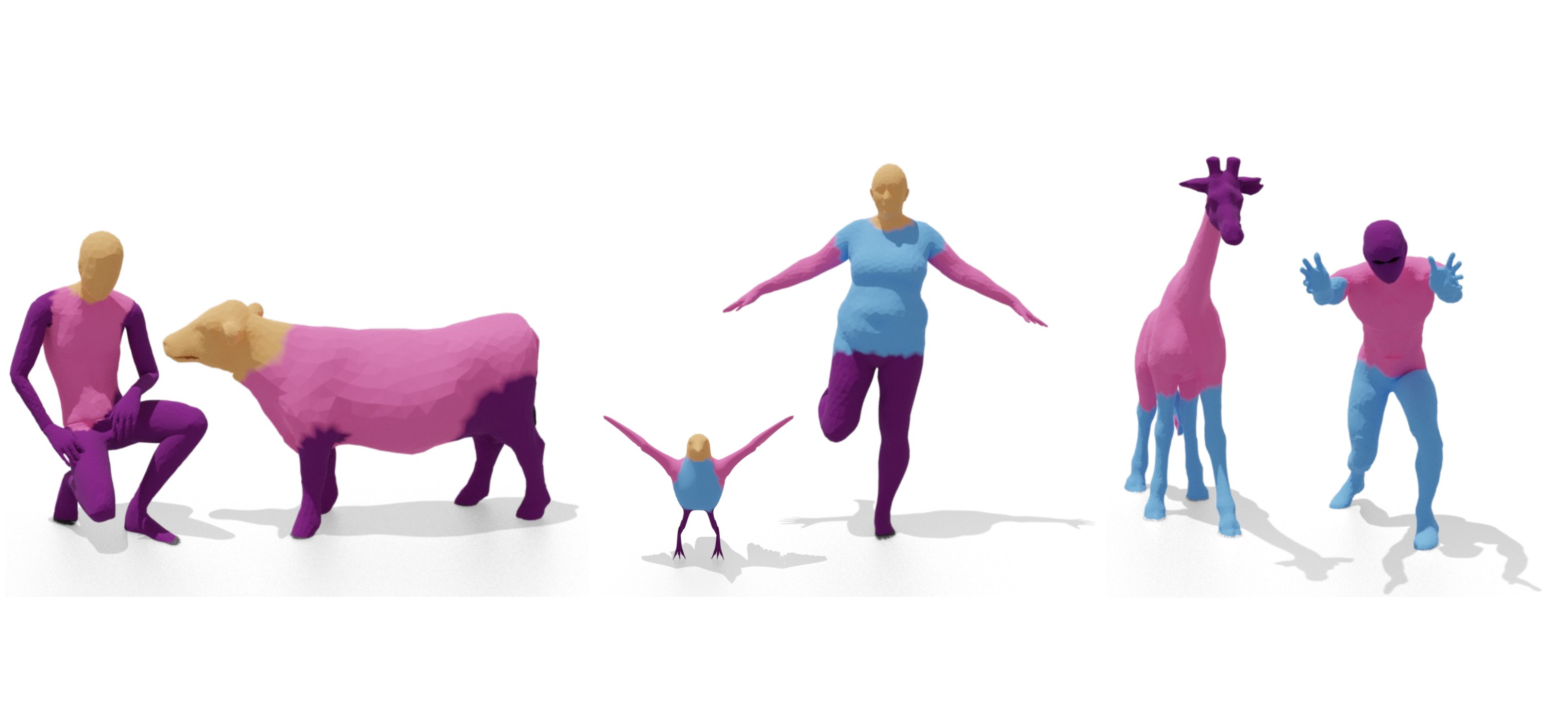}
    \caption{Segmentation transfer}
\end{subfigure}
  \begin{subfigure}{0.49\linewidth}\centering
    \includegraphics[width=0.65\linewidth]{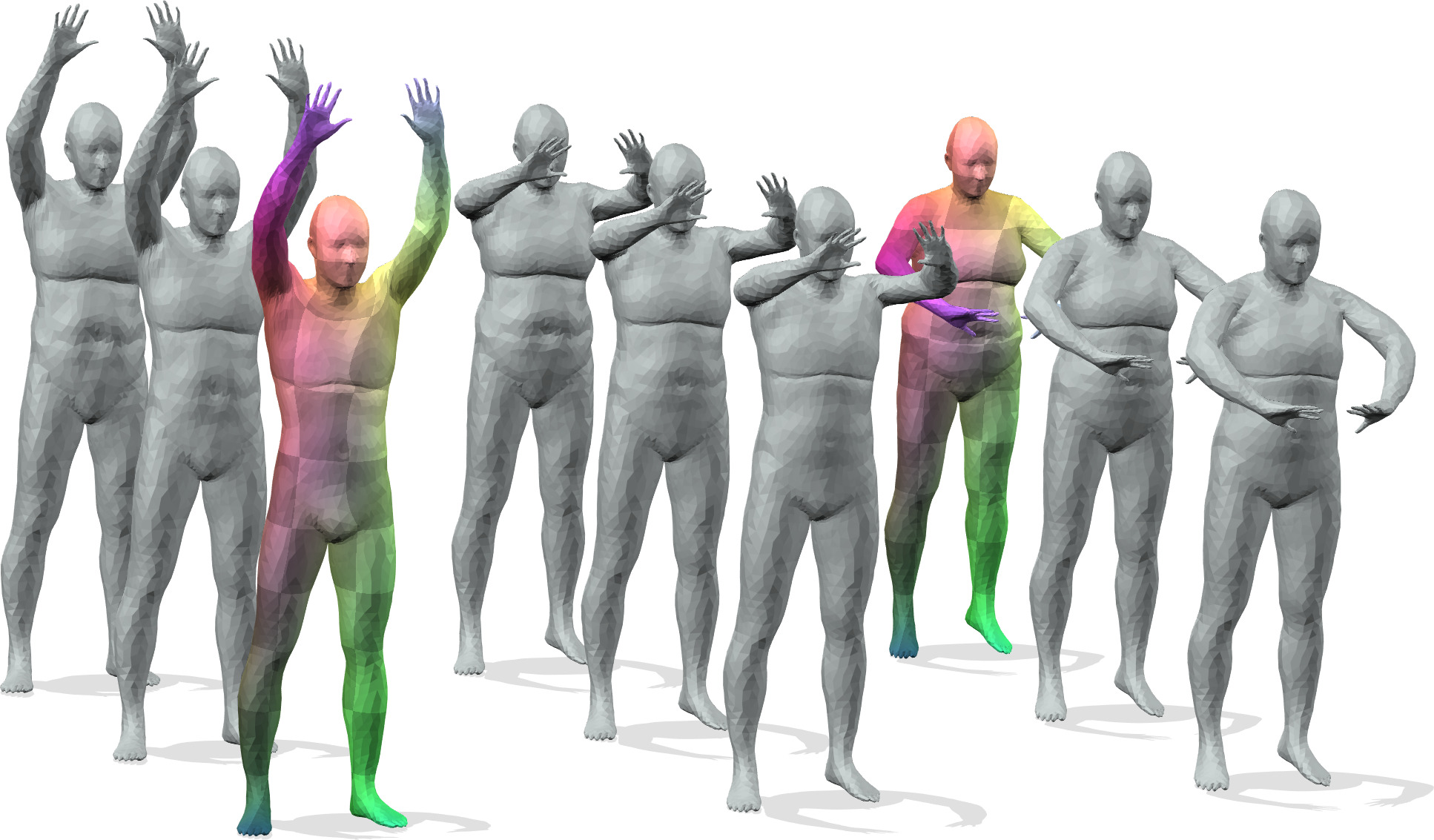}
    \vfill
    \caption{Shape interpolation}
  \end{subfigure}\hfill
\begin{subfigure}{0.49\linewidth}
  \centering
  \vfill
  \includegraphics[width=0.9\linewidth]{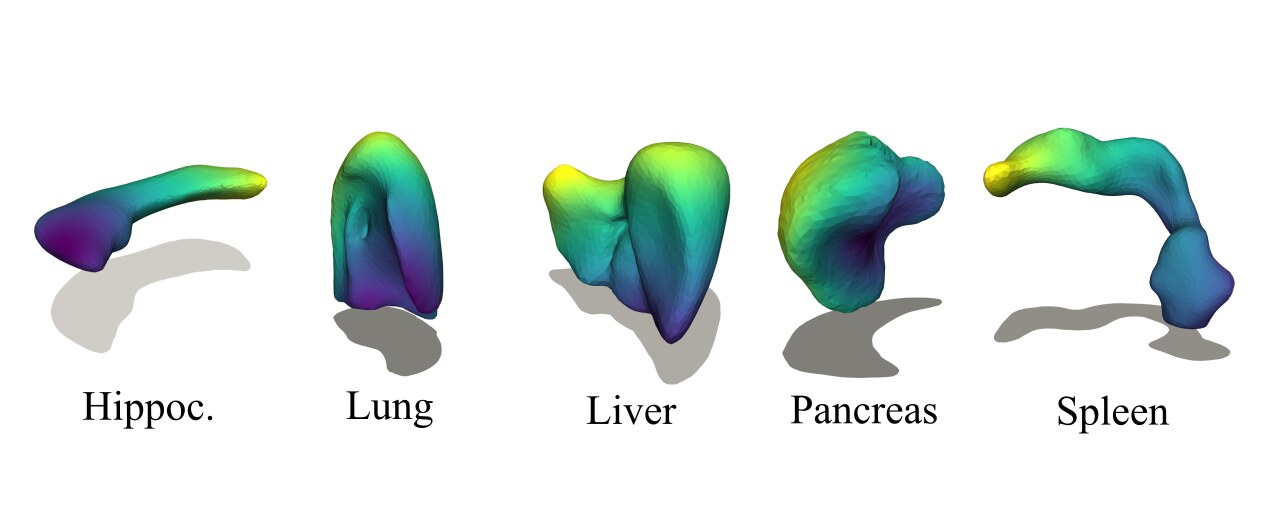}
  \vfill
  \caption{Statistical shape modelling}
\end{subfigure}
\begin{subfigure}{0.49\linewidth}
  \centering
  \includegraphics[width=0.95\linewidth]{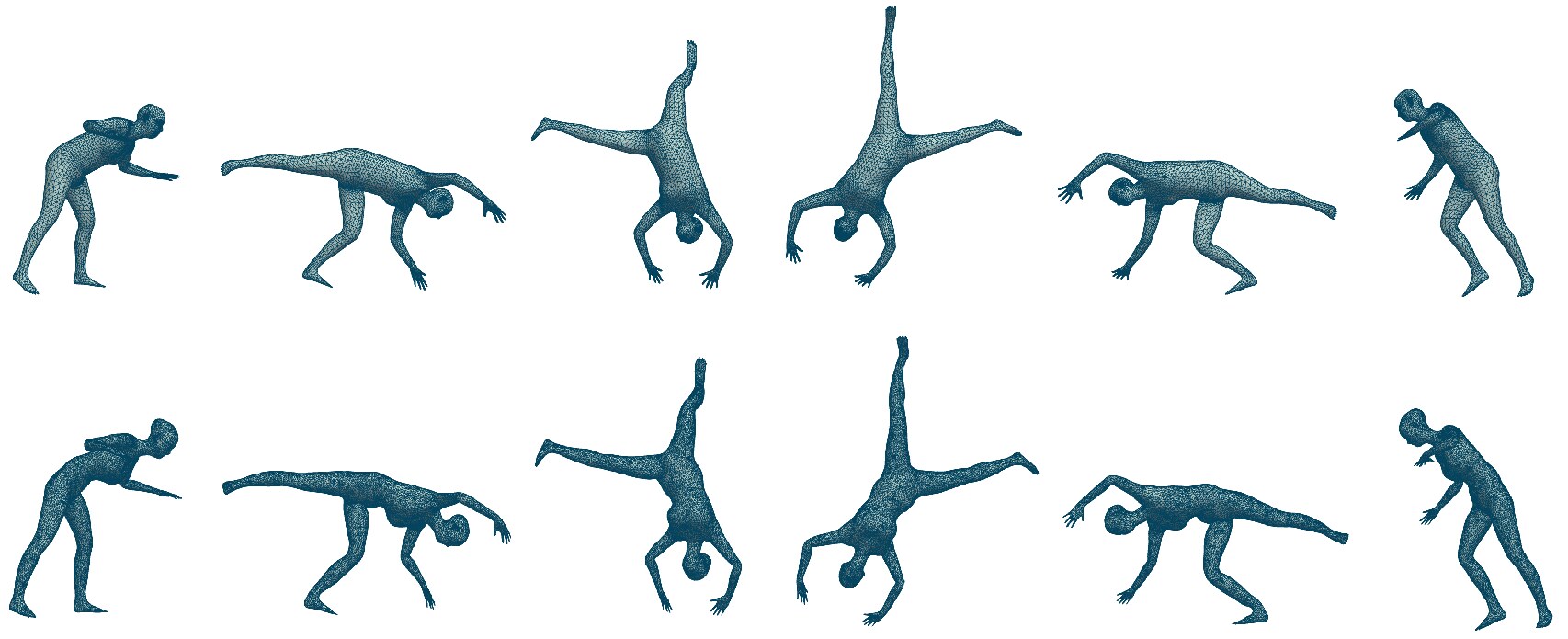}
  \vfill
  \caption{Animation transfer}
\end{subfigure}
  \caption{
        \textbf{Practical applications} of shape correspondence span numerous tasks in graphics and geometry processing.
        \textbf{(a)} Segmentation transfer propagates semantic part annotations across a collection of registered shapes.
        \textbf{(b)} Shape interpolation uses correspondences to generate smooth transitions between source and target shapes.
        \textbf{(c)} Statistical shape modelling uses correspondences to build low-dimensional shape spaces, including anatomical models of organs and tissues.
        \textbf{(d)} Animation transfer allows to propagate motion sequences across registered shapes.
        Figures adapted from \cite{cao_spectral_2024,abdelreheem2023zero,el_amrani_universal_2024,Musoni2021ReposingAR}.
        }  
  \label{fig:practical_applications}
\end{figure*}

Shape correspondence lies at the core of numerous applications involving 3D meshes.
In this section, we highlight some of the most prominent use cases.
Examples of these applications are illustrated in Figure~\ref{fig:practical_applications}.

\paragraph{Information transfer.}

A well-known application of shape correspondence is the transfer of annotations between shapes, which enables numerous tasks in graphics, animation, and robotics.
A classic example is texture transfer, where a texture defined on a source shape is projected onto a target shape, which is often used to visually inspect the quality of correspondences.
Segmentation transfer enables the propagation of part labels across a dataset of registered shapes, allowing a small number of manual annotations to be scaled to a shape collection~\cite{abdelreheem_satr_2023}. 
Deformation transfer propagates articulations, such as pose changes in human bodies, and can extend to complete deformation systems, including rigging properties~\cite{avril2016animation,musoni2021functional} or the mesh structure itself~\cite{melzi2020intrinsic, sumner2004deformation, lakshmipathy2025kinematic}.
Beyond graphics, shape correspondence finds applications in robotics for transferring contact point annotations between objects to guide robotic arm grasping~\cite {zhu_densematcher_2024}.

\paragraph{Shape interpolation.}

Shape interpolation (or morphing) aims to generate a continuous sequence of intermediate shapes that smoothly deform a source geometry into a target.
Accurate correspondences help preserve semantic structure and geometric detail throughout the interpolation process. 
Recent approaches vary in how they incorporate correspondence: some treat it as external supervision or initialization~\cite{sang20254deform, sang2025twosquared}, others solve for correspondence and interpolation jointly~\cite{eisenberger2021neuromorph}, or couple spectral and spatial representations to improve consistency~\cite{cao_spectral_2024}.

\paragraph{Statistical shape modelling and medical applications.}

Statistical Shape Models (SSMs) capture shape variability within a category by learning low-dimensional representations from populations of 3D shapes. 
This process relies on dense correspondences to ensure consistent geometry alignment across the population.

Two broad modelling paradigms exist for building SSMs.
Template-based methods assume shared mesh topology and model deformation within a fixed coordinate system, with prominent examples including SMPL~\cite{loper2015smpl}, MANO~\cite{romero2017mano}, and 3DMM~\cite{egger20203d}. 
Recent extensions adapt this paradigm to animals~\cite{li2024learning} or anatomical skeletons~\cite{Keller_OSSO_CVPR_2022,keller2023skel} by combining template correspondences with learned statistical priors. 
In contrast, groupwise methods jointly estimate correspondences and a shape model to avoid template bias, using encoder–decoder architectures~\cite{iyer2023mesh2ssm} or functional maps~\cite{greenspan_s3m_2023,el_amrani_universal_2024}. 
In both cases, correspondence is the key component enabling statistical analysis.

A major application domain of SSMs is medical data, where correspondence establishes anatomical mappings for disease classification and population-level studies~\cite{adams_point2ssm_2023,iyer2025mesh2ssm++}. 
Functional map-based frameworks have proven effective for unsupervised correspondence estimation when mesh representations are available~\cite{klatzow2022mumatch,greenspan_s3m_2023,el_amrani_universal_2024}. 
In clinically assisted interventions, however, data may arise from partial point clouds acquired via laparoscopic imaging, ultrasound, X-ray~\cite{yang2023learning,gafencu2024shape,gafencu2025us}, or time-of-flight (ToF) depth sensors~\cite{bastian2023disguisor,wang2025beyond}. 
Template-based deformation approaches have improved robustness in applications including abdominal organs~\cite{bongratz2023abdominal}, cortical surfaces~\cite{bongratz2024neural}, and orthopaedic fracture reduction planning~\cite{YibSut_SimtoReal_MICCAI2025}. 
Atlas construction remains another important research direction~\cite{kalaie2025mdeicalAtlas}. 
For a broader overview, we refer to a survey by Zhang et al.~\cite{zhang2025survey} and the extensive MedShapeNet dataset~\cite{li2025medshapenet}.

\paragraph{Shape retrieval.}

Shape retrieval seeks to identify shapes in a database that best match a query despite variations in pose, deformation, or partiality. 
Recent methods increasingly integrate correspondence estimation into the retrieval process. 
Dense maps can define similarity metrics directly between query and candidates~\cite{di2023u}, while keypoint or part-level correspondences support retrieval with subsequent deformation~\cite{zhang2024kp,di2024shapematcher}. 
Other approaches extend correspondence-based metrics to multimodal retrieval settings~\cite{gumeli2022roca,gao2024diffcad}.  
Overall, correspondence helps retrieval focus on geometric rather than purely visual similarity.

\paragraph{Dynamic reconstruction.} 
When reconstructing a dynamic scene into a sequence of 3D frames, the underlying problem implicitly assumes dense correspondence between the frames. 
Instead of finding correspondences between image keypoints, most recent methods rely on a deformation field between a reconstructed template or an initial frame and subsequent frames~\cite{shao2023tensor4d,ebbed2025moangelo}. 
While most methods restrict correspondence to be implicit through the deformation model, it can also be used directly to impose geometric constraints, such as maintaining curvature consistency~\cite {chen2025adaptive}. 

\paragraph{Data alignment.} 
Correspondence is of significant importance for annotating 3D data. For example, it is common to decorate acquisition with an aligned template, providing a semantic meaning to otherwise unordered points~\cite{bhatnagar2022behave, wang20244d}. Correspondence can also be used to unify different datasets in a common semantic frame~\cite{melzi2019shrec}.

\paragraph{Measures and shape analysis.} 
By bringing shapes into correspondence, it is also possible to compare intrinsic properties of objects, especially in terms of measures. For example, correspondence methods are adopted in anthropometry, the science that studies the measurements and proportions of the human body \cite{huang2025shape, zhao2022skin3d}.

\section{Open challenges}
\label{sec:challenges}

Despite the extensive progress made in non-rigid shape correspondence in recent years, several fundamental challenges remain unresolved and limit its applicability to arbitrary real-world data. 

First, existing correspondence methods still struggle with non-trivial partiality and topological changes in the geometry. 
Both of these are commonly encountered in 3D scans that contain occlusions or self-intersections. 
While a few approaches explicitly tackle the partial case (see \Cref{sub:partial-to-full} and \Cref{sub:partial-to-partial}), they require extensive training data or rely on hard assumptions to work. 
Even more challenging is the case of topological changes. 
The spectral properties can change significantly under different topologies, even if the local geometry is similar~\cite{laehner2016topkids}. 
This makes the line of spectral methods incredibly sensitive to topological noise, and most papers do not even show results in this direction. 
In addition, not many datasets, and only very small ones, contain topologically noisy shapes with ground-truth correspondences. 
For this reason, supervised learning methods cannot be applied in this domain, while unsupervised methods struggle due to violations of structural assumptions arising from changes shape topology.

Another major challenge is the lack of large, diverse datasets. 
The most famous and commonly used isometric datasets, FAUST~\cite{bogo_faust_2014} and SCAPE~\cite{anguelov_scape_2005}, are synthetic; they consist only of human shapes and are generated from a fixed template. 
This leads to near-isometric and very clean meshes, resulting in benchmarks that recent methods have begun to saturate. 
On the other hand, there is no large collection of alternatives. 
This is in part due to the challenge of annotating datasets at scale; the largest existing datasets, therefore, still consist of synthetic data. 
As a result, methods are not sufficiently challenged by the lack of realism in noise and data distributions.
The implication that almost all existing data consist of clean shapes, with mostly smooth surfaces and minimal geometric irregularities, is that generalisation to real-world settings is limited. 
The recent dataset BeCoS~\cite{ehm2025beyond} made a step towards realism by simulating extreme partiality using ray casting, but also found that existing methods already struggle under much simpler transformations, such as arbitrary rotations. 
While no standardised benchmarks exist to test such cases, the findings suggest that the current generation of correspondence methods is unlikely to be widely generalisable in real-world scenarios.

\paragraph{Connecting quantitative metrics to downstream applications.}
A further challenge that cuts across all three methodological families is the disconnect between what current benchmarks measure and downstream application requirements.
Most methods are primarily evaluated on geodesic accuracy, yet applications such as deformation transfer~\cite{sumner2004deformation,vigano2025nam}, texture mapping, and statistical shape modelling~\cite{loper2015smpl,magnet2023assessing} also require smooth and neighbourhood-preserving maps.
While recent developments in unsupervised spectral methods enable the curation of datasets with geometric descriptors, the resulting pointwise maps lack such guarantees (see \cref{sec:spectral}).
Combinatorial methods offer a promising way of estimating smooth maps from these descriptors, but remain limited in scalability~\cite{roetzer2025geco,amrani2025highres}.
Closing this gap, either through scalable formulations with geometric guarantees or by employing metrics that go beyond geodesic error, remains an important open problem.
In domain-specific settings such as medical image analysis, a further issue is that purely geometric descriptors may be insufficient or prohibitively difficult to obtain. 
Incorporating prior knowledge from domain experts, such as anatomical landmarks or physiological priors, can improve the ability to produce meaningful correspondences and ultimately drive application impact~\cite{greenspan_s3m_2023,el_amrani_universal_2024,bongratz2023abdominal}.

\paragraph{Potential future directions.}
Problems with partiality, topological changes, and other noise are closely related to implicit assumptions that most methods make about the solution, such as injectivity.
It may be necessary to make a fundamental change in the problem's modelling to overcome these challenges reliably.
This holds for both optimisation- and learning-based approaches, which fall in the same categories and rely on the same basic frameworks. 
To evaluate progress in this direction, new datasets, especially large-scale ones based on real-world data, are needed to capture the diversity of possible data and avoid overfitting to the small data pool.
A potential solution might be to incorporate foundation models from other modalities, leveraging their extensive knowledge of the world to compensate for ill-posed settings and a lack of diverse training data. 
When correspondence methods become robust in real-world scenarios, they will enable a wide range of different applications with noisy data, for example, in medical settings or dynamic reconstruction, that would greatly benefit from correspondence information but cannot provide the clean meshes expected by most methods at the moment.

\section{Conclusions}
\label{sec:conclusions}
The report provides a broad overview of recent advances in non-rigid 3D shape correspondence. 
We discuss three recent branches of techniques which have been emphasised over the last few years: spectral methods based on functional maps, combinatorial formulations with discrete constraints, and deformation-based techniques. 
While the field of non-rigid 3D shape correspondence estimation has seen considerable advances in long-standing paradigms, it is now also fuelled by the recent progress in related computer vision fields and machine learning, including research on foundation models and quantum computing. 

We conclude that further progress depends substantially on the availability of new datasets that cover a wide variety of scenarios, especially for learning-based techniques that rely on large training datasets. 
We also foresee that the outlined open challenges, e.g., ~partial-to-partial shape matching, multi-shape matching, and handling different kinds of noise in the input data, such as topological changes, will keep the current community and the new generation of researchers curious and excited. 
The field is continuing to grow, and we are excited for the next generation of robust techniques and tools, broadly applicable across 3D computer vision, graphics, and related fields in diverse scenarios and applications.

\section{Acknowledgements}
This work is supported by the ERC starting grant no. 101160648 (Harmony); the DFG grant LA 5191/2-1; the UK Royal Society grant NIF/R1/254128; the DFG (German Research Foundation) project number 534951134; the ERC advanced grant no. 884679 (SIMULACRON); JSPS KAKENHI No. 21H05054, 24H00742, and 24H00748; the Federal Ministry of Research, Technology and Space of Germany and the state of North Rhine-Westphalia as part of the Lamarr Institute for Machine Learning and Artificial Intelligence.

\newpage
\bibliographystyle{eg-alpha} 
\bibliography{egbibsample_cleaned}

\end{document}